\documentclass{aa}
\usepackage[varg]{txfonts}
\usepackage{graphicx}
\usepackage{natbib}
\usepackage{xcolor}
\usepackage{ulem}
\usepackage{arydshln}
\usepackage{multirow}
\usepackage[citecolor=blue]{hyperref}
\hypersetup{colorlinks=true, citecolor=blue, urlcolor=blue, linkcolor=blue }
\usepackage{enumitem}
\usepackage{mathrsfs} 
\usepackage{url}

\begin{document}

\title{Multi-phase AGN-driven outflow in the NLSy1 IRAS~17020+4544}
\subtitle{Unveiling dual-feedback and an energy-conserving ionized outflow with MEGARA/GTC integral field spectroscopy}

\author{E. Bellocchi\inst{1,2}, A. L. Longinotti\inst{3}, Q. Salom\'{e}\inst{4},  A. Gil de Paz\inst{1,2},  J. P. Torres-Papaqui\inst{5}, Divakara Mayya\inst{6}, Y. Krongold\inst{3}, A. Castillo-Morales\inst{1,2}, A. Robleto-Or\'{u}s\inst{7,8}, C. Catalán-Torrecilla\inst{1,2}, O. Vega \inst{6}, D. Rosa González\inst{6}}

\institute{
$^1$ Departamento de F\'isica de la Tierra y Astrof\'isica, Fac. CC. F\'isicas, Universidad Complutense de Madrid, Plaza de las Ciencias, 1, Madrid 28040, Spain\\ 
$^2$ Instituto de F\'isica de Part\'iculas y del Cosmos (IPARCOS), Fac. CC F\'isicas, Universidad Complutense de Madrid, E-28040 Madrid, Spain\\
$^3$ Instituto de Astronom\'ia, Universidad Nacional Aut\'onoma de M\'exico, Ciudad Universitaria, Ciudad de M\'exico 04510, M\'exico\\
$^4$ Finnish Centre for Astronomy with ESO (FINCA), University of Turku, Vesilinnantie 5, FI-20014 Turku, Finland \\
$^5$ Departamento de Astronom\'ia, Universidad de Guanajuato Callej\'on de Jalisco S/N, Col. Valenciana CP: 36023 Guanajuato, Gto, M\'exico\\
$^6$ Instituto Nacional de Astrof\'isica, \'Optica y Electr\'onica, Luis Enrique Erro~1, Tonantzintla 72840, Puebla, Mexico\\
$^7$ Instituto de Geolog\'ia y Geof\'isica Benjamin Linder y H\'eroes de Bocay (IGG-BLyHB), Universidad Nacional Aut\'onoma de Nicaragua, Managua (UNAN-Managua), C.P. 663, Managua, Nicaragua\\
$^8$ Centro de Investigaci\'on de Astrof\'isica y Ciencias Espaciales (CIACE), Universidad Nacional Auton\'oma de Nicaragua (UNAN-Managua), C.P. 663, Managua, Nicaragua\\
\vskip1mm
\email{enrica.bellocchi@gmail.com, enricbel@ucm.es}\\ 
}

\date{Received 09 December 2025 / Accepted 13 March 2026}

\abstract
{
The narrow-line Seyfert 1 (NLSy1) galaxy, IRAS~17020+4544, is one of the few known sources exhibiting a multi-phase outflow in the highly ionized and molecular phases consistent with AGN feedback operating in the `energy-conserving' regime.
}
{
We aim to characterize the properties of the ionized warm ionized gas in IRAS~17020+4544 using new optical integral-field spectroscopic (IFS) data, and to assess the presence of outflowing ionized gas and its connection with the other gas phases and its role in the AGN feedback.
}
{
We analyze new optical seeing-limited IFS observations obtained with MEGARA at the Gran Telescopio Canarias (GTC) in both low- (R$\sim$6000; LR) and medium-resolution (R$\sim$12000; MR) modes. We model the H$\alpha$ and [OIII]$\lambda$5007 emission lines using multi-Gaussian fitting to characterize in detail the ionized gas kinematics, particularly that of the ionized outflow, to derive its energetics and compare it with those of the X-ray and molecular phases. Diagnostic diagrams (WHAN, WHaD, and BPT) are used to investigate the dominant ionization mechanism.
}
{
We identify a fast ionized outflow traced by both H$\alpha$ and [OIII] emission lines, with similar extensions (R$_\mathrm{out}$$\sim$1 kpc and $\sim$0.5 kpc, respectively) and velocities (v$_\mathrm{out}$ $\sim$ 1460 and 1240 km s$^{-1}$, respectively). A slower ionized outflow (v$_\mathrm{out}$$\sim$450 km s$^{-1}$) is also detected in the secondary component of the [OIII] line.
The fast outflow follows an `energy-conserving' regime in both H$\alpha$ and the [OIII] lines (from the LR setup), while the slower outflow follows a `momentum-driven' regime. The ionized outflows are enclosed within the molecular outflow detected with NOEMA (R$_\mathrm{CO}$=2.8$\pm$0.3 kpc), and the large momentum boosts derived in both phases suggest efficient AGN feedback, likely dominated by radiatively driven winds (quasar-mode) rather than kinetic (jet-driven) processes. Ionization diagnostics indicate that the outflow is primarily AGN-driven, although a contribution from star formation-driven excitation cannot be ruled out, and some contribution from shocks cannot be excluded on smaller scales.
}
{
Our results support a scenario where the multi-phase outflow in IRAS17020+4544 is AGN-driven and `energy-conserving' in the different (i.e., highly ionized, warm ionized and molecular) phases, efficiently coupling the AGN energy to the host galaxy's interstellar medium. The molecular outflow appears to be the dominant phase, while the ionized phase contributes less to the mass budget and feedback efficiency.
}


\titlerunning{Ionized outflow in IRAS17020+4544}

\authorrunning{Bellocchi et al.}
\maketitle

\section{Introduction}
\label{sect_Intro}

Observed correlations between host galaxy properties and central black hole activity (\citealt{Peterson08, Kormendy13}) suggest the existence of a mechanism linking nuclear-scale black hole behavior to galaxy-scale effects, the so-called `AGN feedback', which is a key component in galaxy evolution models (\citealt{DiMatteo05, Hopkins10}), yet its exact workings and triggers remain unclear.
AGN feedback is generally classified into two main modes: a radiative (or quasar) mode, in which radiation pressure from the accretion disk drives powerful multi-phase winds, and a kinetic (or jet) mode, in which mechanical energy from relativistic jets interacts with the surrounding medium. These processes give rise to AGN-driven winds and jets that launch multi-phase (i.e., highly ionized, warm ionized, neutral and molecular) outflows, constituting a primary feedback mechanism supported by multi-wavelength observations (e.g., \citealt{Cicone18, Fluetsch19, Esposito24}). As in \cite{Zubovas12}, we refer to `wind' as the mildly relativistic (v $\sim$ 0.1c) ejection of accretion disk gas from the vicinity of the SMBH resulting from Eddington accretion, and with `outflow' for the large-scale nonrelativistic flows produced by the interaction between the wind and the galaxy's ambient gas. 
These outflows may provide the connection between the black hole and its host galaxy required to reconcile theory with observations, carrying mass and energy out to larger (galactic) scales (\citealt{Silk98, Hopkins16} and \citealt{Harrison18}, and references therein).

Some models (e.g., \citealt{Faucher12, King15}) propose that feedback begins with a sub-relativistic wind launched near the accretion disk at velocities exceeding several 10$^3$-10$^4$~km~s$^{-1}$ ($\mathrm{v\sim0.01c-0.1c}$), detected as highly ionized and highly blueshifted Fe K-shell absorption lines observed in X-rays as Ultra Fast Outflows (UFOs; \citealt{Cappi06, Tombesi10, Tombesi12}). 
As these winds interact with the interstellar medium (ISM), they produce additional outflowing material, with lower ionization, observed in the optical (e.g., \citealt{Harrison14, Carniani15, Marasco20, Esposito24}) and molecular bands (\citealt{Veilleux13, Feruglio15, Cicone18, Lutz20}).

In the work by \citet{Zubovas12}, the authors present a model that explains how the different outflow phases are connected and how efficiently the wind energy is transmitted to large-scale outflows.
In their model, the transfer of wind energy strongly depends on how the wind interacts with the diffuse ISM of the host galaxy.
The wind is hypersonic and, as it propagates through the ISM, it produces a reverse and a forward shocks separated by a contact discontinuity, which defines a shocked wind and a shocked ambient medium regions on the left and right sides of the contact discontinuity (see Fig.~1 in \citealt{Zubovas12}, Fig.~7 in \citealt{King15}). 
The shocked wind gas then acts like a piston, sweeping up the host ISM at a contact discontinuity that moves ahead of it.
In brief, two possible outcomes arise depending on the cooling time of the shocked wind:
(i) if the shocked wind cools on timescales short compared to the motion of the shock patterns, the shocked wind gas is compressed to high densities and radiates away almost all its kinetic energy. This corresponds to the `momentum-driven' regime, in which the outflow cannot reach large distances due to energy losses;
(ii) if the shocked wind gas does not cool efficiently and instead expands as a hot bubble, the flow remains adiabatic and the wind retains its energy rate due to a momentum boost, which depends on the UFO velocity and the velocity of the gas phase considered (i.e., molecular or ionized). This corresponds to the `energy-conserving' regime, in which the sizes of the shocked wind and shocked ambient medium are much larger than those in the momentum-driven case and the energy is transferred up to larger distances. This latter model can explain the galaxy-wide outflows found, for instance, in quasar objects (e.g., \citealt{Feruglio10, Cicone12, Feruglio15}). 

Following \cite{King05} and \cite{Zubovas12, Zubovas14}, a multi-phase outflow can be produced as the result of thermal instabilities in the outflow that would lead to a two-phase medium and the cool gas would become molecular and form stars. 
In the works by \cite{Richings18a,Richings18b}, they showed, for the first time in their simulations, that the swept up gas from the ambient medium by the outflow is able to cool and form molecules within $\sim$1~Myr, and thus produce large molecular outflow rates (up to 140 M$_\odot$ yr$^{-1}$), supporting the {\it in-situ} molecule formation after cooling and condensation of the hotter gas.

Thus, in principle, comparing the energetics of X-ray and molecular and/or ionized outflows, can provide a test for the energy conservation and the strong coupling between the different phases of the wind (see broader discussion in \citealt{Harrison24}).
Currently, theoretical models and simulations do not include a standardized recipe for properly measuring AGN feedback (e.g., \citealt{Nelson19, Costa20, Ward24}). Observational constraints generally come from a broad range of data covering various AGN types, wavelengths, and spatial and spectral resolutions (e.g., \citealt{Harrison18, Husemann19, Wylezalek20}).

Spatially resolved observations, obtained via millimeter interferometry, optical or near-infrared integral field units (IFUs), are proving mostly adequate to determine the spatial scales and geometry of outflows (e.g., \citealt{Cicone14, Husemann19, Mingozzi19, Davies20, Kakkad20, Venturi23, Harrison24}), at least in local AGN.
Among key findings, surveys like GATOS\footnote{\url{https://gatos.myportfolio.com/home}} (Galactic Activity, Torus, and Outflow Survey), which focus on hard X-ray selected Seyfert galaxies, revealed molecular outflows in ALMA data for sources exhibiting significant nuclear cold gas deficiency (e.g., \citealt{AAH21, GBS21, GB24, Esposito24}). These results support the idea that outflows efficiently expel gas from the central regions.
Regarding ionized gas, the MAGNUM survey (e.g., \citealt{Mingozzi19, Venturi21}), conducted with MUSE/VLT on local Seyfert galaxies hosting low-power radio jets, has shown that besides a clear ionized outflow component aligned with ionization cones and radio jets along the galaxy disk, there exists an additional high velocity, kpc-scale region of excited gas extending perpendicular to this direction. This feature likely results from the jet perturbing the gas in the disk.  
Recent James Webb Space Telescope (JWST) IFU observations have now uncovered similar phenomena in both high-redshift and local AGN. For instance, in XID2028 ($z=1.59$), JWST/NIRSpec revealed an expanding emission-line bubble likely driven by the interaction between the ISM and the radiation-driven outflow and low-power radio jet (\citealt{Cresci23}). 
In the nearby Seyfert 1.5 galaxy NGC~7469, \cite{Armus23}, using JWST/MIRI MRS data, detected blueshifted high-ionization ([Mg V]) lines reaching velocities of $\sim$1700 km s$^{-1}$ cospatial with the starburst ring.  They found that the width of the broad emission and the broad-to-narrow line flux ratios correlate with ionization potential, suggesting the presence of a decelerating, stratified, AGN-driven outflow emerging from the nucleus (see also \citealt{Robleto21}, who characterized the warm ionized outflows in NGC 7469 with MUSE).

Conversely, in the recent years there have been also studies of AGN samples with a well-established outflow component in a single band. For instance, \cite{Salome23} reported on the molecular gas properties and higher star formation efficiency in a collection of 10 Narrow Line Seyfert 1 Galaxies (NLSy1) selected to have a solid detection of X-ray UFO. 
Among their many peculiarities, NLSy1 galaxies are known for their relatively small H$\beta$ FWHM (i.e., $<$2000 km s$^{-1}$; \citealt{Goodrich89}) compared to other Sy1 galaxies, which suggests low black hole masses (M$_\mathrm{BH}$$\sim$10$^{6}$-$10^{8}$ M$_\odot$; e.g. \citealt{Peterson00}). They also typically exhibit strong FeII and weak [OIII] emission (\citealt{Osterbrock85}), as well as pronounced X-ray variability and steep X-ray power-law continua (e.g. \citealt{Panessa11}).
Furthermore, these sources are characterized by significantly higher accretion rates compared to standard Seyfert galaxies (between L/L$_\mathrm{Edd}$ $\sim$0.1-1; \citealt{Boroson92}), indicating an extreme central environment, that may suggest that NLSy1 galaxies are in the early phase of the evolution (i.e., \citealt{Mathur01, VC01}).

 X-ray UFO with complex pattern have become fairly common in NLSy1 objects (\citealt{Xu24}), a fact interpreted as a natural consequence of their high accretion rate that enable the launching of disk winds via radiatively-driven mechanism (\citealt{Matzeu23}). 

There are also uncountable observational studies of individual AGN showing multi-band outflows that are greatly contributing to advance our understanding of the outflow phenomenology and the interplay of the different outflowing gas phases (e.g., \citealt{Feruglio15, Morganti15, Fluetsch21, Longinotti23, Esposito24}).

The study presented herein is somewhat one of this kind. The radio-loud NLSy1 galaxy IRAS17020+4544\footnote{This galaxy, classified as a Luminous Infrared Galaxy (LIRG) and lies above the star-forming Main Sequence (MS; \citealt{Salome23}).} (IRAS17 throughout) was the first AGN where a multi-component UFO was identified using XMM-Newton observations (\citealt{Longinotti15}), along with a complex pattern of slower X-ray velocity absorbers that were interpreted as the signature of the UFO  shocking the ISM (\citealt{Sanfrutos18}). Multi-band follow-up studies have uncovered the presence of outflows in every single bandwidth: Very Long Baseline Interferometry (VLBI) studies (\citealt{Giroletti17}) revealed elongated synchrotron emission on $\sim$10~pc scales possibly indicating that this low-power radio jet contributes to shock-driven interactions with the ISM. \cite{Longinotti18} reported a powerful `energy-conserving' molecular outflow in single-dish observations obtained by the Large Millimeter Telescope (LMT). The UV study carried out on HST-COS data by \cite{Mehdipour23} identified Ly$\alpha$ absorption at 23400 km~s$^{-1}$ as the counterpart of the X-ray UFO, a rare finding interpreted as gas entrainment in the shocked wind. 
Interferometry observations with the NOrthern Extended Millimeter Array (NOEMA) provided the first spatially resolved view of the host galaxy of this interesting AGN:  based on these data, \cite{Salome21} revealed an interaction with a satellite galaxy located up to $\sim$8 kpc north of the nucleus, and \cite{Longinotti23} reported a biconical outflow structure in CO gas extending out to a radius of $\lesssim$3~kpc, with velocities ranging from 1000 to 1800~km~s$^{-1}$, and fully consistent with `energy-conserving' regime, confirming therefore the result obtained by the LMT.

Within the framework of the MATRIOSKA\footnote{This project stands for “Multiphase nAture of ulTra-fast outflows in naRrow lIne seyfert 1: the Optical Survey and Kinematic Analysis.” See more details at \url{https://guaix.ucm.es/home/science/nearby-galaxies}.} project, we aimed to characterize the ionized gas phase of the outflowing component in IRAS17 using the {\it Multi-Espectr\'ografo en GTC de Alta Resoluci\'on para Astronom\'ia} (MEGARA; \citealt{AGdP18, Carrasco18}) mounted on the Gran Telescopio Canarias (GTC).
Our observations enabled a detailed spatially resolved (2D) kinematic study of the warm ionized gas, traced by the H$\alpha$-[NII]$\lambda\lambda$6548,6583 and H$\beta$-[OIII]$\lambda\lambda$4959,5007 complexes. This study allowed us to assess the presence of the ionized outflowing gas, its connection with other phases (i.e., whether it still follows the previously reported `energy-conserving' regime for the molecular component), and its overall role in AGN feedback.

This paper is organized as follows. In Sect.~\ref{obs_data}, we present the observations and data reduction. In Sect.~\ref{sect_data_analysis}, we analyze the ionized gas kinematics. In Sect.~\ref{sect_results}, we present the kinematic results of the ionized gas phase traced by different emission lines (i.e., H$\alpha$ and [OIII]), deriving key physical parameters of the outflow. In Sect.~\ref{sect_discussion}, we discuss our results, attempt to connect the different outflow phases, and assess the global impact of the outflow on the host galaxy. In Sect.~\ref{sect_summary} we summarize our main findings and present our conclusions. In App.~\ref{more_info_AV}, we provide details on the derivation of the visual extinction. In App.~\ref{more_maps}, we present additional kinematic maps derived for other emission lines observed in the different setups. Finally, in App.~\ref{app_companion}, we report new results on the ionized-gas emission from the companion galaxy, which is thought to be interacting with IRAS17.

Throughout this paper, we assume a $\Lambda$DCM cosmology with H$_0$ = 70 km s$^{-1}$ Mpc$^{-1}$, $\mathrm{\Omega_M} = 0.3$, and $\mathrm{\Omega_\Lambda} = 0.7$. Using the E. L. Wright Cosmology calculator, which is based on the prescription given by \cite{Wright06}, we used a a luminosity distance, D$_\mathrm{L}$, of 274.3 Mpc, and an angular diameter distance, D$_\mathrm{A}$ = 243.5 Mpc, as derived in \cite{Salome21}. This corresponds to a scale of 1.181 kpc/\arcsec at a redshift $z = 0.0612$.

\section{Observations \& data reduction}
\label{obs_data}

\subsection{MEGARA/GTC observations}
\label{obs_megara}

The data were gathered as part of two different programs and at different epochs (GTC8-20AMEX and GTC1-24AMEX; PI:~Longinotti A.; see Table~\ref{Input_obs}) using the MEGARA\footnote{\url{https://www.gtc.iac.es/instruments/megara/media/MEGARA_cookbook_1I.pdf} } instrument. 
MEGARA is a high-resolution optical spectrograph mounted on the 10.4 m GTC telescope with two observing modes: (i) the integral field unit (IFU), or large compact bundle (LCB), mode, which covers a field of view (FoV) of 12.5 $\times$ 11.3 arcsec$^2$ (i.e., $13.6 \times 14.5$~kpc$^2$ at the distance of our object) with a spatial resolution of 0.62\arcsec, and (ii) the multi-object spectroscopy (MOS) mode, which allows to observed up to 100 objects (8 for the sky and 92 for the observations) in a region of 3.5 arcmin $\times$ 3.5 arcmin around the IFU bundle. Both the IFU and MOS capabilities of MEGARA provide intermediate-to-high spectral resolutions (R=$\lambda$/FWHM$\sim$6000, 12000 and 20000 for the LR, MR and HR modes, respectively). 
In this work we used the MEGARA-IFU, which uses 567 fibers in a hexagonal tessellation. Additional 56 fibers are also used to simultaneously measure sky-background spectra with the observation of the scientific target to subtract them from the observations afterwards. 
MEGARA has 18 different volume-phase holographic (VPH) transmission gratings that allow covering a range from 3650 to 9700 \AA\ with the three modes. 

\begin{figure}
\hskip-3mm\includegraphics[width=0.5\textwidth]{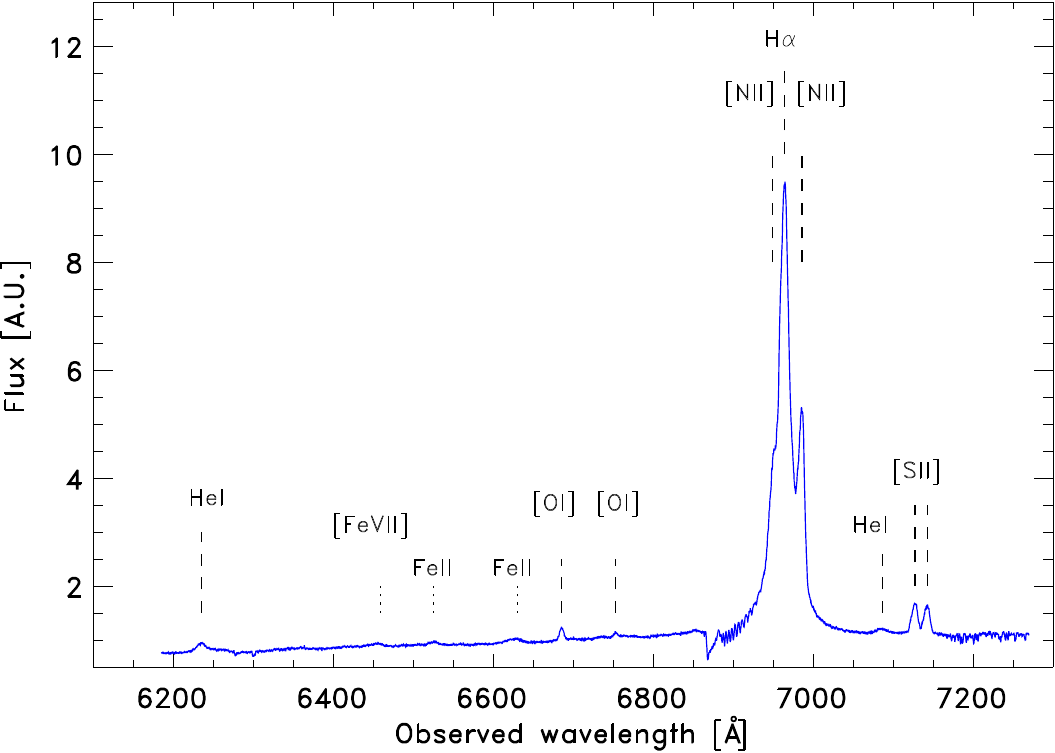}
\caption{
Observed integrated spectrum of the inner region (R $\sim$~1.9~kpc $\approx$~1.6\arcsec) obtained with the LR-R setup. Several emission lines are detected, including the prominent H$\alpha$-[NII] complex, the fainter [SII] and [OI] doublets, as well as very faint [FeVII], FeII and HeI lines.
}
\label{total_int_spectrum}
\end{figure}

For our study, we obtained both low- (LR-V and LR-R) and medium-resolution (MR-G) data, which allowed us to cover the spectral ranges 5140-7300 \AA\ and 4960-5445 \AA, respectively.
In particular, the VPH gratings used have the following characteristics: (i) LR-V: wavelength range of 5140-6170~\AA\ with an average resolution R=6080 (i.e., 0.27~\AA\ pix$^{-1}$); (ii) LR-R: wavelength range of 6100-7300~\AA\ with an average resolution R=6100 (i.e., 0.32 \AA\ pix$^{-1}$); (iii) MR-G: wavelength range of 4970-5445~\AA\ with an average resolution R=12035 (i.e., 0.13~\AA\ pix$^{-1}$). These gratings cover relevant spectral features, such as the H$\alpha$-[NII] complex, the [OI]$\lambda\lambda$6300,6363, and [SII]$\lambda\lambda$6717,6731 doublets in the LR-R setup (see the nuclear spectrum in Fig.~\ref{total_int_spectrum}), while H$\beta$-[OIII]$\lambda\lambda$4959,5007 complex in the LR-V and MR-G setups. 
Due to the lower signal-to-noise ratio (SNR) obtained in the previous MR-G observations, we recently requested additional MEGARA IFU data in the LR-V setup to spatially resolve and better characterize the ionized gas using the [OIII]$\lambda\lambda$4959,5007 doublet emission. Indeed, although the LR-V data have lower spectral resolution than the MR-G setup and the H$\beta$ emission line falls at the edge of the blue part of the spectrum (making it unusable), these data provide higher S/N per spaxel for the [OIII] doublet, enabling the detection of fainter components.  

\subsection{Data reduction}

MEGARA data were reduced following the MEGARA Data Reduction Pipeline (DRP; v0.12.0; \citealt{Pascual22}), entirely written in a python environment. 
The data reduction was performed according to the instructions provided in the MEGARA cookbook (i.e., \citealt{Castillo20}\footnote{\url{https://zenodo.org/records/3932063}}). 
We briefly describe the main steps of the data reduction process, which include: (1) the sky and bias subtraction; (2) the flat field correction; (3) the spectra tracing and extraction; (4) correction of fiber and pixel transmission; (5) wavelength calibration and (6) flux calibration (using the spectrophotometric standard star observed each night). For a in-depth description of the data reduction we refer the reader to the work by \cite{Chamorro23}. 

We finally transformed the final raw-stacked spectra (RSS\footnote{The `Final Raw Stacked Spectra' are the end product of the data reduction process, with sky emission already subtracted. They are 2D images containing spectra from 623 fibres, whose spaxels have a hexagonal shape.}) file to a standard IFS data cube by applying a regularization grid to obtain a resample spaxel size of 0.4\arcsec\ (corresponding to a physical size of 0.472 kpc).
The point spread function (PSF) of the MEGARA data cube can be estimated from the FHWM of the 2D profile brightness distribution of the standard star used for flux calibration (HR8634 and HR4963), which gives a FWHM of 1.2\arcsec and 0.9\arcsec, for the grisms LR-R and MR-G, and LR-V, respectively.

\begin{table*}[!h]
\centering
\caption{MEGARA VPH gratings and observation details used in this work.}
\label{Input_obs}
\resizebox{\textwidth}{!}{%
\begin{tabular}{cccccccccc} 
\hline\hline\noalign{\smallskip}  
Program &\multicolumn{1}{c}{Grism} & Spectral coverage & R. L. D. & R & Observing Date & t$_{\rm exp}$ & Airmass & Seeing & Atm. Conditions \\
 & 	&	 [\AA] & [\AA\ pix$^{-1}$] &  & & [s] &  & [\arcsec] &  \\
(1) & (2) & (3) & (4) & (5) & (6) & (7) & (8) & (9) & (10)\\
\hline\noalign{\smallskip} 
GTC8-20AMEX & LR-R  (VPH675$\_$LR) &  6100-7300 	& 0.32 	& 6100	&	23 June 2020 & 3$\times$1000 & 1.31 & 1.2& Clear \\
GTC8-20AMEX & MR-G (VPH521$\_$MR) &  4970-5445	& 0.13	&12035	&	23 June 2020 & 3$\times$1000 & 1.16 & 1.2 & Clear \\
GTC1-24AMEX & LR-V (VPH570$\_$LR) &   5140-6170 & 0.27 	& 6080	&	11 May 2024 & 6$\times$1120 & 1.05 & 0.9 & Clear \\
\hline\noalign{\smallskip} 	
\end{tabular}
}
\begin{minipage}{18cm}
{\bf Notes}: Column (1): Observing program; Column (2): VPH grating ID; 
Column~(3): Spectral coverage of the setup; 
Column~(4): Reciprocal linear dispersion; 
Column~(5): Spectral resolution;
Column~(6): Observing date; 
Column~(7): Exposure time; 
Column~(8): Airmass measured at the beginning of the observation; Column~(9): Seeing (FWHM); Column (10) Atmospheric conditions of the night. \\ 
\end{minipage}
\end{table*}

\begin{figure}[!h]
\centering
\includegraphics[width=0.4\textwidth]{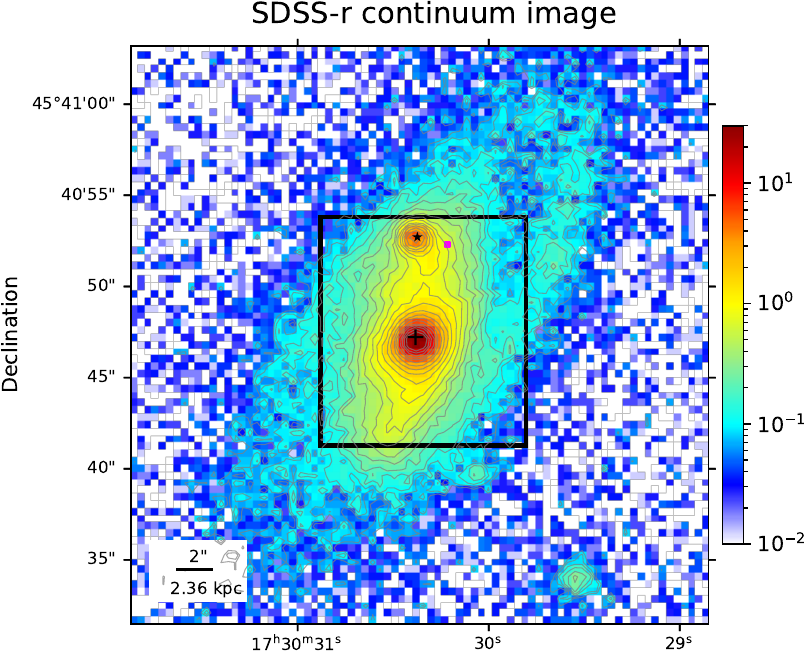}\vskip3mm
{\includegraphics[width=0.405\textwidth]{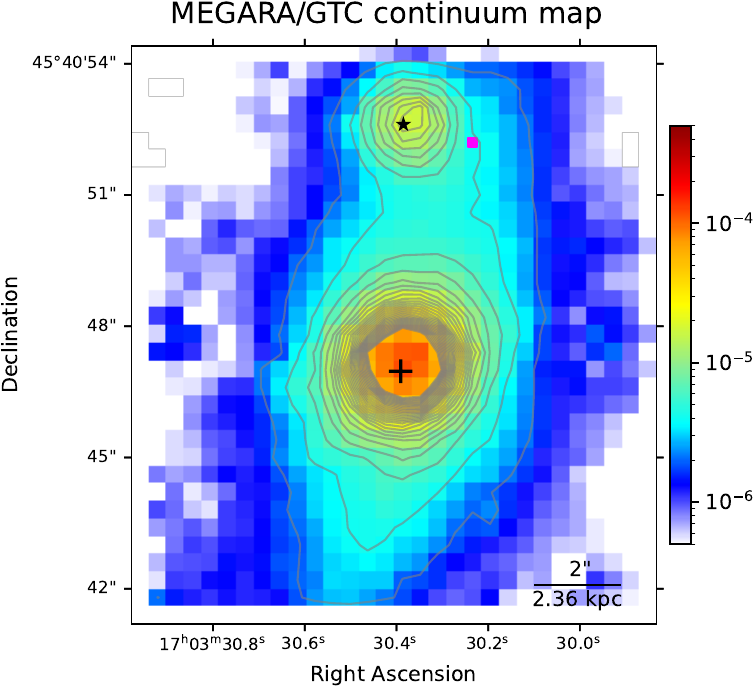}}  
\caption{{\it Top:} SDSS-r continuum image of IRAS17. The MEGARA-IFU FoV is indicated by a black rectangle. A foreground star (marked with a black star) is visible to the north along with the companion galaxy (indicated by a magenta square) detected by \cite{Salome21}. The flux intensity map is shown in nanomaggy units (1 nanomaggy = 3.631 $\mu$Jy). {\it Bottom:} Continuum image obtained with MEGARA-IFU within the wavelength range 6770-6840~\AA. The flux intensity map is shown in Jy. North is up and east is to the left in both images. The continuum peak emission is marked with a black cross in both panels and corresponds to the AGN position, as supported by its coincidence with the optical AGN position measured by Gaia, the radio emission detected with e-MERLIN at 1.5 GHz (\citealt{Longinotti23}), and the CO peak emission (\citealt{Salome21}).
}
\label{cont_maps}
\end{figure}


\section{Data analysis}
\label{sect_data_analysis}

\subsection{Optical continuum maps}

IRAS17 has been extensively studied at millimeter (\citealt{Longinotti18, Longinotti23}) and X-ray wavelengths (\citealt{Longinotti15, Sanfrutos18}), but it has received comparatively little attention in the optical regime (e.g., \citealt{deGrijp92}). 
In Fig.~\ref{cont_maps}, we show the continuum image obtained with the SDSS instrument in the $r$-band, which was used as a reference for the astrometric alignment, since its spatial resolution is similar to that of the MEGARA data, allowing an accurate registration of the features.

The optical continuum image derived from the MEGARA-IFU data cube was extracted over the observed wavelength range 6770-6840~\AA\ (Fig.~\ref{cont_maps}, bottom panel). In the SDSS image, we can clearly identify the nuclear emission of IRAS17, its spiral arms, and the presence of a foreground star in the northern part of the FoV.
Furthermore, a faint emission feature located just below the star, together forming a ‘comma-shaped’ structure, is visible in the SDSS image, although it is only marginally detected in the MEGARA continuum map, where the contours appear elongated toward the southwest.
We associate this feature with the newly identified companion galaxy reported by \citet{Salome21}, which resembles a tidal-tail-like structure that appears to connect the companion with IRAS17, suggesting a past interaction between them.

Throughout this paper, we consider the peak of the continuum emission as the photometric center (AGN position) of the galaxy. This choice is supported by the spatial coincidence between the optical continuum peak and the optical AGN position measured by Gaia, the radio emission detected with e-MERLIN at 1.5 GHz (\citealt{Longinotti23}), and the CO peak emission (\citealt{Salome21}). All these measurements are consistent within their respective uncertainties.

\subsection{Multi-component fit and map generation}
\label{multi_fit_line}

In our analysis, we did not subtract the underlying stellar emission of the host galaxy in the MEGARA spectra, as the absence of typical stellar feature in the spectra (and the limited wavelength coverage in the MR-G setup) prevent effective modeling of the stellar continuum. As in Sy1 galaxies, we have minimal starlight contamination, since the continuum is largely dominated by the AGN. 
However, since the main goal of this work is to characterize the kinematics of the ionized outflow in the galaxy, this limitation does not impact our key results. As discussed in \citet{Bellocchi19a}, while continuum subtraction is important for accurately recovering line fluxes as well as the derivation of the equivalent widths, it is not crucial for kinematic measurements. Therefore, we directly fit the observed emission-line profiles to derive the kinematic properties.

The emission features in each spectral band, such as the H$\alpha$-[NII] complex in the LR-R grism and the H$\beta$-[OIII] complex in the MR-G and LR-V grisms, were analyzed using Gaussian profile fitting (see example in Fig.~\ref{nuclear_3c_LRR_MRG}). This was performed with the IDL routine {\tt MPFITEXPR} \citep{Markwardt09}, which employs a Levenberg-Marquardt least-squares minimization to optimize the fit parameters in each spatial element (spaxel).
To properly reproduce the observed line profiles in both complexes, we employed a multi-component Gaussian fitting approach (e.g., \citealt{Bellocchi13, Bellocchi19b}). This method allowed us to distinguish between a primary (narrow) component and a secondary (broader) one based on their velocity dispersions. In our case, the secondary emission component can even dominate in flux over the systemic one (i.e., H$\alpha$ emission).
In some spaxels, located in the innermost region, a tertiary component was required. This tertiary  component is the broadest one, and it clearly traces a more extreme (i.e., fast) ionized outflow in both the H$\alpha$ and [OIII] lines (see Sect.~\ref{sect_results} for details). Thus, in each spaxel, when multiple Gaussian components are fitted, the different components (i.e., primary (P), secondary (S) and tertiary (T)) are labelled based on their velocity dispersion, from the narrowest to the broadest, with the narrowest component generally tracing the systemic emission.
As a first step, we performed initial fits using a single Gaussian profile per line (hereafter referred to as the one-component, or 1c, model). However, in the central regions (R$<$2 kpc), where line profiles are more complex, multi-component fits (two or three Gaussians per line, 2c or 3c) were clearly required, first based on visual inspection, to adequately reproduce the observed features. We only considered spaxels with S/N$>$4, a threshold that holds for each individual component.
The S/N was computed by dividing the integrated flux in the velocity channels by the corresponding noise, estimated from the standard deviation of the data-model residuals of the fit around each line (within a range about 60 \AA\ wide).

To avoid overfitting, we adopted the $\varepsilon$-criterion introduced by \citet{Cazzoli22}, which compares the standard deviation of a line-free continuum region ($\varepsilon_{cont}$) near the line of interest with the standard deviation of the residuals under the fitted line profile ($\varepsilon_{line}$). For the H$\alpha$-[NII]  complex, the residuals were computed across all three lines to properly account for blending in spaxels with broad profiles. In the case of the [OIII] doublet, we excluded the H$\beta$ line and calculated the residuals using only the two [OIII] lines. 
We applied a threshold of $\varepsilon_{line}/\varepsilon_{cont}\geq1.5$ to justify the inclusion of additional components. This criterion, supported by visual inspection of asymmetries in the line profiles, motivated the use of a second component and, in some inner spaxels, a tertiary one, especially when residuals from the 2c fit remained significant (i.e., still applying the condition $\varepsilon_{line}/\varepsilon_{cont}\geq1.5$).

The line widths $\sigma$ of each line is constrained to be broader than the instrumental resolution ($\sigma_{\mathrm{INS}}$) and the observed velocity dispersion ($\sigma_{\mathrm{obs}}$) was corrected for instrumental broadening to derive the intrinsic velocity dispersion, $\sigma_{\mathrm{line}}$, by subtracting $\sigma_{\mathrm{INS}}$ in quadrature, following the relation $\sigma_{\mathrm{line}} = \sqrt{\sigma_{\mathrm{obs}}^2 - \sigma_{\mathrm{INS}}^2}$ (with $\sigma_{\mathrm{INS}} \sim$ 22.2~km s$^{-1}$ for the LR setups and $\sim$ 11.2 km s$^{-1}$ for the MR setup).

For each emission line, we derived flux intensity, velocity field,\footnote{We corrected for the heliocentric velocity, using $v_{\rm helio}$$\sim$~-5.6~km~s$^{-1}$ for the LR-R and MR-G setups, and $v_{\rm helio} \sim 2.4$~km~s$^{-1}$ for the LR-V setup.} and intrinsic velocity dispersion maps (see Figs.~\ref{fig_LR_R_setup}, \ref{fig_LR_R_setup_NII}, \ref{fig_Hb_MR_LR_new}, \ref{fig_OIII_MR_LR_new}, \ref{LRV_setups_maps} and panels in App.~\ref{more_maps}).
A comprehensive analysis of the kinematic maps as well as their physical interpretation, is presented in the next sections.

\subsubsection{LR-R setup for the H$\alpha$-[NII] complex}

\begin{figure}[!h]
\centering
\includegraphics[height=0.35\textwidth]{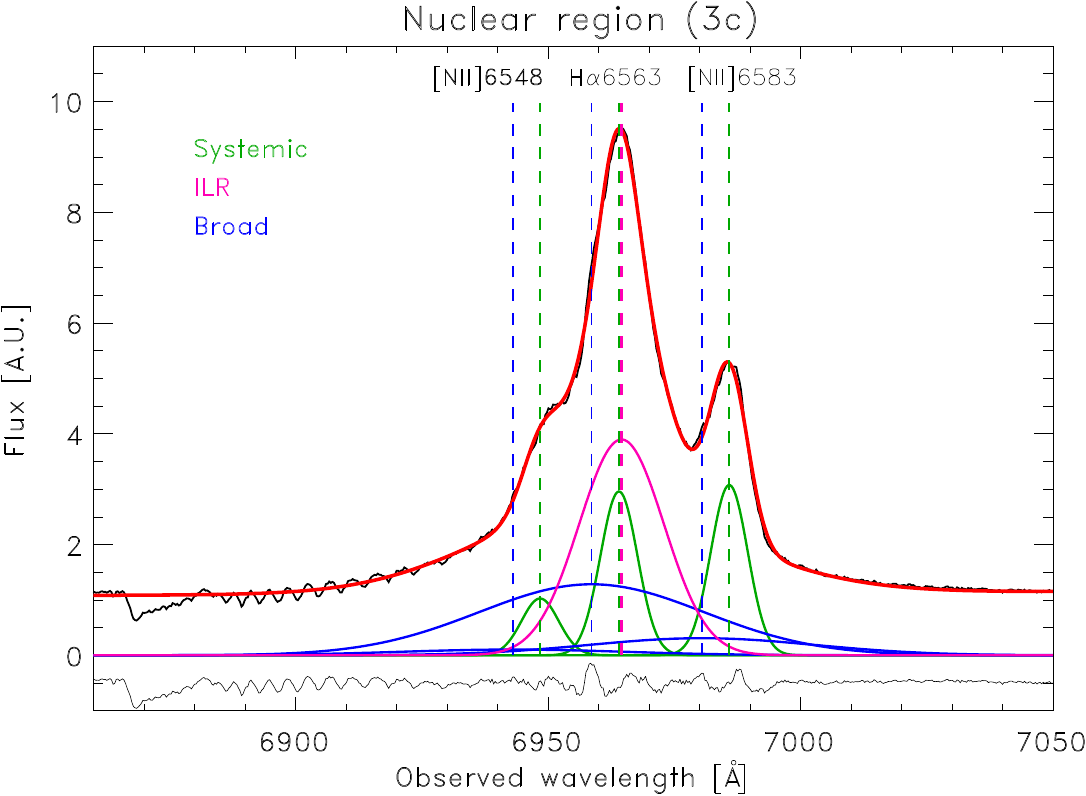}
\vskip5mm\hskip-3mm
\includegraphics[height=0.35\textwidth]{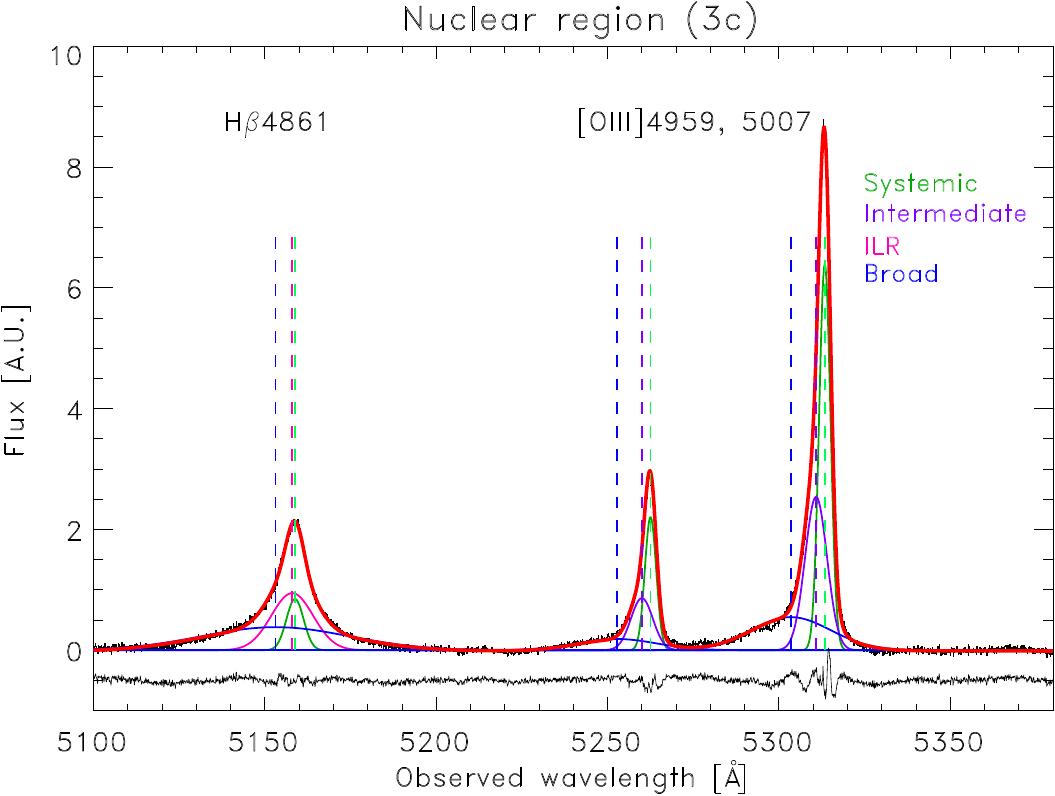}
\caption{Nuclear integrated spectra (within the PSF region) of the H$\alpha$-[NII]$\lambda\lambda$6548,6583 ({\it top panel}) and H$\beta$-[OIII]$\lambda\lambda$4959,5007 complexes ({\it bottom panel}), fitted with three kinematic components. For the Balmer lines, the green, magenta, and blue curves represent the systemic, ILR, and broad (outflowing) components, respectively. For the [OIII] lines, the green, purple, and blue curves represent the systemic, intermediate (slow outflow), and broad (fast outflow) components.
}
\label{nuclear_3c_LRR_MRG}
\end{figure}

For the LR-R setup, we generally fitted the H$\alpha$-[NII] complex with two kinematic components (i.e., primary and secondary), tying the velocity and velocity dispersion of H$\alpha$ and both [NII] lines to minimize degeneracies and mitigate the effect of telluric absorption near [NII]6548\footnote{We found that telluric absorption barely affects ($\lesssim$10\%) the [NII]6548 emission line.}. For the [NII] doublets, we fixed the [NII]$\lambda\lambda$6583/6548 flux ratios to $\sim$3 (e.g., \citealt{Osterbrock06}).
In the innermost region of the FoV, however, an additional intermediate-width H$\alpha$ component was required to account for the emission from the intermediate-broad line region (ILR), as expected in a NLSy1 galaxy (see Fig.~\ref{nuclear_3c_LRR_MRG}, top panel). In this region, we associated the ILR with the secondary component, whereas the broadest and most blueshifted component was identified as the ionized outflow (see Table~\ref{tab_interpretation}).

\subsubsection{MR and LR setups for the H$\beta$-[OIII] complex}
\label{MR_LR_}

\begin{figure}[!h]
\centering
\includegraphics[width = 0.48\textwidth]{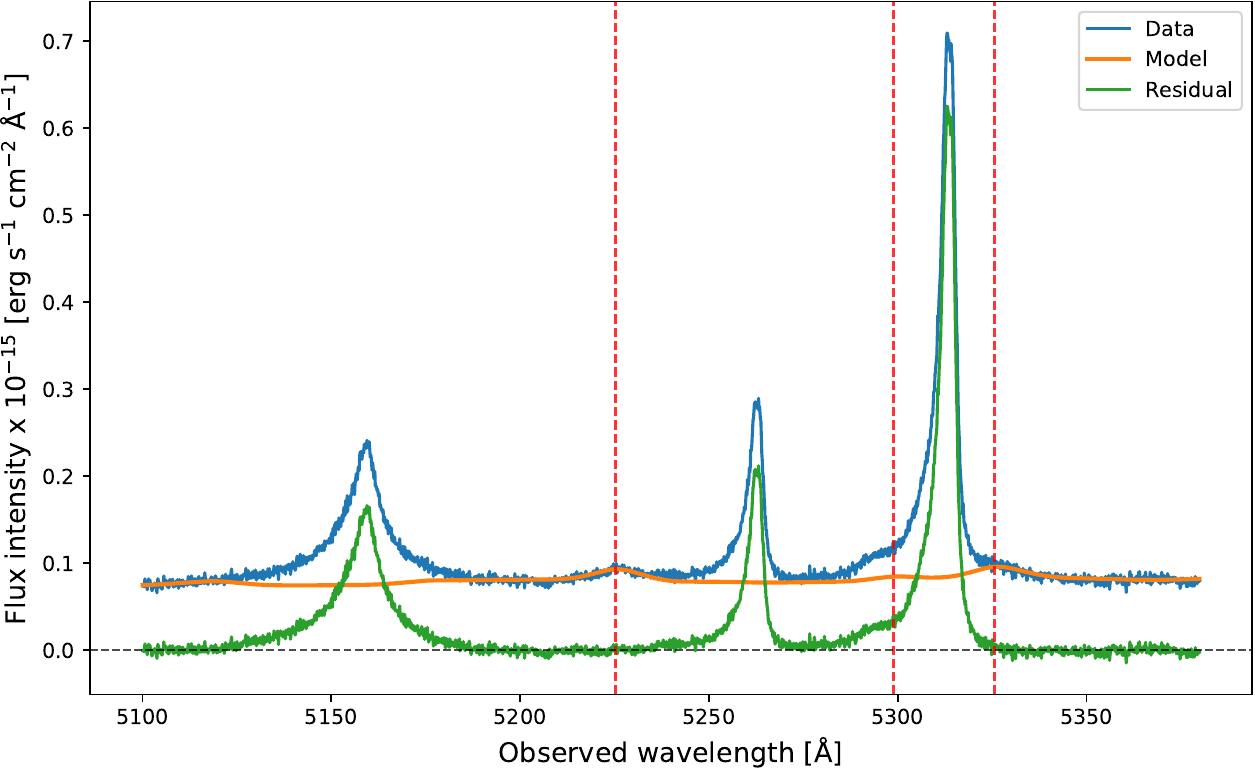}
\caption{FeII template subtraction from the observed H$\beta$-[OIII] complex at the position of the emission peak in the MR-G setup (within a single spaxel of 0.4\arcsec$\times$0.4\arcsec). The observed spectrum is shown in light blue, the FeII template in orange, and the residuals (pure H$\beta$-[OIII] complex) in green. The red dashed vertical lines mark the iron emission lines in this wavelength range (i.e., rest-frame wavelengths: $\lambda_{\mathrm{Fe\,II}} = 4923.92$, 4993.35, and 5018.45~\AA).}
\label{Fe_template_OIII}
\end{figure}

For the MR-G and LR-V setups, we first subtracted the FeII emission template (see Fig.~\ref{Fe_template_OIII}) by fitting a template for the multiple iron lines, adding an AGN power-law component ($\mathrm{f_\lambda \propto \lambda^\beta}$). We used the new template constructed by \cite{Park22}, based on a spectrum of the NLSy1 galaxy Mrk~493 obtained with the Hubble Space Telescope (e.g., \citealt{Greene07}).
Compared to the canonical FeII template of I Zw 1 constructed by \citet{VC04} (see also \citealt{TorresPapaqui20}), Mrk~493 exhibits narrower broad-line widths, lower reddening, and a less extreme Eddington ratio, $\lambda_{\mathrm{Edd}}$ (i.e., $\lambda_{\mathrm{Edd}}\sim 2.5$ for I~Zw~1 {\it versus} $\sim0.5$ for Mrk~493), more similar to that of IRAS17 ($\lambda_{\mathrm{Edd}}\sim0.7$). The fitting method consists of scaling and velocity-broadening the FeII template to match the original MEGARA spectrum (see also \citealt{Greene05} for details). Figure~\ref{Fe_template_OIII} shows an example of the FeII template fitting in the MR-G setup.

After subtracting the FeII emission template, we fitted the H$\beta$-[OIII] complex independently in the MR-G setup\footnote{In the LR-V setup, the H$\beta$ line could not be analyzed reliably.}, motivated by the fact that these lines may trace gas with different kinematics (see also \citealt{Cazzoli22, Peralta23, Esposito24, HM24}). In this approach, the kinematics of the H$\beta$ components were allowed to vary independently from those of the [OIII] doublet, as tying their kinematics was previously tested and found to provide not very good fits to the three lines, because H$\beta$ may trace partially different gas phases not present in [OIII], such as the ILR (see Fig.~\ref{nuclear_3c_LRR_MRG}).

In particular, we modeled the [OIII]$\lambda\lambda$4959,5007 doublet with three distinct components: a narrow primary component tracing the galactic rotation (systemic component), and intermediate and broad components. The kinematics of these lines were tied, and their flux ratio was fixed according to atomic physics (i.e., [OIII]$\lambda\lambda$5007/4959$\sim$3; \citealt{Osterbrock06}).

H$\beta$, observed in the MR-G setup\footnote{In the LR-V setup this was not possible because the data quality of the H$\beta$ emission is insufficient for a reliable fit (see Sect.~\ref{obs_megara}).}, was also fitted with three components following the H$\alpha$ kinematic structure (i.e., a narrow primary component, an intermediate-width component, and a broad component associated with the outflow). 
Upper limits on the H$\beta$ component widths were set to the average H$\alpha$ widths of the corresponding components to ensure physically consistent broad-line properties (see Table~\ref{tab_interpretation}).

\begin{table}[!h]
\centering
\caption{Kinematic interpretation of the multi-Gaussian components.}
\label{tab_interpretation}
\begin{tabular}{cccc} 
\hline\hline\noalign{\smallskip}  
  Line & Gaussian component &  Kinematic \\
  & & Interpretation \\
(1) & (2) &  (3)  \\
\hline\noalign{\smallskip} 
H$\alpha$, H$\beta$	 & primary & systemic \\
H$\alpha$, H$\beta$ & secondary	& ILR(+turbulence)\\
H$\alpha$, H$\beta$ & tertiary  		& (fast) outflow\\ 
\hline\noalign{\smallskip} 
$[\mathrm{NII}]6583$ & primary	 &  systemic \\
$[\mathrm{NII}]6583$ & secondary	 &  outflow(+turbulence)\\
\hline\noalign{\smallskip} 	
$[\mathrm{OIII}]5007$ & primary	 &  systemic \\
$[\mathrm{OIII}]5007$ & secondary	 &  (slow) outflow\\
$[\mathrm{OIII}]5007$ & tertiary	 &  (fast) outflow\\
\hline\noalign{\smallskip} 	
$[\mathrm{SII}]6731$ & primary	 &  systemic\\
\hline\noalign{\smallskip} 	
$[\mathrm{OI}]6300$ & primary	 &  systemic\\
\hline\hline\noalign{\smallskip} 	
\end{tabular}
\begin{minipage}{8.7cm}
{{\bf Notes}: Column (1): Emission line considered. Column (2): Gaussian component used in the fit. Column (3). Kinematic interpretation of each component for the different lines (see text for details).}
\end{minipage}
\end{table}

\section{Main results: gas kinematics}
\label{sect_results}

\begin{figure}[!h]
\centering
\includegraphics[width=0.5\textwidth]{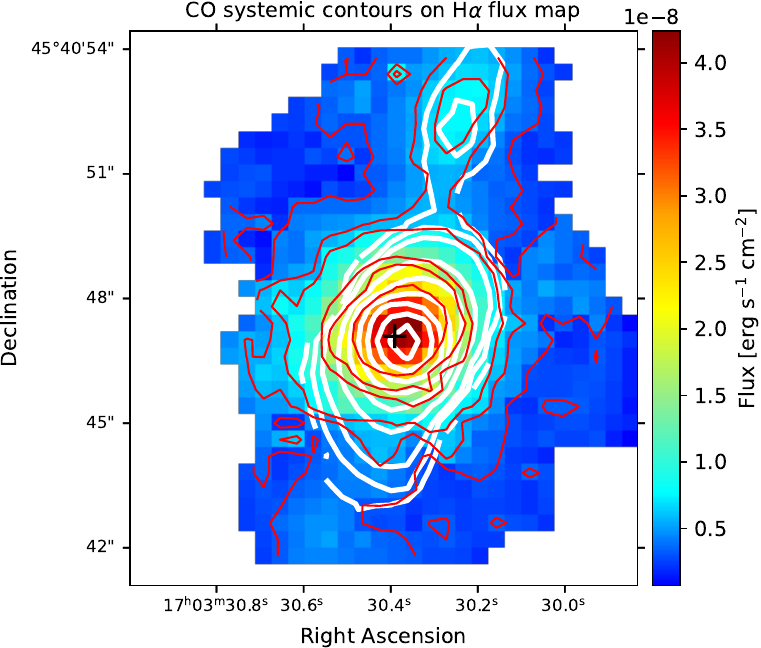}
\caption{H$\alpha$ flux map of the primary component (red contours), with overlaid $^{12}$CO(1-0) contours of the systemic emission (white) from \citet{Salome21}. The molecular and ionized gas in IRAS17 and its companion galaxy, traced on the north-west side by the CO contours, exhibit a remarkable spatial overlap.}
\label{Ha-CO_flux}
\end{figure}

As recently discussed by \citet{Salome21} and \citet{Longinotti23}, two sources can be identified within the MEGARA FoV alongside our target, IRAS17:
(i) a foreground star to the north, clearly visible in all optical continuum images (Fig.~\ref{cont_maps}); and (ii) a companion galaxy, located slightly to the north-west of the star, first detected via CO molecular gas emission by \citet{Salome21}, which is not immediately apparent in the optical image presented here (Fig.~\ref{cont_maps}).
Our MEGARA/GTC data further reveal that the companion galaxy also exhibits H$\alpha$ emission in the same region where the CO(1-0) emission was initially detected (Fig.~\ref{Ha-CO_flux}; see next section for details).

\subsection{H$\alpha$-[NII] complex}
\label{description_Ha}

Within the LR-R setup, we detected the bright H$\alpha$ and [NII] emission-line complex.
The H$\alpha$ emission was fitted using three components in the inner region (see spectra in Fig.~\ref{nuclear_3c_LRR_MRG}, top), each corresponding to distinct kinematic features (see Fig.~\ref{fig_LR_R_setup}). At this wavelength, we identified the following kinematic components for IRAS17:
~
\begin{enumerate}
\item A {\tt primary} component, corresponding to the systemic component of the galaxy, traces its rotation and shows a peculiar velocity dispersion map with a pronounced enhancement over a large area (i.e., a `butterfly' or `X-shaped' region), surrounded by lower-dispersion regions;

\item A {\tt secondary component}, characterized by a complex velocity field. Within this component, we can identify an inner region with velocities close to the systemic value, possibly associated with the intermediate broad line region (ILR hereafter, according to its velocity dispersion value, see Sect.~\ref{second_section}), and a redshifted region toward the north, apparently connected to the companion galaxy. This secondary component also shows an enhanced velocity dispersion with a `butterfly' pattern (as for the systemic component) found around the ILR, while the companion galaxy shows a relatively low velocity dispersion;

\item A {\tt tertiary} component, interpreted as the ionized outflow in H$\alpha$, based on its significant velocity shift relative to the systemic component ($\Delta$v = $\mathrm{v_T}$ - $\mathrm{v_P}$$\sim$-250 km s$^{-1}$) and very broad line widths ($\sigma$$\sim$900 km s$^{-1}$).
\end{enumerate}
~
As mentioned in the previous section, the kinematics of the [NII] doublet\footnote{{Although [NII] was fitted with only two Gaussian components in the LR-R setup, the secondary [NII] component can be separated into two distinct components, depending on the region considered. Specifically, we identify a secondary broad component surrounding the ILR and a tertiary, very broad component associated with the ionized outflow, as traced by the tertiary H$\alpha$ emission. For this reason, in Fig.~\ref{fig_LR_R_setup_NII} we show three [NII] components corresponding to the three H$\alpha$ components (i.e., primary, secondary, and outflow).}} were constrained to match those of the H$\alpha$ line. The main difference between the [NII] and H$\alpha$ panels (Figs.~\ref{fig_LR_R_setup} and \ref{fig_LR_R_setup_NII}) is that the inner region of the secondary component, associated with the ILR, is absent in the forbidden-line tracer [NII].

\subsubsection{H$\alpha$ primary (systemic) component} 

The primary (systemic) component covers almost the entire MEGARA FoV, extending primarily in the north-south direction. Its H$\alpha$ emission highlights the base of the spiral arms extending toward the northwest and southeast, along with the bright central emission from our target. 
The H$\alpha$ flux peaks at the same location as the R-band continuum emission, as shown in Fig.~\ref{fig_LR_R_setup} (top left panel). The continuum emission reveals the foreground star located to the north, very close to the H$\alpha$ emission from the companion galaxy.

The velocity field exhibits the typical signature of a purely rotating disk (i.e., a `spider-like' pattern) with a velocity shear, $\mathrm{v_{shear}}$, of $\sim$211 km s$^{-1}$ (see Table~\ref{tab_Ha}, similar to the rotational velocity derived for the molecular phase by \citealt{Salome21}). The velocity shear is defined as half the difference between the median values of the lowest and highest 5\% of the velocity distribution: $\mathrm{v_{shear}}$=$\mathrm{\frac{1}{2}(v_{max}^{5\%} - v_{min}^{5\%}})$.
The systemic (heliocentric) velocity, derived at the continuum peak position, is 18331 km s$^{-1}$, corresponding to a redshift of $z = 0.0611$, in excellent agreement with the value derived from the molecular emission by \citet{Salome23} ($z = 0.0612$; giving a redshift difference between these two tracers of 0.3 km s$^{-1}$).
At the position of the companion galaxy, the velocity field is slightly redshifted by $\sim$200 km s$^{-1}$, consistent with the velocity offset observed in the CO(1-0) emission (i.e., $\sim$210 km s$^{-1}$).

\begin{figure}[!h]
\centering
\includegraphics[width=0.5\textwidth]{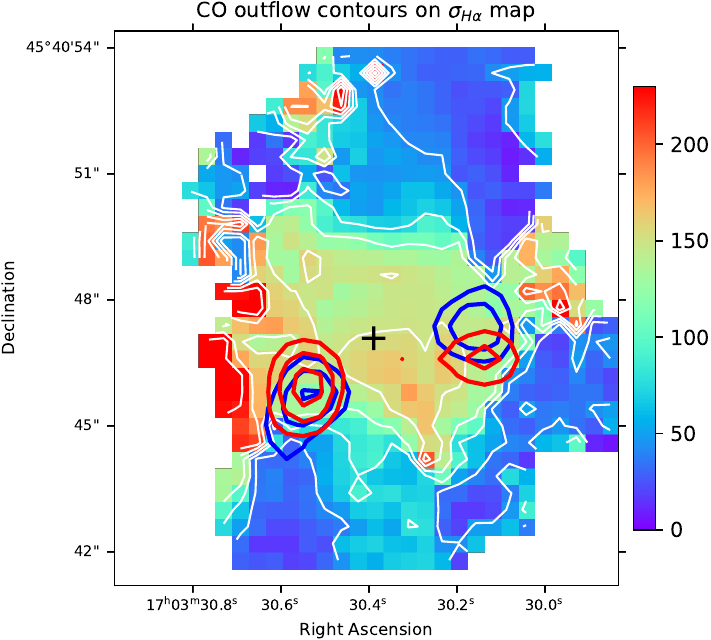}
\caption{H$\alpha$ velocity dispersion map of the primary component (white contours) with overlaid CO(1-0) contours of the molecular outflow (blue and red contours trace the approaching and receding molecular outflowing gas, respectively) from \cite{Longinotti23}.}
\label{Ha_COoutflow}
\end{figure}

The velocity dispersion map reveals an `X-shaped' (or `butterfly'-shaped) structure, showing quite high velocity dispersion values ($\sigma$$\sim$200 km s$^{-1}$) in the central region (R$\lesssim$~4 kpc) and perpendicular to the galaxy's bar. The bar is known to be aligned in the N-S direction (see \citealt{Ohta07}), and we suggest that the stream of velocity dispersion values with $\sigma$ between 50 and 100 km s$^{-1}$ observed along the N-S direction may trace this structure (e.g., \citealt{Hernandez05, Erroz15}). The `butterfly' structure previously observed in the LINER galaxy NGC~1052 by \cite{Cazzoli22} (see also \citealt{HM24}) may be associated with an outflow interacting with the surrounding ISM and/or the radio jet-ISM interaction, producing a region of enhanced velocity dispersion (i.e., turbulence). As shown in Fig.~\ref{Ha_COoutflow}, the `butterfly' region is clearly spatially coincident with the molecular outflow traced in CO (i.e., \citealt{Longinotti23}).

According to the observed (i.e., uncorrected for inclination) dynamical ratio, v/$\sigma$, between the velocity shear and the mean velocity dispersion (see Table~\ref{tab_Ha}), we found $v/\sigma \sim 2.3$, which clearly indicates that the primary (systemic) component is dominated by rotation.

Furthermore, we compare the H$\alpha$ flux emission of the primary component with the molecular emission from the CO(1-0) transition observed with NOEMA \citep{Salome23}, as shown in Fig.~\ref{Ha-CO_flux}. We found good agreement between the two at the positions of our source, IRAS17, and the companion galaxy.
Although the interpretation of the H$\alpha$ kinematic maps is not straightforward, our results suggest that the companion galaxy appears to extend over a larger area in H$\alpha$ emission than in the molecular component. According to the H$\alpha$ emission (see also Sect.~\ref{second_section}), we identified an extended structure toward the south-east, resembling a tidal tail. This feature is characterized by redshifted velocities and low velocity dispersion values ($\sigma \lesssim 50$ km s$^{-1}$; top right panel in Fig.~\ref{fig_LR_R_setup}), may be connected to IRAS17.

\subsubsection{H$\alpha$ secondary component} 
\label{second_section}

The secondary component extends over a similar FoV as the primary component but is significantly brighter. Within this kinematic component, we can clearly distinguish the regions corresponding to the three sources: IRAS17, the foreground star, and the companion galaxy (see Fig.~\ref{fig_LR_R_setup}).
The emission from the star has been removed from the map (see grey flux contours in Fig.~\ref{fig_LR_R_setup} toward the north).
In this component, the emission from IRAS17 is the main contributor to the H$\alpha$ flux, while the companion galaxy appears fainter than in the systemic flux map. 
Toward the north, an arm-like structure is visible, specially in the velocity field and velocity dispersion maps, likely associated with the companion galaxy. This structure is characterized by higher velocities (redshifted by $\sim$200 km s$^{-1}$) and lower velocity dispersion values ($\sigma \sim$100 km s$^{-1}$). Its contribution in the corresponding flux map is hard to discern.

The secondary component overall exhibits complex kinematics, with both the velocity field and velocity dispersion maps showing clear signs of turbulence and possible interaction effects. The width of this component is broader ($>$250 km s$^{-1}$) than that of the systemic component ($<$250 km s$^{-1}$), with a velocity shear of approximately 170 km s$^{-1}$ across the FoV (see Table~\ref{tab_Ha}).

However, this component is not characterized by rotation, because no rotational velocity pattern is observed and finding a v/$\sigma$$\sim$ 0.5, thus being dispersion-dominated. 
In the central region, we observed a median velocity very close to the systemic median velocity (i.e., with a velocity shift of $\sim$15 km s$^{-1}$). In contrast, the companion exhibits a (positive) velocity shift of $\sim$110~km s$^{-1}$ with respect to the systemic component. 
According to the features observed in the velocity dispersion map of the secondary component, we can identify three distinct regions:
 \begin{enumerate}[label=\textbf{\alph*.}]
    \item \texttt{Low $\sigma$} ($\sim$110 km s$^{-1}$): Located in the north, overlapping the companion’s position.
    \item \texttt{Intermediate $\sigma$} ($\sim$280-380~km s$^{-1}$): Spanning a $\sim$28~arcsec$^2$ area (i.e., R~$\sim$3.0\arcsec$\sim$3.5~kpc; light green region in the map) around the central region, underlying the high-$\sigma$ (i.e., `butterfly') region.
    \item \texttt{High $\sigma$} ($\sim$600 km s$^{-1}$): Corresponding to the `X-shaped' (or `butterfly') region, also seen in the primary component.
\end{enumerate}

The intermediate-$\sigma$ component (i.e., represented by the Gaussian shown in magenta in Fig.~\ref{nuclear_3c_LRR_MRG}) is found in the region where the tertiary component (tracing the outflow) is added in the fit. In this area, the velocity field resembles that of the primary component, and the velocity dispersion values are indicative of an ILR.
This interpretation is supported by the fact that the line widths are not as broad as those typically observed in Sy1 objects, where FWHM $\gtrsim$ 2000 km s$^{-1}$ (corresponding to $\sigma \gtrsim$ 850 km s$^{-1}$).
In contrast, we observe a characteristic $\sigma$$\sim$280-380~km s$^{-1}$ (FWHM $\sim$700-900 km s$^{-1}$), consistent with typical values reported for ILRs (FWHM in the range 700-1200 kms$^{-1}$; \citealt{Crenshaw07, Adhikari16} and references therein). Furthermore, its velocity field closely matches the systemic velocity ($\Delta$v$\sim$0), reinforcing our interpretation. As presented in \cite{Adhikari16}, the ILR may represent an inner extension of the Narrow Line Region (NLR).

In contrast, the high-$\sigma$ region reveals significant turbulent motions surrounding the ILR, possibly induced by the outflow or representing the contribution from the ionization cones. 
Figure~\ref{fig_LR_R_setup} (bottom panel) shows representative H$\alpha$-[NII] complex spectra extracted from two locations within the `butterfly' region, toward the NE and SW directions of the outflow. Both primary and very broad components are required to properly reproduce the total emission. In these regions, the broad H$\alpha$ component dominates the flux emission.

Regarding the kinematics of the companion galaxy, the H$\alpha$ emission in the secondary component does not align with that of the CO emission, appearing nearly perpendicular. Specifically, the CO emission forms a bridge aligned along the NW-SE direction, while the ionized gas appears to wrap around the molecular structure, moving toward the N-SW and offset by approximately 40 degrees.

\subsubsection{H$\alpha$ tertiary (outflow) component}

We identified the emission of the tertiary component with the ionized outflow due to its kinematic properties.
It extends over a roughly circular area of approximately $\sim$6 arcsec$^2$ (observed radius of 1.4\arcsec $\sim$1.65 kpc), corresponding to $\sim$8.3 kpc$^2$. This component is spatially resolved beyond the PSF region (see Sect.~\ref{size_outflow} for further details).

The velocity field shows a velocity shear of v$_{\text{shear}}^{T}$$\sim$140~km~s$^{-1}$ and a median velocity dispersion of $\sigma^T$$\sim$956 km s$^{-1}$ (FWHM$\sim$2250 km s$^{-1}$).
These velocity shifts are consistent with ionized outflows observed in local galaxies, whereas the velocity dispersion values are significantly higher than those typically found in local star-forming or AGN-dominated galaxies. Instead, they are more comparable to values observed in high-z sources, such as those in the SUPER survey and quasar samples (e.g., \citealt{Kakkad20}, using [OIII] lines; FWHM = 600-3000~km s$^{-1}$).
The velocity shift relative to the primary component, defined as $\Delta$v = v$_{\text{T}}$ - v$_{\text{P}}$, ranges from -329.6 to -168.3 km s$^{-1}$, with a mean value of -252.5 km s$^{-1}$ (median: -243.9 km s$^{-1}$; see Table~\ref{tab_Ha}).

Although the geometry of the ionization cones and the outflow is difficult to define, under the assumption of a biconical outflow roughly perpendicular to the disk, the redshifted (receding) component may be partially obscured by the disk, while the blueshifted (approaching) component is less affected. In the case of IRAS17, the outflow velocity field shows blueshifted velocities with respect to the systemic component across the entire FoV, suggesting that the receding part is completely obscured by the disk. A few south-eastern spaxels show the most extreme blueshifted velocities together with low velocity dispersions. This may indicate projection effects and obscuration, possibly associated with an interaction between the ionized outflow and denser ambient material in this region, which could lower the observed velocity dispersion.

In Sect.~\ref{Params_outflow}, we will derive the physical and kinematic properties of the ionized outflow, such as its mass, kinetic energy, momentum, and velocity v$_{\text{out}}$ based on this component.

\begin{figure*}[!h]
\centering
\includegraphics[width=0.85\textwidth]{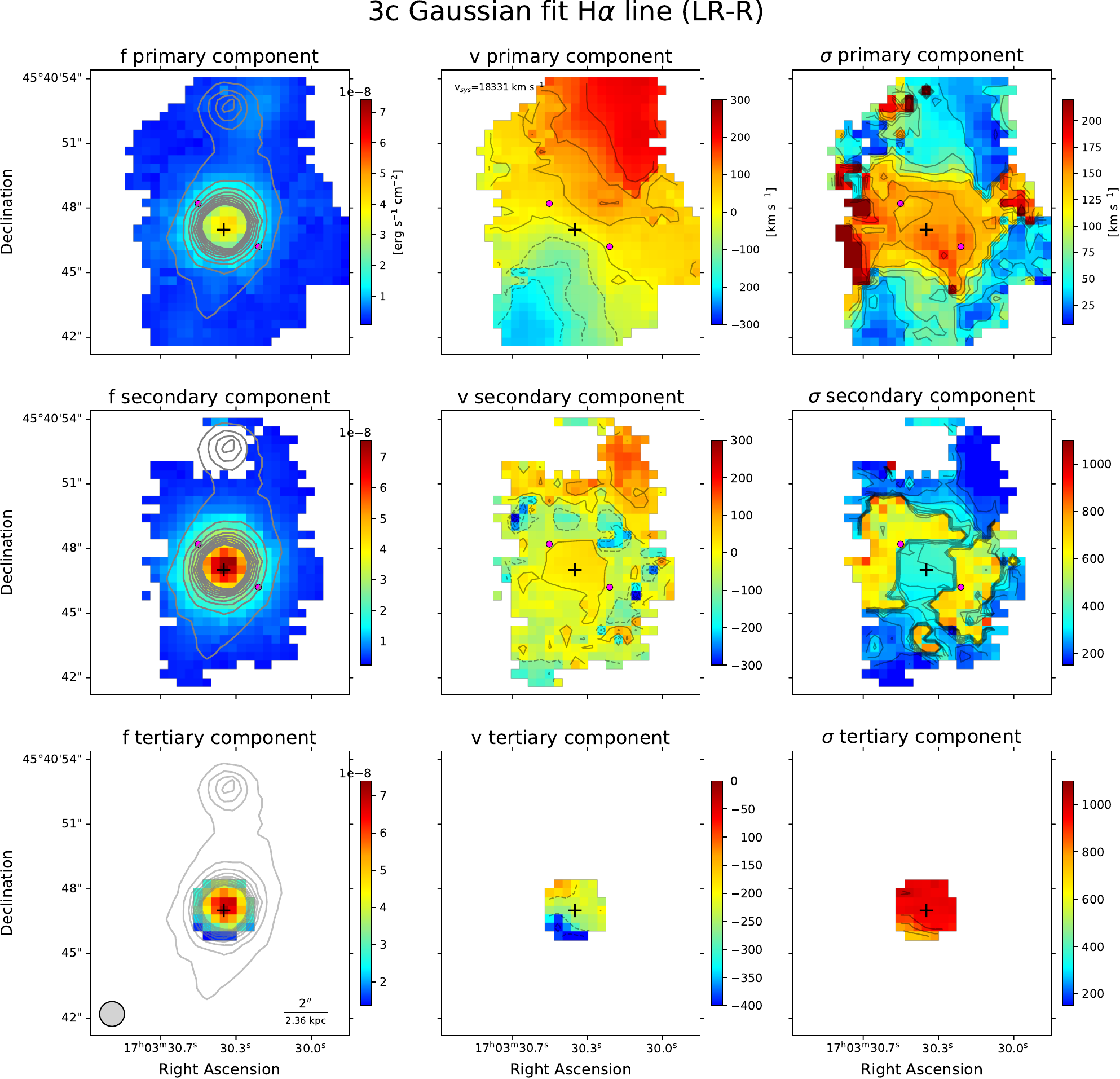}
\vskip3mm
\includegraphics[height=0.23\textwidth]{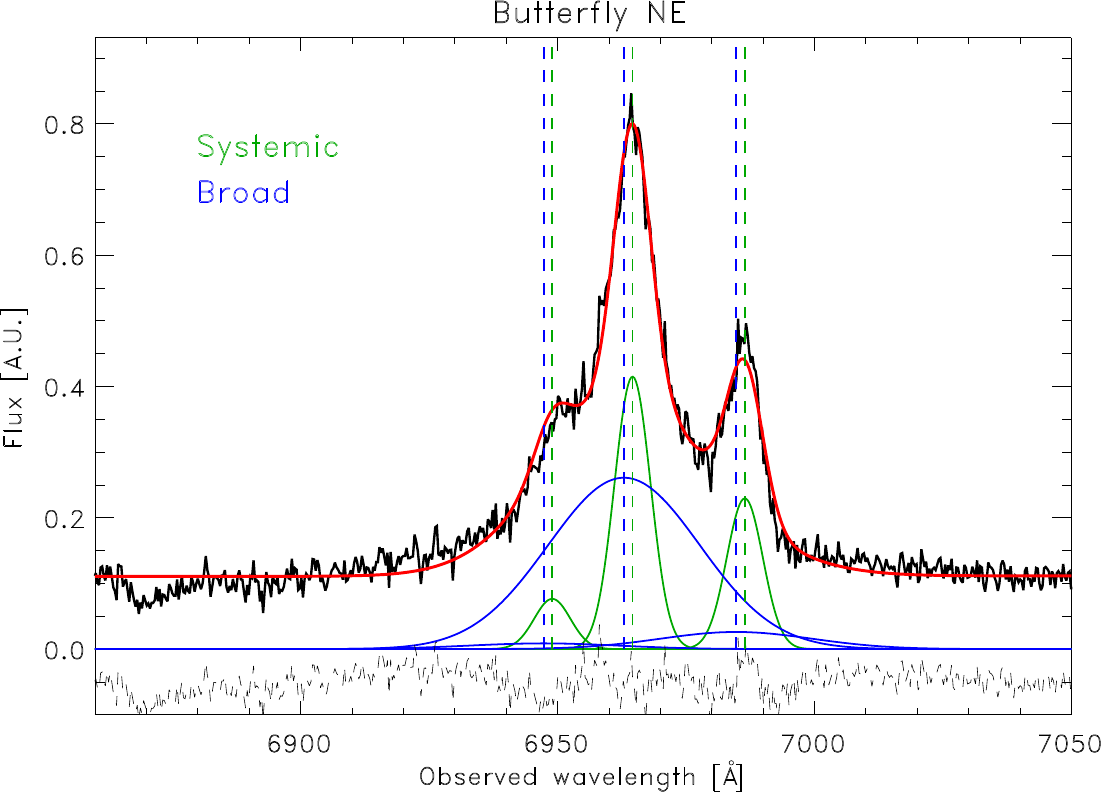}
\includegraphics[height=0.23\textwidth]{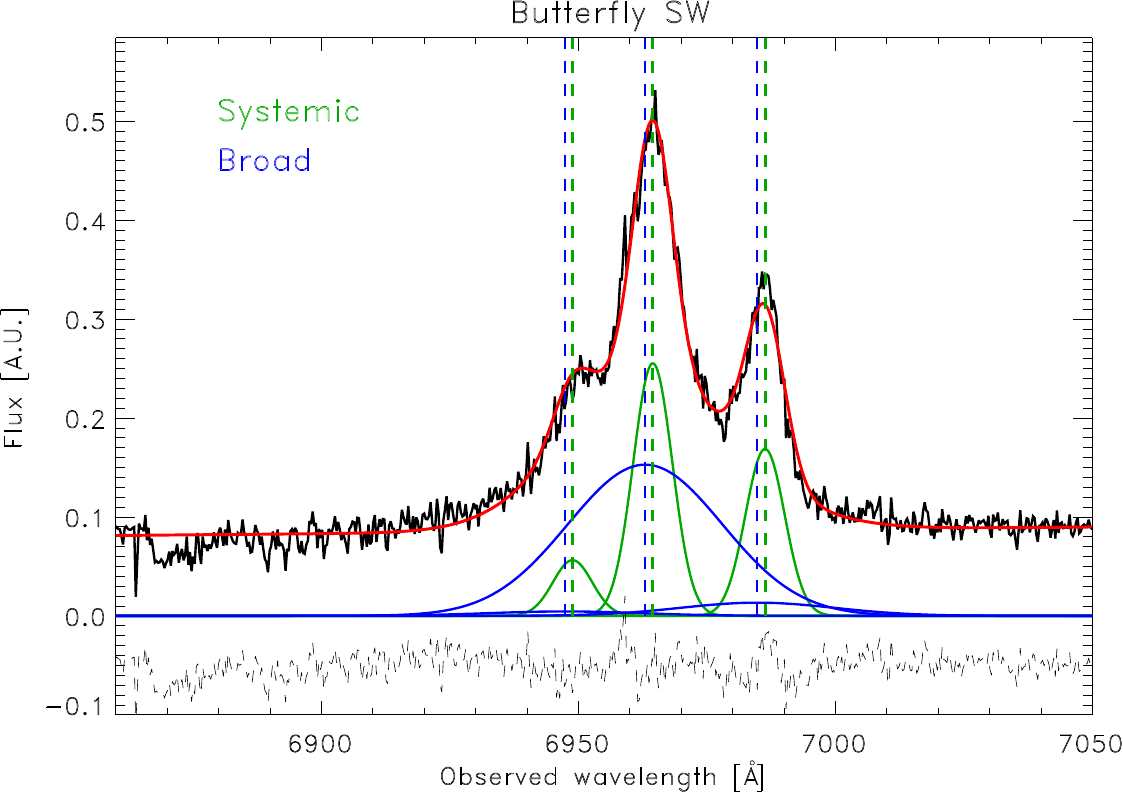}
\caption{Kinematic maps for the ionized component as traced by the H$\alpha$ line.  
{\it From left to right, top to bottom:} flux intensity, velocity field and (intrinsic) velocity dispersion maps of the primary (systemic), secondary and tertiary components. 
The velocity maps of each component have been corrected for the systemic velocity, v$_{\rm sys}$, derived at the AGN position (i.e., at the continuum intensity peak, marked by a black cross in all the maps). 
Grey contours in the flux intensity systemic maps trace the R-band continuum emission, highlighting the position of the foreground star. 
The magenta circles in the primary and secondary maps mark two individual spaxels in the NE and SW directions within the `butterfly' region, from which the spectra shown in the bottom panel are extracted. The spectra are fitted with two components, and the broad [NII] emission is detected at S/N$>$4 (i.e., 5.5 in the NE and 4.1 in the SW).
The light gray circle in the bottom left panel is the MEGARA PSF (FWHM). For all the maps, north is up and east to the left. 
}
\label{fig_LR_R_setup}
\end{figure*}

\begin{figure*}[!h]
\centering
\includegraphics[width=0.85\textwidth]{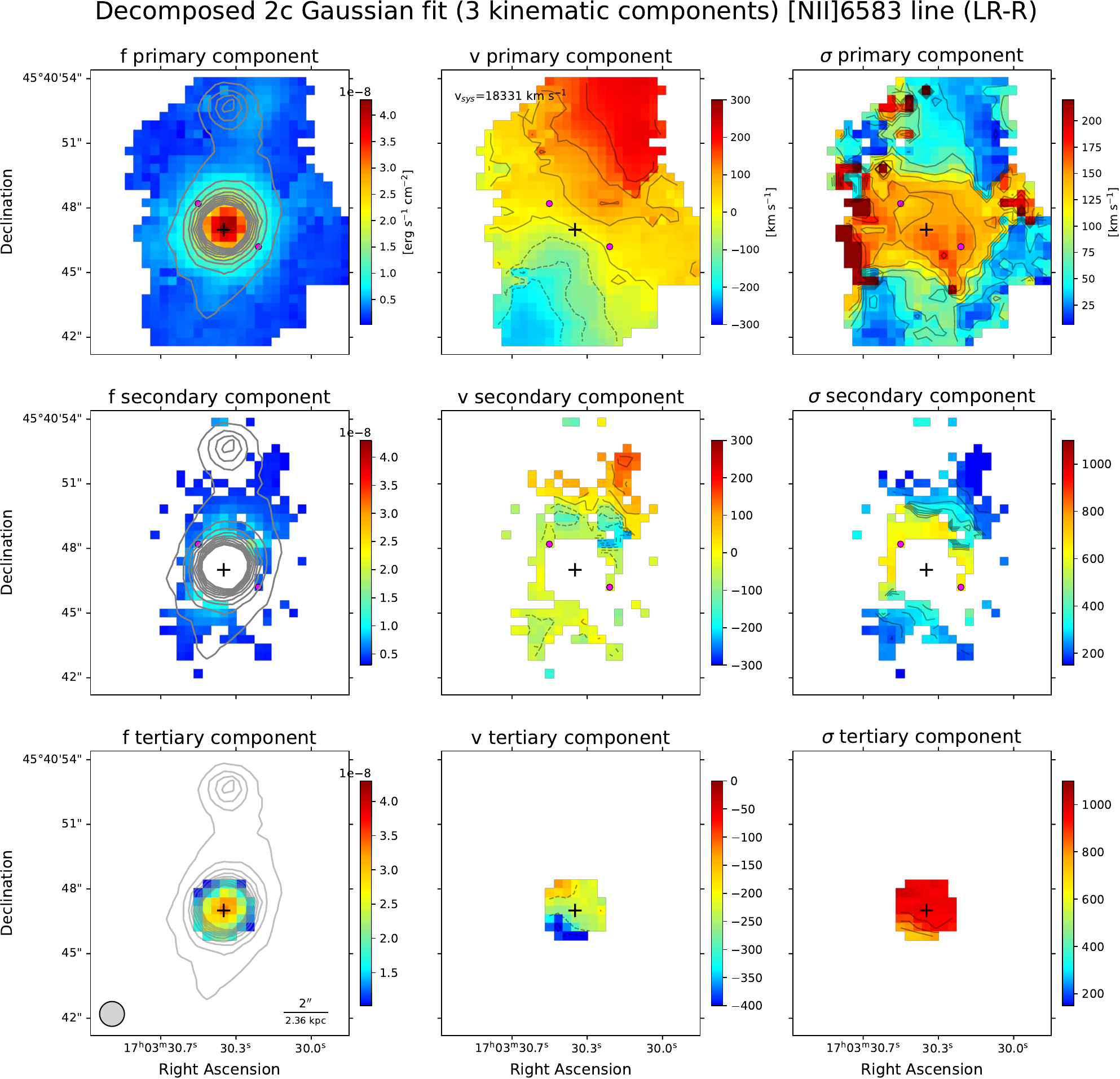}
\caption{Same figure caption as in Fig.~\ref{fig_LR_R_setup} but for the [NII]6583 line. The main difference between the [NII] and H$\alpha$ (Fig.~\ref{fig_LR_R_setup}) panels is that the inner region of the [NII] secondary component, corresponding to the ILR in the H$\alpha$ map, is absent in the forbidden [NII] line.
}
\label{fig_LR_R_setup_NII}
\end{figure*}

The main difference between the [NII] and H$\alpha$ (Fig.~\ref{fig_LR_R_setup}) panels is that the inner region of the [NII] secondary component, corresponding to the ILR in the H$\alpha$ map, is absent in the forbidden [NII] line.

\subsection{H$\beta$-[OIII] complex in the MR-G and LR-V setups}
\label{Oxygen_section}

Thanks to the MR-G and LR-V setups, we were able to characterize in detail the kinematics of the H$\beta$ and [OIII] lines.In both setups, three kinematic components were required to fit the observed [OIII] emission (i.e., primary, secondary and tertiary) while three components were needed for the H$\beta$ line in the MR-G setup, and only one component was considered in the LR-V (see Sect.~\ref{obs_megara} and Table~\ref{tab_interpretation}). 
The secondary and tertiary components in [OIII] have been associated with a `modest' and a `fast' outflows, respectively, as the kinematics of the former are less extreme than that of the latter (see following subsections).

As noted in Sect.~\ref{MR_LR_}, we linked the H$\beta$ profile to the H$\alpha$ kinematics\footnote{{Simultaneously fitting the H$\beta$-[OIII] and H$\alpha$-[NII] complexes in IRAS17 by tying their kinematic parameters is not an optimal strategy due to both observational and physical reasons: (i) the H$\alpha$ and H$\beta$ lines were observed with different instrumental setups (LR-R and MR-G), which have distinct spectral resolutions (R = 6000 and 12000, respectively), potentially leading to small zero-point differences in the wavelength calibration; (ii) variations in continuum normalization across different spaxels can further complicate the fitting process, particularly when comparing lines observed under different conditions.}}, which enables us to characterize this fainter emission using three components: a narrow primary component, an intermediate secondary component associated with the ILR, and a broad tertiary feature associated with the outflow. Owing to the lower S/N of H$\beta$ and [OIII] lines, their primary components are detected over a much smaller spatial extent than that of H$\alpha$.

\begin{figure*}
\centering
\includegraphics[width=0.85\textwidth]{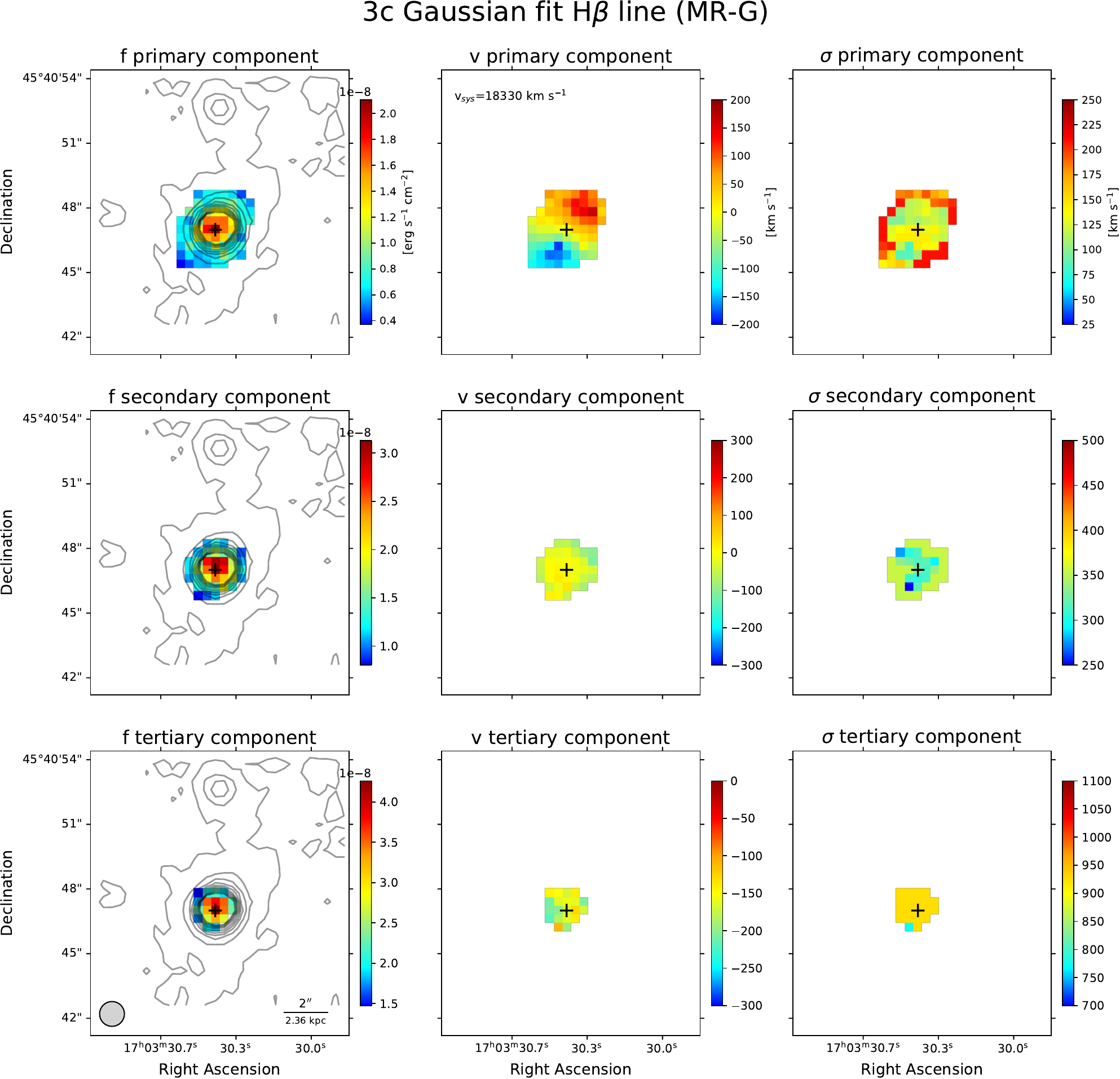}
\caption{Kinematic maps for the ionized component as traced by the H$\beta$ line according to the MR-G setup. See Fig.~\ref{fig_LR_R_setup} for details about the panels.
}
\label{fig_Hb_MR_LR_new}
\end{figure*}

\begin{figure*}
\centering
\includegraphics[width=0.85\textwidth]{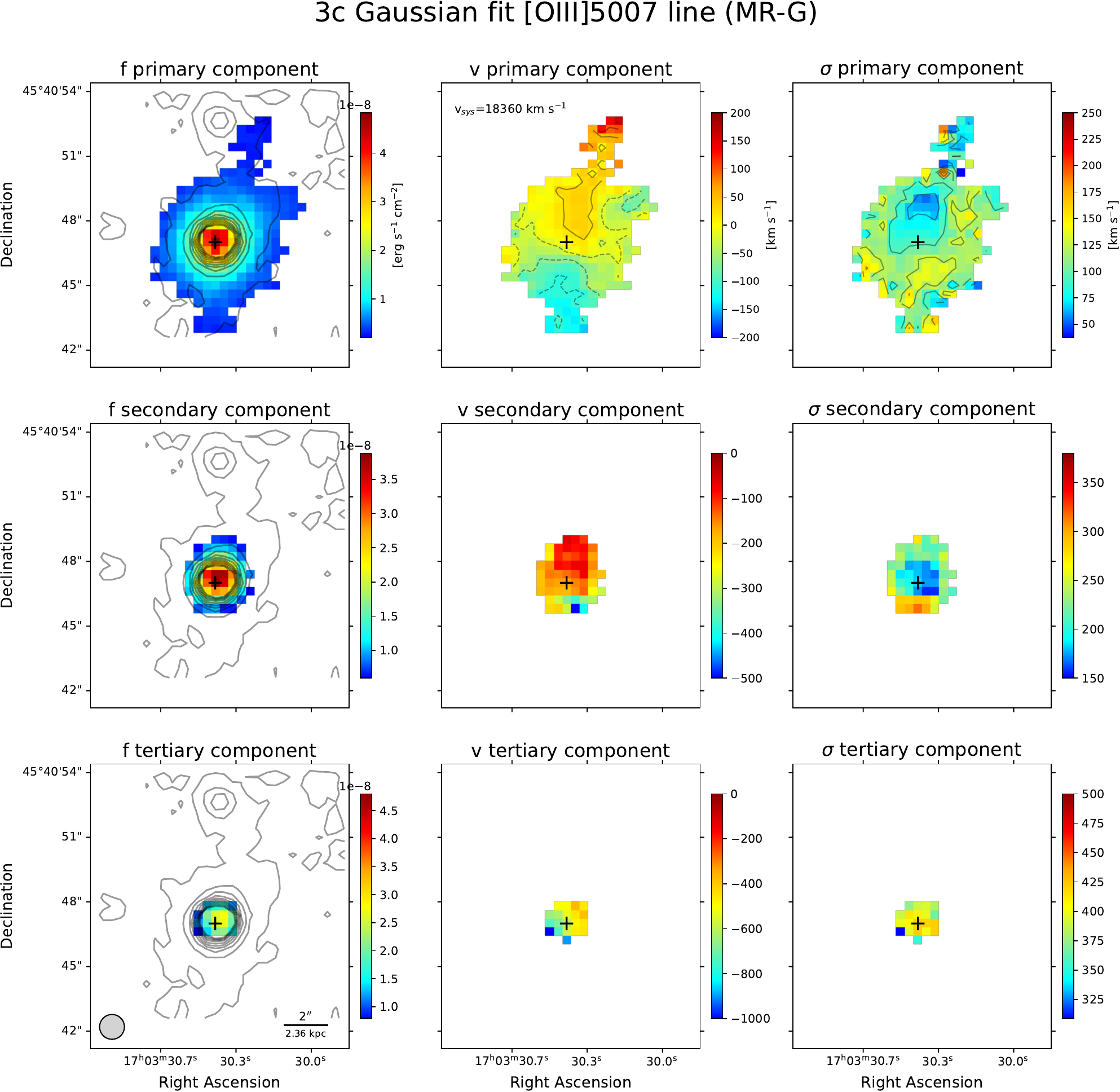}
\caption{Kinematic maps for the ionized component as traced by the [OIII]5007 line according to the MR-G setup. See Fig.~\ref{fig_LR_R_setup} for details about the panels.
}
\label{fig_OIII_MR_LR_new}
\end{figure*}

\begin{figure*}[!h]
\centering
{\includegraphics[width=0.85\textwidth]{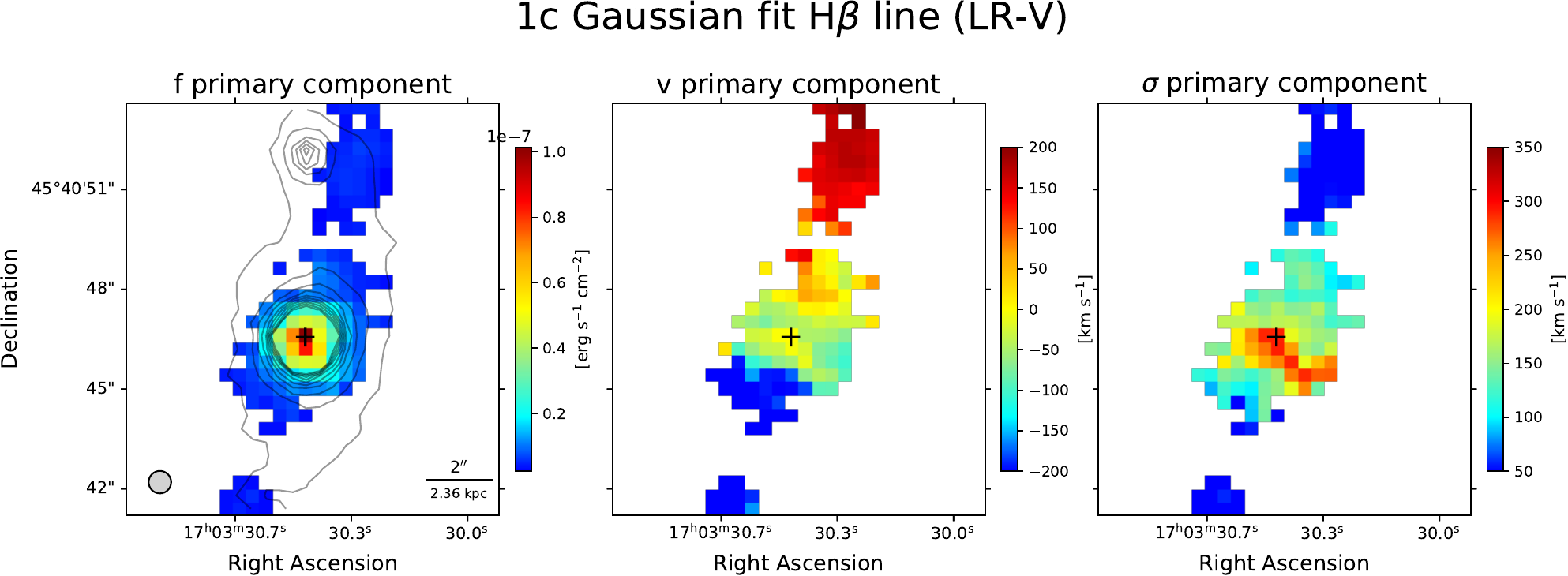}}
\vskip5mm
\includegraphics[width=0.85\textwidth]{{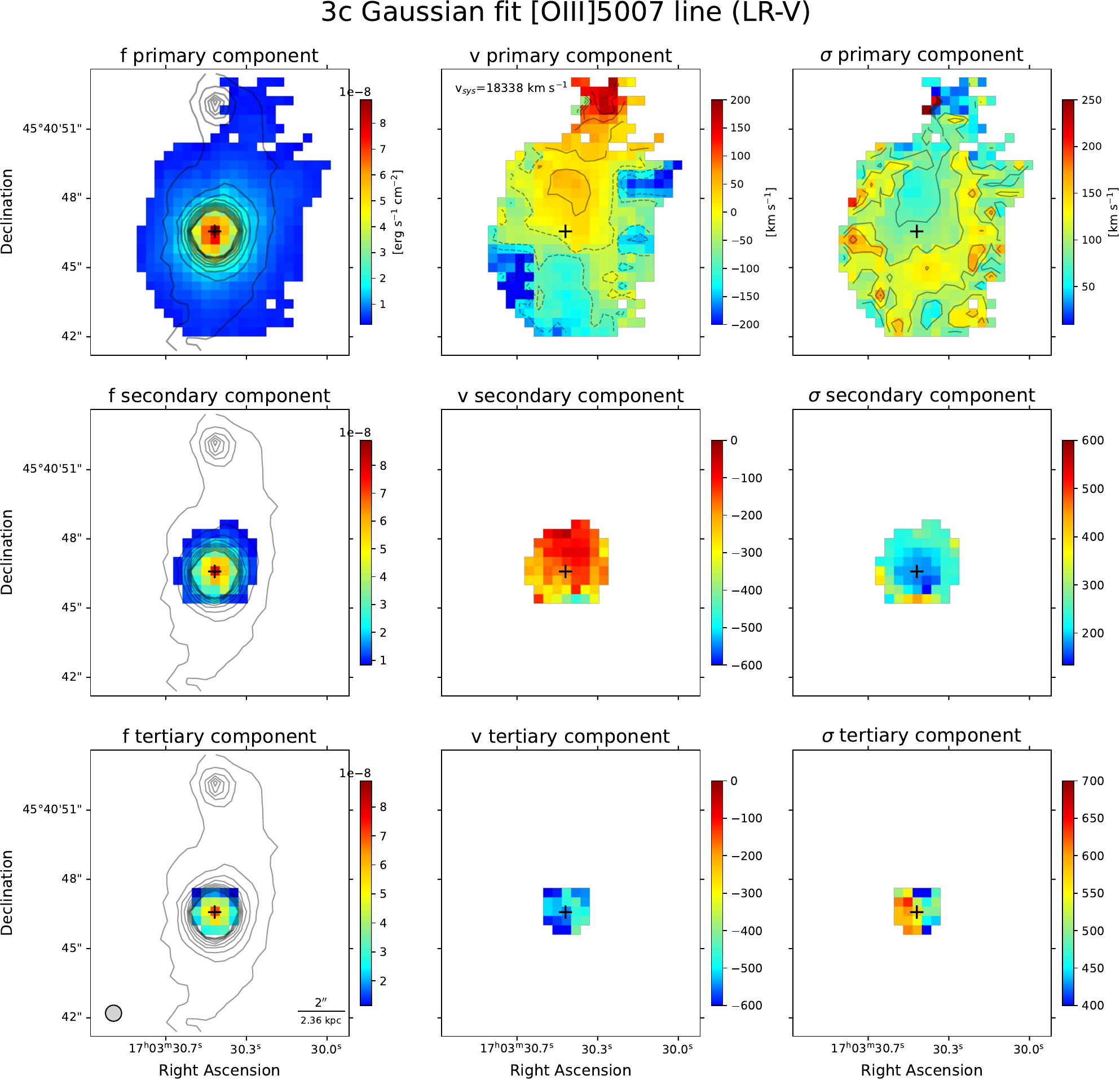}}
\caption{Kinematic maps for the ionized component as traced by the H$\beta$ (top) and [OIII]5007 (bottom) complex as derived in the LR-V setup. 
See Fig.~\ref{fig_LR_R_setup} for details about the panels.
}
\label{LRV_setups_maps}
\end{figure*}

In Figs.~\ref{fig_Hb_MR_LR_new}, \ref{fig_OIII_MR_LR_new}, and \ref{LRV_setups_maps} the kinematic maps of the H$\beta$-[OIII]5007 complex obtained using the MR-G and LR-V setups are shown.
The [OIII] emission appears a bit more compact in the MR-G setup than in LR-V setup and this could be due to the longer exposure time (more than $\times$2) used in the LR-V mode compared to MR-G, as well as better seeing conditions, which allowed us to better trace fainter emission found at larger radii. A proper characterization of the outflow extension in the different lines will be discussed in Sect.~\ref{size_outflow}. Nevertheless, the main kinematic results are in agreement across both setups (see Tables~\ref{tab_MR} and \ref{tab_LR}) and three components were needed to properly fit the observed emission in both cases.

As mentioned in Sect.~\ref{MR_LR_}, the H$\beta$ emission was modeled with a three-component fit in the MR-G setup (Fig.~\ref{fig_Hb_MR_LR_new}). In the LR-V setup, however, the line lies at the blue edge of the spectral range, close to the sensitivity limit, making the fit less reliable. Although some extended emission is detected in LR-V (but not in MR-G), we do not draw conclusions from it. Therefore, we adopt the kinematic results derived from the MR-G data, as they are more robust.

\subsubsection{H$\beta$-[OIII] primary (systemic) components}

\begin{figure}[!h]
\centering
\includegraphics[width=0.5\textwidth]{{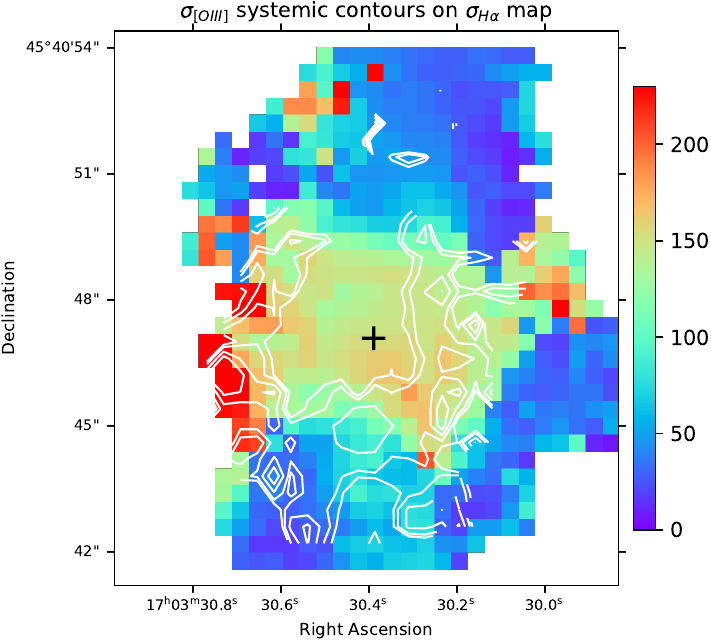}}
\caption{H$\alpha$ velocity dispersion map of the primary component, with the [OIII] velocity dispersion contours (white) of the primary component overlaid, obtained from the LR-V setup.}
\label{Ha_OIII_compare}
\end{figure}

The [OIII] primary (systemic) component extends in a smaller region than that covered by the H$\alpha$ line (see Fig.~\ref{fig_OIII_MR_LR_new} and Sect.~\ref{size_outflow} for further details), although it still overlaps with the IRAS17 emission, and the companion galaxy shows detectable emission.
The main kinematic features are consistent across both the MR and LR setups, including a rotation-dominated velocity field and a velocity dispersion map that reveals a ring-like structure with enhanced $\sigma$ values. However, we find a clear difference between the kinematic maps derived from the [OIII] and H$\alpha$ emission lines, particularly in the velocity dispersion distribution. As shown in Fig.~\ref{Ha_OIII_compare}, the regions of enhanced $\sigma$ traced by the [OIII] emission coincide with the edges of the H$\alpha$ `butterfly' structure, surrounding the nuclear outflow.

The H$\beta$ primary component closely follows the global H$\alpha$ kinematics and is consistent with the systemic component.

\subsubsection{H$\beta$-[OIII] secondary components}

The secondary component observed in [OIII] has been identified with an outflow. In particular, for this component in the MR-G mode, we derived a mean (median) velocity offset (with respect to the spatially resolved systemic velocity map) of $\Delta$v= -174.0$\pm$9.0 km s$^{-1}$ (median: $-157.6$ km s$^{-1}$), and a mean (median) velocity dispersion of 223.4$\pm$5.3 km s$^{-1}$ (median: 221.7 km s$^{-1}$). The less bluehifted\footnote{It is worth noting that the velocity shift, $\Delta$v, of the secondary and tertiary components only show negative values ($\Delta$v$<$0) with respect to the systemic component, so the entire emission of the outflow is blueshifted with respect to the systemic component.} region is located to the north-west, while the most blueshifted region lies to the south-east. The less blueshifted side displays a lower velocity dispersion, whereas the most blueshifted side exhibits higher $\sigma$, consistent with the interpretation that the latter suffers less extinction. This behavior is consistent with an orientation-dependent obscuration scenario: the most blueshifted component likely traces the near side of the outflow, which is less affected by extinction, allowing us to observe gas closer to the nucleus and with higher turbulence. In contrast, the less blueshifted components are plausibly associated with gas on the far side of the outflow (or partially embedded in the galactic disk), where line-of-sight obscuration preferentially suppresses emission from the most turbulent regions, leading to a lower observed velocity dispersion. These kinematic differences therefore reflect orientation and obscuration effects within a dusty circumnuclear environment.

The same pattern is found for the [OIII] line observed in the LR-V setup, for which we also derived similar kinematic values (see Table~\ref{tab_LR}).
Both [OIII] outflows identified by the secondary components are resolved, according to the PSF values of each setup.

The secondary H$\beta$ component follows the global H$\alpha$ kinematics and is identified with the ILR, consistent with the H$\alpha$ emission (se Table~\ref{tab_interpretation}).

\subsubsection{H$\beta$-[OIII] tertiary components}

In both setups, the tertiary component of the [OIII] line appears more compact than its secondary component, while exhibiting more extreme kinematics, consistent with a faster outflow. In particular, the [OIII] tertiary component is unresolved in the MR-G setup, whereas it is resolved in the LR-V setup (see Sect.~\ref{size_outflow}). The kinematics of this outflow shows typical median values of velocity offset $\Delta$v=-540 km s$^{-1}$ and velocity dispersion $\sigma$$\sim$400 km s$^{-1}$ in the MR-G setup, as derived from the integrated emission within the PSF region.
Therefore, for the resolved emission in the LR-V setup, we derived a median velocity offset of $\Delta \mathrm{v}$$\sim\!-536~\mathrm{km\,s^{-1}}$, and a median velocity dispersion of $\sim 520$~km~s$^{-1}$ (see Table~\ref{tab_LR}).
The kinematic results obtained for both setups and emission lines are in mutual agreement and are reported in Tables~\ref{tab_MR} and \ref{tab_LR}.

The H$\beta$ tertiary component (detected in the MR-G setup) traces the global H$\alpha$ kinematics and  is likewise associated with the fast ionized outflow.

\section{Deriving the physical properties of the outflow(s)}
\label{Params_outflow}

\subsection{Constraining the outflow(s) intrinsic size}
\label{size_outflow}

We derived the size of the ionized outflow using the H$\alpha$ and [OIII] lines by applying a fixed percentage of the peak flux to each outflow map, in order to avoid any dependence on the SNR of the individual lines.
This method uses a flux threshold (e.g., 10\% or 50\% of the peak flux) to eliminate the S/N dependency of the size measurement. From these measurements, we computed an equivalent circular radius by converting the area enclosed by the isophote into the radius of a circle with the same area: $R_\mathrm{eq}$=$\sqrt{\frac{\mathrm{Area}}{\pi}}.$ In our analysis, we adopted the 10\% peak-flux isophote as the primary measure of the outflow size, as it captures the full spatial extent of the faint, extended emission and is less sensitive to flux concentration in the central regions compared to the 50\% peak-flux isophote. We counted the number of pixels above this threshold and calculated the total area as $\mathrm{Area} = N_\mathrm{pix} \times \mathrm{pixscale}^2$. This approach considers only the number of pixels above the threshold, not the sum of their fluxes.

To derive the intrinsic size of the outflow, we accounted for the PSF size measured at 10\% of the peak emission, given by PSF$_\mathrm{10\%}~=~1.82 \times \mathrm{PSF_{FWHM}}$. 
Using the 10\% flux cutoff, we obtained half the PSF (PSF$_\mathrm{10\%}/2$) of 1.09\arcsec\ for the LR-R and MR-G setups, and 0.82\arcsec\ for the LR-V setup.

The intrinsic outflow sizes derived using this method for the H$\alpha$ and [OIII] components are in good agreement, with values $\lesssim 1$~kpc for both tracers. Specifically, for the H$\alpha$ outflow we obtained an intrinsic size of $0.98 \pm 0.32$~kpc, while for [OIII] in the MR-G setup we measured $0.70 \pm 0.41$~kpc for the secondary component. The tertiary [OIII] component is unresolved in MR-G, for which we derived an upper limit to its size based on the PSF, $<$1.29~kpc.
In the LR-V setup, both components are resolved: for the secondary component, we measured a size of 0.70$\pm$0.26~kpc, while for the tertiary component we obtained 0.47$\pm$0.25~kpc.

All these values have been used to derive the physical outflow parameters presented in Tables~\ref{out_props_Ha}, \ref{out_props_OIII_MRG_2c}, \ref{out_props_OIII_MRG_3c}, \ref{out_props_OIII_LRV_2c} and \ref{out_props_OIII_LRV_3c} (see Sect.~\ref{outflow_physics}).

\subsection{Outflow electron density derivation}
\label{n_e_derivation}

\begin{figure}[h]
\centering
\hskip-1mm\includegraphics[width=0.42\textwidth]{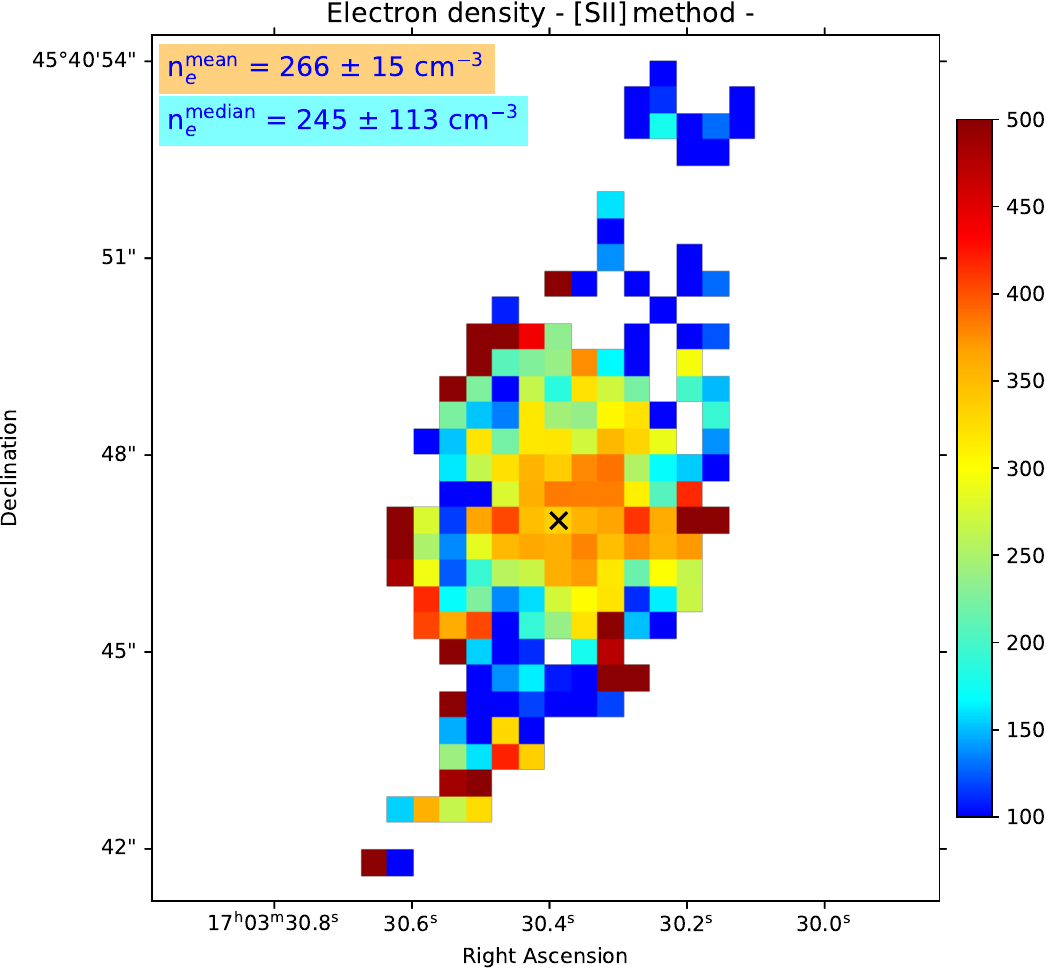}
\vskip3mm
\includegraphics[width=0.43\textwidth]{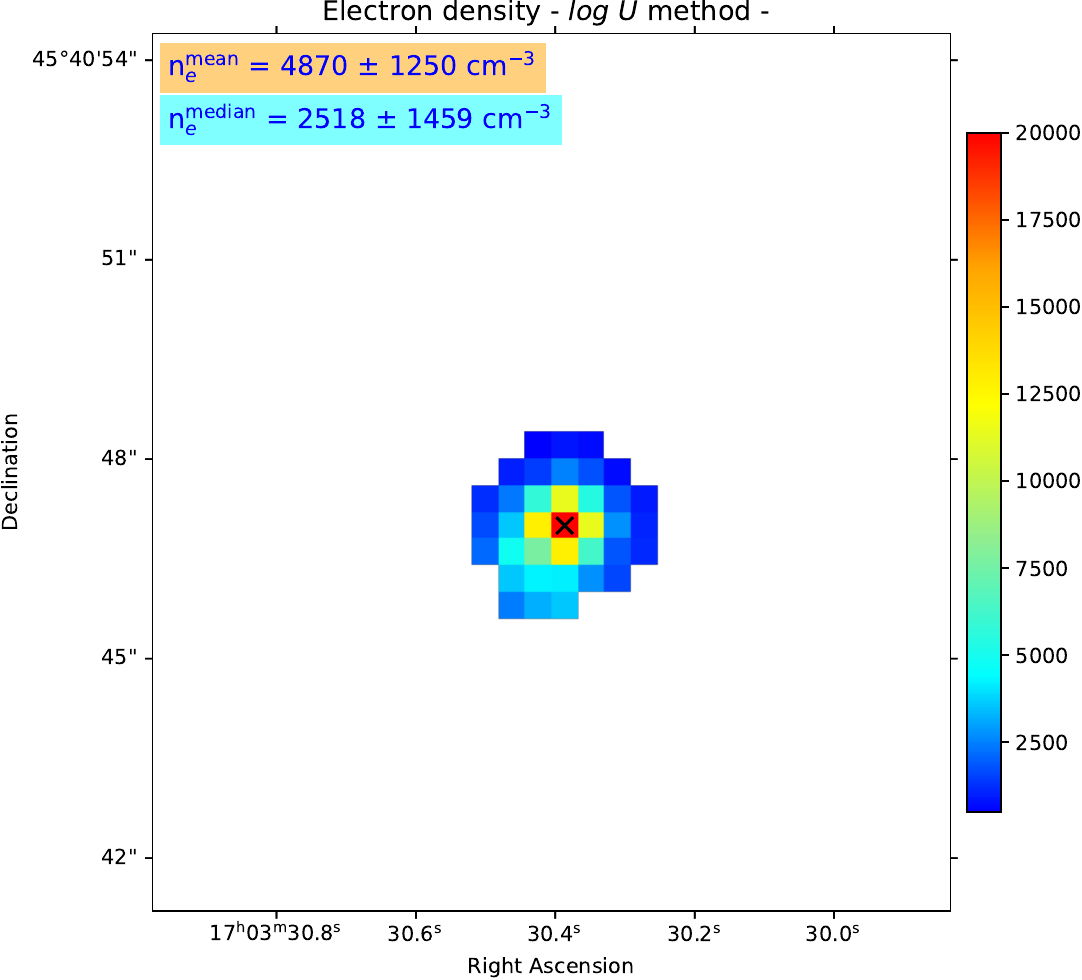}
\caption{
Electron density (n$_\mathrm{e}$) maps in units of cm$^{-3}$ derived using the [SII] (top) and $log~U$ (bottom) methods. The [SII] method is applied to the total [SII]$\lambda\lambda$6717,6731 line profile, while the $log~U$ map sums the flux contributions from the secondary and tertiary components. Typical mean and median values, together with the standard deviation and the median absolute deviation, respectively, are shown in the insets.
}
\label{fig_ne_BNM}
\end{figure}

The electron density of ionized gas, particularly in both the systemic (disk) and outflowing components of galaxies, is a crucial parameter for understanding the physical conditions of the ionized ISM and for accurately determining the properties of outflows like its mass, mass outflow rate and kinetic energy. However, several works (i.e., \citealt{Harrison18, Kakkad18}) pointed out, the electron density, n$_\mathrm{e}$, is the parameter with the largest uncertainties in these calculations.
In literature, different methods can be used to derive this parameter (\citealt{Davies20}). 
A typical method largely used is based on the [SII]~doublet ratio, [SII]$\lambda$6717/$\lambda$6731 (e.g., \citealt{Osterbrock06, Kakkad18, Rodriguez19, Venturi21}; hereafter, {\it [SII]~method}). This method suffers from saturation at high density and cannot probe high densities because collisional de-excitation dominates above 10$^4$~cm$^{-3}$: this ratio is, thus, only sensitive to densities in the range 50$<$n$_\mathrm{e}$$<$5000~cm$^{-3}$, giving lower electron densities by down to an order of magnitude than those obtained with other methods (see \citealt{Davies20}).
Thus, the values obtained from this method could lead to an overestimation of the mass of the outflow, as well as of the mass outflow rate. 
In our case, we could not isolate the outflow emission in the [SII]$\lambda\lambda$6716,6731 doublet by fitting each line with a single-component (1c) model (see Fig.~\ref{LR_R_S2_O1}, top panel).
Therefore, as a first approximation, we used the flux ratio derived from the 1c fit (primary component) to calculate the electron density in each spaxel following the prescription of \cite{Sanders15}:
~ 
\begin{equation}
n_\mathrm{e} = \frac{cR - ab}{a - R},
\end{equation}
~
where R = [S II]$\lambda$6717/$\lambda$6731 and a=0.4315, b=2107, c=627.1. 
According to this method, we derived a mean (median) electron density n$_\mathrm{e}=266\pm15~(245)$ cm$^{-3}$ across the entire FoV (see the spatially resolved n$_\mathrm{e}$ map in Fig.~\ref{fig_ne_BNM}, top), while in the inner region (i.e., the area covered by the outflow) we obtained a higher value of n$_\mathrm{e}$ $\sim$350 cm$^{-3}$. The electron density profile is centrally peaked and decreases with increasing radius, although a few spaxels show relatively high values ($>$500 cm$^{-3}$), possibly due to lower S/N in the emission, which leads to more uncertain results.
As we know, this method tends to underestimate the electron density because a significant fraction of the [S II] emission preferentially originates from the outer, less-dense regions of the ionized outflow (see \citealt{Kakkad18}). 
Furthermore, the lack of a detectable broad component in the [SII] emission lines prevents a reliable kinematic decomposition of this doublet, making the [SII]-based electron density estimate potentially unrepresentative of the outflowing gas. In particular, the derived value should be regarded as a lower limit, which would imply unrealistically high outflow masses if directly adopted. 
For this reason, we also derived the electron density of the outflow using the method described by \cite{Baron19}, which derived the n$_\mathrm{e}$ from the ionization parameter value, $U$ (i.e., hereafter, {\it log~U~method}; see also \citealt{Davies20, Peralta23, Esposito24}). This method is more sensitive to the high-density, fully ionized regions, where H$\alpha$ and [OIII] emission lines originate. It derives the n$_\mathrm{e}$ parameter using the ratio between the emission lines such as H$\alpha$/[NII], and [OIII]/H$\beta$, typically strong in AGN. 
According to this method we can derive the ionization parameter $U$, defined as the number of ionizing photons per atom: $U$ = $\mathrm{Q_H/(4\pi ~{\it r}^2~n_H~c}$), where Q$_\mathrm{H}$ is the rate of hydrogen-ionizing photons (s$^{-1}$ units), $r$ is the distance from the ionizing source, n$_\mathrm{H}\sim$~n$_\mathrm{e}$ is the hydrogen density, and $c$ is the speed of light. 
The electron density can be written as a function of the AGN luminosity (i.e., L$_\mathrm{bol} = 10^{44.5}$ erg s$^{-1}$ from \citealt{Longinotti15}), the distance from the AGN ($r$) and the ionization parameter ($U$):
~
\begin{equation}
\begin{aligned}
\hskip5mm \mathrm{n_e} \approx 3.2 \left(\frac{\mathrm{L_{bol}}}{10^{45} \, \mathrm{erg \, s^{-1}}} \right) \left( \frac{r}{1 \, \mathrm{kpc}} \right)^{-2} \left( \frac{1}{U} \right) \, \mathrm{cm^{-3}}.
\end{aligned}
\label{eq:el_density}
\end{equation}

The $U$ parameter is defined according to the following expression:
\begin{equation}
\begin{aligned}
\log U = &\, -3.766 
+ 0.191 \log\left(\frac{[\mathrm{O\,III}]}{\mathrm{H}\beta}\right) 
+ 0.778 \log^{2}\left(\frac{[\mathrm{O\,III}]}{\mathrm{H}\beta}\right) \\
& - 0.251 \log\left(\frac{[\mathrm{N\,II}]}{\mathrm{H}\alpha}\right) 
+ 0.342 \log^{2}\left(\frac{[\mathrm{N\,II}]}{\mathrm{H}\alpha}\right).
\end{aligned}
\label{eq:ioniz_par}
\end{equation}
~
In our case, we derived the ionization parameter, $U$, and the electron density, n$_\mathrm{e}$, for the outflowing component by summing the flux contributions of the secondary and tertiary components. In particular, we used the secondary and tertiary components for the H$\alpha$, H$\beta$, and [OIII] lines, while for the [NII] line we only considered the secondary component associated with the outflow. 
Under these assumptions, we derived a median electron density, n$_\mathrm{e}$$\approx$2500 cm$^{-3}$ (Fig.~\ref{fig_ne_BNM}, bottom), which we adopt for the derivation of the outflow parameters (see Tables~\ref{out_props_Ha}, \ref{out_props_OIII_MRG_2c}, \ref{out_props_OIII_MRG_3c}, \ref{out_props_OIII_LRV_2c} and \ref{out_props_OIII_LRV_3c}). This value is roughly a factor of $\sim10$ larger than that derived using the [SII] method, and is in agreement with those found in the outflow component in nearby AGN systems (e.g., \citealt{Rose18, Singha22}). It is worth mentioning that, as noted in several works, the electron density can vary across the FoV; therefore, assuming a constant electron density can introduce an additional systematic uncertainty (e.g., \citealt{Kakkad18}).

For comparison, we also derived the outflow parameters adopting an electron density of 500 cm$^{-3}$, comparable to our [SII]-based estimate, and representative of local active galaxies (e.g. U/LIRGs, LINERs, type 2 AGN in the MAGNUM survey; \citealt{Fluetsch21, Venturi21, HM24}).

The classical [SII]$\lambda$6716/$\lambda$6731 ratio typically yields lower electron densities, characteristic of the diffuse ionized medium. In contrast, applying the \citet{Baron19} method results in higher densities, as it combines multiple line diagnostics with photoionization modeling. The discrepancy reflects the different sensitivity regimes: the [SII] ratio becomes insensitive above a few $\times$10$^3$ cm$^{-3}$, tracing mainly extended, low-density gas, whereas the \citet{Baron19} approach probes denser clumps that contribute significantly to the emission but are underrepresented in the [SII]-based estimate. Together, both measurements indicate a multiphase ionized medium, where compact, high-density knots coexist with more diffuse gas.

\begin{table*}[!h]
\centering
\caption{Kinematic properties as derived from the H$\alpha$ line from LR-R setup.}
\label{tab_Ha}
\begin{tabular}{lcccccc} 
\hline\hline\noalign{\smallskip}  
{ Gaussian fit} \&	 & v$_\mathrm{shear}$ & 	$\Delta$v$_\mathrm{mean}$ & $\sigma_\mathrm{mean}$	 & $\sigma_\mathrm{mean}^c$  & v/$\sigma$ \\ 
 Line-Component& (km s$^{-1}$) & (km s$^{-1}$) & (km s$^{-1}$)  & (km s$^{-1}$)  &   \\ 
\hskip7mm(1) & (2) &  (3)& (4)  & (5) & (6) & \\ 
\hline\noalign{\smallskip} 
3c, H$\alpha$-P	 & 211.4$\pm$8.8	& --   & 90.1$\pm$2.4 (70.8)	& 151.5$\pm$1.3 (149.0)  & 2.35$\pm$0.16 \\ 
\hline
\hskip5mm H$\alpha$-S 	& 169.8$\pm$60.7	& -39.2$\pm$5.4 (-44.4) 	&	 379.5$\pm$9.6 (318.2) & 354.8$\pm$4.0 (357.0)	&  0.45$\pm$0.15 	\\ 
\hline
\hskip5mm H$\alpha$-T 	& 140.1$\pm$17.1 & -252.5$\pm$6.4 (-243.9) 	& 927.0$\pm$11.2 (956.0) &  927.0$\pm$11.2 (956.0) &  0.15$\pm$0.14 \\	
\hline\hline\noalign{\smallskip} 
2c, [NII]-P	 & 211.4$\pm$8.8	& --   & 	 89.7$\pm$9.5 (70.6)  & --  & 2.36$\pm$0.35\\
\hskip5mm$\mathrm{[NII]}$-S 	& 227.8$\pm$69.3	& -66.8$\pm$6.1 (-76.7)	&	 434.8$\pm$20.9 (326.7) & (= [H$\alpha$-3])	&  0.53$\pm$0.19 	\\ 
\hline\hline\noalign{\smallskip} 	
1c, [SII]-P	 & 211.8$\pm$8.5	&   -- & 	 93.2$\pm$9.7 (110.9)  & 142.1$\pm$20.3 (142.1)$^c$  & 2.27$\pm$0.32 \\
\hline\hline\noalign{\smallskip} 	
1c, [OI]-P  & 123.9$\pm$17.2	&   -- & 	 114.4$\pm$3.1 (114.1)  & 120.4$\pm$3.0 (117.1)$^c$  & 1.08$\pm$0.12 \\ 
\hline\hline\noalign{\smallskip} 	
\end{tabular}
\begin{minipage}{18.5cm}
{{\bf Notes}: Column (1): Total number of Gaussians used for each line, Line and Component considered. Column (2): velocity shear, defined as half of the difference between the median value of the 5 percentile at each end of the velocity distribution: v$_\mathrm{shear}$ = $\frac{1}{2}$ (v$_\mathrm{max}^{5\%}$-v$_\mathrm{min}^{5\%}$). Column (3): mean (median) velocity offset between the secondary or tertiary component and the primary component (i.e., $\Delta$v = v$_\mathrm{i}$ - v$_\mathrm{sys}$, for $i$ = S or T). Column (4): mean (median) velocity dispersion. Column (5): mean (median) velocity dispersion derived within the area covered by the outflow. Column (6): (observed) dynamical ratio, defined as the ratio between the v$_\mathrm{shear}$ (Column 2) and $\sigma_\mathrm{mean}$ (Column 4). All these kinematic values include the contribution from the companion galaxy.
}
\end{minipage}
\end{table*}

\begin{table*}[!h]
\centering
\caption{Kinematic properties as derived from the H$\beta$-[OIII] complex from the MR-G setup.}
\label{tab_MR}
\begin{tabular}{lccccc} 
\hline\hline\noalign{\smallskip}  
{ Gaussian fit} \&	 & v$_\mathrm{shear}$ & 	$\Delta$v$_\mathrm{mean}$ & $\sigma_\mathrm{mean}$	&  v/$\sigma$ \\
Line-Component & (km s$^{-1}$) & (km s$^{-1}$) & (km s$^{-1}$)   &   \\
\hskip7mm(1) & (2) &  (3)& (4)  & (5)\\
\hline\noalign{\smallskip} 
3c, H$\beta$-P	 & 159.8$\pm$18.5	& --   		& 151.5$\pm$4.8 (145.0)	  & 1.06$\pm$0.13   \\
\hskip5mm H$\beta$-S 	& 82.7$\pm$41.6	& -37.3$\pm$18.8 (-52.9) 	&	 331.8$\pm$5.4 (354.5) 	&  0.25$\pm$0.13	\\
\hskip5mm H$\beta$-T   &  47.2$\pm$14.5  & 	-175.8$\pm$17.5 (-189.5)	&  917.0$\pm$7.3 (924.5)     & 0.05$\pm$0.02   \\
\hline\hline\noalign{\smallskip}
3c,~[OIII]-P  & 110.7$\pm$36.4	& --   					& 105.8$\pm$1.6 (108.8)	& 1.05$\pm$0.34	\\
\hskip5mm [OIII]-S   & 185.5$\pm$40.2	 & -174.0$\pm$9.0 (-157.6) 	&  223.4$\pm$5.3 (221.7) 	&  0.83$\pm$0.18\\
\hskip5mm [OIII]-T   &  284.7$\pm$5.6  & -580.6$\pm$31.5 (-537.1)	&  394.0$\pm$6.1 (398.7)     & 0.72$\pm$0.02  \\
\hline\hline\noalign{\smallskip} 
\end{tabular}
\begin{minipage}{18.5cm}
{{\bf Notes}: Same description as in Table~\ref{tab_Ha}.}
\end{minipage}
\end{table*}

\begin{table*}[!h]
\centering
\caption{Kinematic properties as derived from the H$\beta$-[OIII] complex from the LR-V setup. }
\label{tab_LR}
\begin{tabular}{lcccccc} 
\hline\hline\noalign{\smallskip}  
{ Gaussian fit} \&	 & v$_\mathrm{shear}$ & 	$\Delta$v$_\mathrm{mean}$ & $\sigma_\mathrm{mean}$	&  v/$\sigma$ \\
Line-Component & (km s$^{-1}$) & (km s$^{-1}$) & (km s$^{-1}$)   &  \\
\hskip7mm(1) & (2) &  (3)& (4)  & (5) \\
 \hline\noalign{\smallskip} 
1c,~H$\beta$-P	 & 243.1$\pm$25.3	& --   		& 119.4$\pm$6.2 (122.9)	 & 2.04$\pm$0.24 \\
\hline\hline\noalign{\smallskip}
{ 3c,~[OIII]-P}	 & 168.2$\pm$36.5	& --   					& 100.0$\pm$1.7 (99.5)		& 1.68$\pm$0.37  \\
\hskip5mm{ [OIII]-S} 	& 185.3$\pm$39.3	& -177.8$\pm$7.2 (-164.7) 	& 242.9$\pm$12.8 (237.2) 	&  0.76$\pm$0.17\\
\hskip5mm{ [OIII]-T} 	&  123.1$\pm$18.1   &  -537.6$\pm$14.1 (-535.7)	& 526.9$\pm$10.8 (519.8)      & 0.23$\pm$0.03 \\
\hline\hline\noalign{\smallskip} 
\end{tabular}
\begin{minipage}{18.5cm}
{{\bf Notes}: Same description as in Table~\ref{tab_Ha}.}
\end{minipage}
\end{table*}

\subsection{Deriving outflow parameters using the H$\alpha$ line}
\label{outflow_physics}

In order to understand which is the impact of the outflow in this galaxy, we first need to derive its main properties such as the mass of the outflow, M$_\mathrm{out}$, the mass outflow rate, $\mathrm{\dot{M}_{out}}$, the kinetic energy, E$_\mathrm{kin}$, the kinetic power, $\mathrm{\dot{E}_{kin}}$ and the momentum rate, $\mathrm{\dot{P}_{out}}$. These quantities have been derived as in \cite{Fiore17} (see also \citealt{Cresci17} and \citealt{Venturi23}, and references therein).
We first derived the mass of the outflow from the extinction corrected luminosity as derived from the integrated total H$\alpha$ flux of the tertiary component. 
This is a more robust estimation of the mass compared to that obtained using the [OIII] line because its luminosity does not depend on gas metallicity either on the energy of the ionizing photons (e.g., \citealt{Carniani15, Venturi21}). 
To correct the H$\alpha$ luminosity for the extinction we used the \cite{Calzetti00} attenuation law (assuming the typical R$_\mathrm{V}$=4.05 for starbursts) and an intrinsic ratio ($\mathrm{H\alpha/H\beta)_0= 3.1}$ (for an electron temperature of T$_\mathrm{e}=10^4$ K; \citealt{Osterbrock06}). We derived a (median) value of A$_\mathrm{V}$ = 2.73 mag for the outflow (see App.~\ref{more_info_AV} for details). 
We then computed the mass of the ionized outflow using the relation:
~
\begin{equation}
\mathrm{M_{out}} = 6.1 \times 10^8 \left( \frac{\mathrm{L_{H\alpha}}}{10^{44}\hskip1mm\text{erg} \hskip2mm\text{s}^{-1}} \right) \left( \frac{500\hskip1mm\text{cm}^{-3}}{\mathrm{n_e}} \right) \mathrm{M_\odot} ,
\label{eq:mass_ion_eq}
\end{equation}
~
where n$_\mathrm{e}$ has been derived as described in the previous section. 
We assumed two different values (i.e., n$_\mathrm{e} = 500$ and 2500 cm$^{-3}$; see Sect.~\ref{n_e_derivation} for details) and the outflow parameters obtained from using these values are shown in Table~\ref{out_props_Ha}. In both cases, we obtained similar outflow masses, in the range $\sim$0.6-3.2~$\times$ 10$^7$~$\mathrm{M_\odot}$.
M$_\mathrm{out}$ is the ionized gas outflow mass integrated within the outflow region. 

A key parameter for the characterization of the outflow kinematics is the maximum outflow velocity, v$_\mathrm{out}$, derived as in \cite{Rupke05}: $\mathrm{v_{out} = |\Delta v| + \frac{FWHM_{out}}{2} = |\Delta v| + 1.18\hskip1mm\sigma_{out}}$, where $\mathrm{\Delta v = v_{T} - v_{P}}$ is the maximum velocity difference between the outflowing (i.e., $\mathrm{v_{T}}$) and primary (i.e., $\mathrm{v_{P}}$) components, and $\sigma_\mathrm{out}$ is the median velocity dispersion of the outflow.

We then estimated the mass outflow rate, $\mathrm{\dot{M}_{out}}$, at a given radius, $\mathrm{R_{out}}$, using the formula from \citet{Fiore17} (see also \citealt{Lutz20}):
~
\begin{equation}
\mathrm{\dot{M}_{out} = C \cdot \frac{M_{out} \hskip2mm v_{out}}{R_{out}},}
\label{eq:mass_out}
\end{equation}
~
where M$_\mathrm{out}$ has been derived in Eq.~\ref{eq:mass_ion_eq}  and  R$_\mathrm{out}$ is the size of the outflow as derived in Sect.~\ref{size_outflow} (i.e., $\mathrm{R(H\alpha)} = 0.98 \pm 0.32$ kpc).
The $C$ factor depends on the adopted outflow history.
A value of $C=1$ assumes that the outflow is the result of a single explosive event in which clouds were ejected and continued to expand at a constant mass outflow rate until the present. In some cases, authors have assumed $C=3$, which implies that the outflow is continuously replenished by clouds ejected from the galactic gaseous disk (see \citealt{Lutz20} for further details; see also \citealt{Fiore17, HM24}). This assumption corresponds to a constant volume density of the outflowing gas and a mass outflow rate decreasing to zero over time. 
However, this scenario is at odds with the presence of a UFO in our object, as the continuously replenished outflow implied by $C=3$ assumes a slowly declining mass outflow rate fed by the galactic disk, whereas the UFO indicates a recent, highly energetic ejection directly from the AGN, inconsistent with a long-lived, disk-fed outflow.
Therefore, in our case, we considered the same assumption used in \cite{Marasco20} ($C=1$) who assumed a constant mass outflow rate during the flow time, defined as $\mathrm{R_{out}/v_{out}}$, which leads to a decreasing density of the outflowing gas. This approach is consistent with the `time-averaged thin-shell' model (e.g., \citealt{Rupke05}), and is widely used to derive outflow rates in both ionized and neutral gas phases (e.g., \citealt{Heckman15, Gonzalez17, CCT20, Marasco20}).
This assumption ($C = 1$) also represents a more conservative approach for deriving the outflow parameters. Since the $C$ factor is constant, it does not affect the main conclusions of this work, although we have taken the different approaches into account to compare our results with previous studies (see \citealt{Longinotti23}).
Under these assumptions, we derived a mass outflow rate between $\sim$10-50  M$_\odot$ yr$^{-1}$ using n$_\mathrm{e} = 2500$ or 500 cm$^{-3}$, respectively, as reported in Table~\ref{out_props_Ha}.

We then computed the kinetic energy, E$_\mathrm{kin}$, and the kinetic power, $\mathrm{\dot{E}_{kin}}$, of the outflow as in \cite{Rose18} (also \citealt{Santoro20}):
~
\begin{equation}
\mathrm{E_{\text{kin}} = \frac{1}{2}  \sigma_{out}^2 M_{out}}, 
\label{eq:en_kin}
\end{equation}
~
\begin{equation}
\mathrm{\dot{E}_{\text{kin}} = \frac{\dot{M}_{out} }{2} (v_{out}^2 + 3 \sigma_{out}^2)}. 
\label{eq:kin_power}
\end{equation}
~
The corresponding momentum rate of the outflow is derived as $\mathrm{\dot{P} = \dot{M}_{out} \hskip1mm v_{out}}$, which yields a few $\times$10$^{35}$ cm g s$^{-2}$. When we compare the momentum rate of the outflow with the radiative momentum rate of the AGN, defined as $\mathrm{\dot{P}_{rad} = L_{bol}/c} = 1.7 \times 10^{34}$ cm g s$^{-2}$ for our source, we obtained a ratio $\mathrm{\dot{P}_{H\alpha}/\dot{P}_{rad}}$ between $\sim$5 and 27 (depending on the assumed $\mathrm{n_e}$ value; see Table~\ref{out_props_Ha}). Under our assumptions, we derive $\mathrm{\dot{P}_{H\alpha}/\dot{P}_{rad}} = 5.3 \pm 3.2$.

\begin{figure}
\centering
\includegraphics[width=0.47\textwidth]{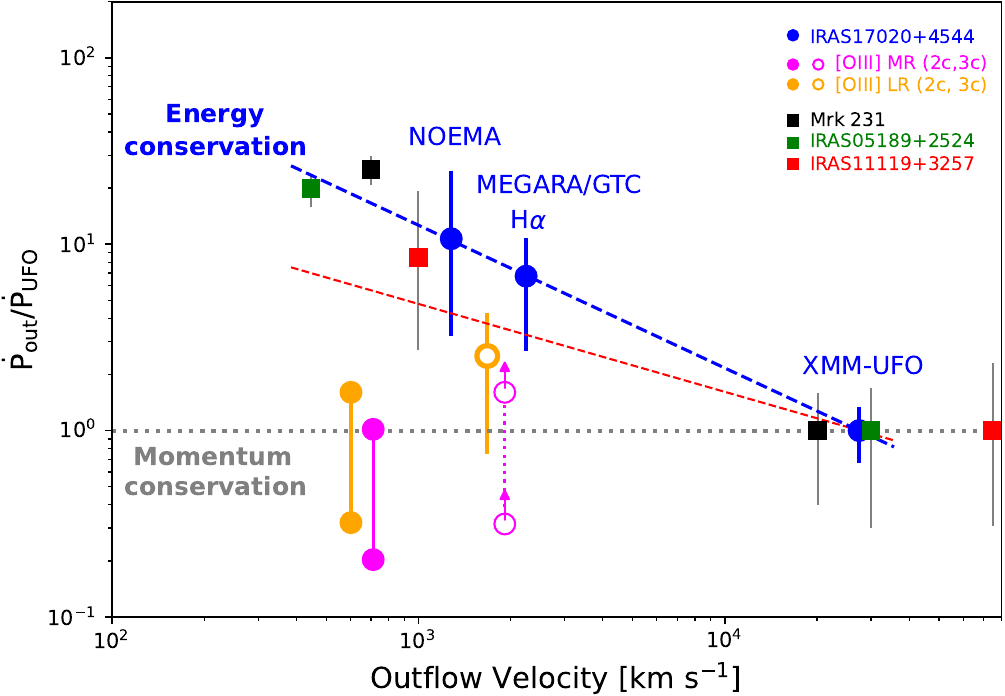}
\caption{
Force of the molecular, ionized, and X-ray phases of the outflow, $\mathrm{\dot{P}_{out}}$, in IRAS17, normalized to the source's X-ray momentum rate, $\mathrm{\dot{P}_{UFO}}$ (i.e., $\mathrm{\dot{P}_{UFO}/\dot{P}_{rad} }= 1.87\pm0.62$; see \citealt{Longinotti18}), is shown as a function of the outflow velocities.  The results for IRAS17 derived from NOEMA, MEGARA (i.e., H$\alpha$ using $\mathrm{n_e} = 2500$ cm$^{-3}$) and XMM-Newton are shown with blue circles. To compare results across the three phases, we used the same formula for v$_\mathrm{out}$ as in \cite{Longinotti23} (i.e., $\Delta$v + 2$\times\sigma_B$) and normalized their $\mathrm{\dot{M}_{out}}$ values assuming $C=1$.
Filled circles in orange and magenta represent the results derived for 2c component of the [OIII] emission (i.e., slow outflow) in the LR and MR setups, respectively. Two values are plotted for each setup (connected by a solid line), corresponding to the two assumed electron densities, $\mathrm{n_e} = 500$ and 2500 cm$^{-3}$ (see text for details). 
Empty circles, following the same color code as for the 2c component, indicate the 3c component of the [OIII] emission (i.e., fast outflow) in both setups. 
The fast [OIII] outflow is spatially resolved in the LR setup (empty orange circle), while it remains unresolved in the MR setup (empty magenta circles), for which we show the two results (lower limits) derived assuming the two n$_\mathrm{e}$ values. 
Results for three other well-known sources for which the `energy-conserving' regime has been observed, Mrk~231, IRAS~05189+2524 and IRAS11119+3257 (all ULIRGs, shown using black, green and red squares, respectively), are also included. Their data are extracted from \cite{Feruglio15}, \cite{Lutz20} and \cite{Tombesi15}, respectively (see also \citealt{Marasco20} and references therein).
The blue and red dashed lines represent the two linear fits marking the `energy-conserving' outflow regime: the former obtained using CO, H$\alpha$, and X-ray data, and the latter using these three points plus the [OIII] result from the LR-V setup (empty orange circle). The grey dotted line indicates the prediction for `momentum-conserving' outflows (i.e., $\mathrm{\dot{P}_{out}/\dot{P}_{UFO} \sim 1}$).
}
\label{fig_energy_all}
\end{figure}

According to our results, we can now compare the momentum rate values of the outflow (i.e., $\mathrm{\dot{P}_{out}/\dot{P}_{rad}}$) obtained for the different gas phases (molecular, ionized and highly ionized, derived, respectively, in \citealt{Longinotti23}, this work and \citealt{Longinotti15}) as a function of the outflow velocity (see Fig.~\ref{fig_energy_all}). This plot is useful for comparing the energy transfer predicted by different models between UFOs and galaxy-scale outflows, under the assumption that the molecular phase carries most of the outflow mass (i.e., $\mathrm{\dot{P}^{tot}_{out} \sim \dot{P}^{mol}_{out}}$ ; \citealt{Tombesi15, Feruglio15}).

In order to compare our results with those from previous works (i.e., \citealt{Longinotti23}), we needed to homogenize the formulas used in this study with those adopted in earlier ones. In particular, \cite{Longinotti23} used the formula proposed by \citet{Rupke13} for the derivation of the outflow velocity, which assumes v$_{\text{out}}$ = $\Delta$v + 2\hskip1mm$\sigma_{\text{out}}$, while in this work we followed the definition by \citet{Rupke05}. The former gives larger outflow velocity than the latter used in this work. In the case of MEGARA data, we would have derived larger outflow velocity by a factor of $\sim$1.5.
On the other hand, the $C$ factor used by \cite{Longinotti23} to derive the mass outflow rate is 3, whereas in our case we assumed $C = 1$. Both these assumptions increase the velocity and the energetics of the outflow, making the outflow to appear more extreme. Since we preferred to be more conservative in the derivation of the outflow physical parameters, we assumed the $C=1$ as first assumption maintaining the (higher) v$_\text{out}$ formula by \cite{Rupke13} for a direct comparison of the results.

We finally reported the results by \cite{Longinotti18} and \cite{Longinotti23} adjusted according to the $C=1$ assumption for a direct comparison with our results (i.e., $\mathrm{\dot{M}_{out}}$ and $\mathrm{\dot{P}_{out}}$; see Table~\ref{out_props_Ha}). The H$\alpha$ outflow, as measured by MEGARA, confirms the presence of the `energy-conserving' regime for the ionized gas phase, previously identified for the molecular emission, in which a momentum boost is required as a function of the outflow velocity (i.e., $\mathrm{\dot{P} \propto v_{out}^{-1}}$). The ionized gas retains a significant fraction of the energy from the inner wind, suggesting that AGN feedback is dynamically important and can have a strong impact on galaxy evolution. The results are shown in Fig.~\ref{fig_energy_all}.

\subsection{Deriving outflow parameters from the [OIII] line}

We also determined the mass of gas contained in the outflow using the [OIII] lines, which can be expressed with the relation (see \citealt{Carniani15, Fiore17, Venturi23}): 
\begin{equation}
\mathrm{M^{[OIII]}_{out}= 8.0 \times 10^7 \left( \frac{L_{[OIII]}}{10^{44} \, \text{erg} \, \text{s}^{-1}} \right) \left( \frac{500 \, \text{cm}^{-3}}{<n_e>} \right) \frac{\mathcal{CF}}{10^{[O/H]-[O/H]_\odot}}M_\odot },
\end{equation}
where <n$_\mathrm{e}$> is the electron density averaged over the ionized outflow volume (i.e., $<$n$_\mathrm{e}$$>$=$\int{n_e f dV}$/$\int{ f dV})$ and $\mathcal{CF}$=$\mathrm{<n_e>^2}$/$\mathrm{<n_e^2>}$ is the so-called `condensation factor'. 
Under the simplifying assumption that all ionizing gas clouds have the same density, we set $\mathcal{CF} = 1$, thereby eliminating the dependence of the outflow mass on the filling factor of the emitting clouds. The [O/H] represents the oxygen abundance relative to the solar value, [O/H]$_\odot$. We assumed that the metallicity of the outflowing material is always assumed solar ([O/H] = [O/H]$_\odot$).
The $\mathrm{L_{[OIII]}}$ is the outflow luminosity corrected for the extinction (see App.~\ref{more_info_AV}). 
Since the [O III] lines are highly sensitive to both temperature and ionization, they are poor tracers of the outflow mass. Therefore, this method can, at best, provide only an order-of-magnitude estimate of the outflow properties.
Finally, in order to estimate the total outflow mass using [OIII] lines, we considered the approximation used by \cite{Fiore17}, for which $\mathrm{M_{tot}^{out} = 3\times M^{[OIII]}_{out}}$ (see also \citealt{Peralta23}).
In order to compute the energetics of the ionized outflow, we considered the same equations used for the H$\alpha$ line (see Eqs.~\ref{eq:mass_out}, \ref{eq:en_kin}, \ref{eq:kin_power}).
The outflow mass derived using the [OIII] line in the LR-V setup is higher than that obtained with the MR-G configuration by a factor of two. This difference can be attributed to the longer exposure time in LR-V (twice that used for the MR-G grism) and better seeing conditions, which result in higher sensitivity to lower surface brightness, mostly found in the outer regions of the galaxy (see Figs.~\ref{fig_OIII_MR_LR_new} and \ref{LRV_setups_maps}).
Nevertheless, the kinematic values derived for the outflow are consistent (within the errors) between the two setups.

In the [OIII] emission, we identified two distinct outflows. The secondary outflow component is less kinematically extreme yet more spatially extended (R$_\mathrm{out}$$\sim$0.7 kpc) than its tertiary counterpart in both the MR-G and LR-V setups. The tertiary outflow component is compact (R$_\mathrm{out}$$\sim$0.5 kpc) and kinematically extreme, characterized by large velocity shifts and broad line widths. It remains unresolved at the spatial resolution of the MR-G setup, consistent with the nuclear base of the outflow, but is resolved in the LR-V setup.
In contrast, the lower-ionization lines show little evidence of outflowing gas: [NII] exhibits only a faint broad component, while [S II] shows none. This indicates that these lines are intrinsically weak in the highly ionized outflowing gas. Consequently, high-ionization lines such as [OIII] provide the clearest tracers of the outflow, and the limited broad emission in [NII] and [SII] reflects their low intrinsic flux rather than limitations in spectral decomposition on a spaxel-by-spaxel basis. Consequently, [OIII], along with H$\alpha$, provides the most robust tracer of the outflow kinematics in this source.

In Tables~\ref{out_props_OIII_MRG_2c} to \ref{out_props_OIII_LRV_3c}, we present the results obtained for the [OIII] line with both the setups.

\begin{table*}[h]
\caption{Outflow properties derived from the H$\alpha$ emission based on MEGARA data (tertiary component).}
\label{out_props_Ha}
\resizebox{\textwidth}{!}{%
\begin{tabular}{ccccccccc} 
\hline\hline\noalign{\smallskip}  
Instrument 	& v$_\mathrm{out}$ & M$_\mathrm{out}$	 & $\mathrm{\dot{M}_{out}}$ & 	E$_\mathrm{kin}$ & $\mathrm{\dot{E}_{kin}}$	 & $\mathrm{\dot{P}_{out}}$  & $\mathrm{\dot{P}_{out}/\dot{P}_{rad}}$  & Assumptions \\
 	&	(km s$^{-1}$) & (M$_\odot$) & (M$_\odot$ yr$^{-1}$) &  (erg) & (erg s$^{-1}$) & (cm g s$^{-2}$) & 		& 	\\
 	&	 &  $\times$10$^7$ &  &  $\times$10$^{56}$& $\times$10$^{43}$ &$\times$10$^{35}$  & 		& 	\\
 (1) & (2) &  (3)& (4)  & (5) & (6) & (7) & (8) & (9)\\
 \hline\noalign{\smallskip} 
MEGARA 		&   1461.0$\pm$155.9    &3.2$\pm$0.4  & 49.2$\pm$18.3  & 2.9$\pm$0.8  &  7.6$\pm$3.1 & 4.5$\pm$1.8   & 26.7$\pm$10.3 & RV05; n$_\mathrm{e}$=500 \\
{\bf MEGARA }     &   1461.0$\pm$155.9    &0.6$\pm$0.4  & 9.8$\pm$6.7 & 0.6$\pm$0.4  &  1.5$\pm$1.0 & 0.9$\pm$0.6    & 5.3$\pm$3.2 & RV05; n$_\mathrm{e}$=2500 \\
\hline\noalign{\smallskip} 
$^\star${MEGARA } &    2247.3$\pm$248.9      & 3.2$\pm$0.4   & 75.7$\pm$28.2    & 2.9$\pm$0.8  &  18.6$\pm$7.6 & 10.7$\pm$4.2   & 61.3$\pm$24.5  &RV13;  n$_\mathrm{e}$=500 \\
$^\star${MEGARA } &    2247.3$\pm$248.9  & 0.6$\pm$0.4  & 15.1$\pm$10.3    & 0.6$\pm$0.4  &  3.7$\pm$2.6   & 2.1$\pm$1.5   & 12.6$\pm$8.7  &RV13;  n$_\mathrm{e}$=2500\\
\hline\hline\noalign{\smallskip}
LMT [L18]		& -660 [-1600] & 15.4$\pm$4.9 		& 	480 [1162] & -- & --  & 20 	& 117 [11-670]& RV13\\
\hline\noalign{\smallskip}
$^\star${NOEMA [L23]} &  	1280$\pm$480	& 	10.0$\pm$3.5 	 & 46 [21-78]  & -- & -- &  $\sim$3.4 [1-8]	& { $\sim$20 [6-46]}	& RV13 ($\alpha$=0.5)\\
\hline\noalign{\smallskip}
$^\star${NOEMA [L23]}  &  1280$\pm$480	& 15.8$\pm$5.4	 & $\sim$73 [34-122]  & -- & -- &  $\sim$5 [1.5-12]	&  $\sim$31 [9-72]	& RV13 ($\alpha$=0.8) \\
\hline\hline\noalign{\smallskip}
XMM-Newton [L15]& 	$\sim$30000 & & & &  & 	&	1.87$\pm$0.62\\
\hline\hline\noalign{\smallskip}
\end{tabular}
}
\begin{minipage}{18.5cm}
{\small
{\bf Notes}: 
Col.~(1): Instrument considered. The references L15, L18, and L23 correspond to \citet{Longinotti15}, \citet{Longinotti18}, and \citet{Longinotti23}, which report the derivation of the outflow parameters based on XMM-Newton, LMT, and NOEMA data, respectively. The original and scaled results from these datasets are also shown. In boldface we highlight the main MEGARA results derived under our assumptions (see text). The star symbol~($^\star$) marks the NOEMA and MEGARA results that can be directly compared, since they used the same v$_\mathrm{out}$ formula (see Col.~9). The NOEMA values were normalized using $C = 1$ (see text for details). 
Col~(2): velocity of the outflow derived according to the formula described in Col~(9). Col~(3-7): mass of the outflow, mass outflow rate, kinetic energy of the outflow and kinetic power of the outflow obtained following Eqs.~\ref{eq:mass_out}, \ref{eq:en_kin}, \ref{eq:kin_power} (see text for details). The molecular radius derived in \cite{Longinotti23} is R$_\mathrm{CO}$=2.8$\pm$0.3~kpc. The intrinsic ionized outflow radius derived in this work is $\mathrm{R_{out}(H\alpha)}$=0.98$\pm$0.32 kpc.
Col~(8): ratio between the momentum rate of the outflow $\mathrm{\dot{P}_{out}}$ and $\mathrm{\dot{P}_{rad}=L_{bol}/c=1.7\times10^{34}}$ g cm s$^{-2}$ is the radiation force of the AGN, being L$_\mathrm{bol}$=5.2 $\times$10$^{44}$ erg s$^{-1}$ (\citealt{Giroletti17, Longinotti18}). 
Col~(9): Assumptions considered to derive the outflow parameters with the different instruments: (i) to derive the ionized outflow velocity, two formulas are used: $\mathrm{v_{out} = \Delta v + 1.18\hskip1mm\sigma_B}$ from \cite{Rupke05} (RV05), and $\mathrm{v_{out} = \Delta v + 2\hskip1mm\sigma_B}$ from \cite{Rupke13} (RV13); (ii) to derive the ionized outflow mass, two electron densities (n$_\mathrm{e}$ = 500 and 2500 cm$^{-3}$) are considered for the MEGARA data; (iii) to derive the molecular outflow mass, two CO-to-H2 conversion factor values, $\alpha_\mathrm{CO}$, are considered in \cite{Longinotti23}: $\alpha_\mathrm{CO}$ = 0.5 and 0.8 M$_\odot$ (K km s$^{-1}$ pc$^2$)$^{-1}$.
}
\end{minipage}
\end{table*}

\begin{table*}[!h]
\centering
\caption{Outflow properties for secondary (resolved) component of the [OIII] line derived from the MR-G setup.}
\label{out_props_OIII_MRG_2c}
\resizebox{\textwidth}{!}{%
\begin{tabular}{ccccccccc} 
\hline\hline\noalign{\smallskip}  
Instrument	& v$_\mathrm{out}$ & M$_\mathrm{out}$	 & $\mathrm{\dot{M}_{out}}$ & 	E$_\mathrm{kin}$ & $\mathrm{\dot{E}_{kin}}$	 & $\mathrm{\dot{P}_{out}}$  & $\mathrm{\dot{P}_{out}/\dot{P}_{rad}}$  & Assumptions \\
 	&	(km s$^{-1}$) & (M$_\odot$) & (M$_\odot$ yr$^{-1}$) &  (erg) & (erg s$^{-1}$) & (cm g s$^{-2}$) & 		& 	\\
 	&	 &  $\times$10$^6$ &  &  $\times$10$^{54}$& $\times$10$^{42}$ &$\times$10$^{34}$  & 		& 	\\
 (1) & (2) &  (3)& (4)  & (5) & (6) & (7) & (8) & (9)\\
\hline\noalign{\smallskip} 
MEGARA    &   550.8$\pm$78.9   & 6.7$\pm$0.8  & 5.4$\pm$2.1  & 2.6$\pm$1.1  &  0.71$\pm$0.3	& 1.9$\pm$0.8 & 1.1$\pm$0.5 & RV05; n$_\mathrm{e}$=500\\
{\bf MEGARA}   &   550.8$\pm$78.9   & 1.3$\pm$0.8 & 1.1$\pm$0.8  &  0.5$\pm$0.4 &  0.14$\pm$0.1 & 0.4$\pm$0.3  & 0.2$\pm$0.2 & RV05; n$_\mathrm{e}$=2500\\
\hline\noalign{\smallskip}
$^\star$MEGARA     &    712.0$\pm$101.2   & 6.7$\pm$0.8  & 7.0$\pm$2.7 & 2.6$\pm$1.1 &  1.4$\pm$0.6 & 3.2$\pm$1.3    & 1.9$\pm$0.8& RV13; n$_\mathrm{e}$=500\\
$^\star$MEGARA  &    712.0$\pm$101.2   & 1.3$\pm$0.8  & 1.4$\pm$1.0    & 0.5$\pm$0.4 &  0.3$\pm$0.20 & 0.6$\pm$0.4   & 0.4$\pm$0.3 &RV13; n$_\mathrm{e}$=2500\\
\hline\hline\noalign{\smallskip}
\end{tabular}
} 
\begin{minipage}{18.5cm}
\small
{{\bf Notes}: Same notes as in Table~\ref{out_props_Ha} but for the [OIII] line. The intrinsic ionized outflow size derived using this setup is $\mathrm{R_{out}([OIII])}=0.70\pm0.41$ kpc.}
\end{minipage}
\end{table*}
~
\begin{table*}[!h]
\centering
\caption{Outflow properties for tertiary (unresolved) component of the [OIII] line derived from the MR-G setup. }
\label{out_props_OIII_MRG_3c}
\resizebox{\textwidth}{!}{%
\begin{tabular}{ccccccccc} 
\hline\hline\noalign{\smallskip}  
Instrument	& v$_\mathrm{out}$ & M$_\mathrm{out}$	 & $\mathrm{\dot{M}_{out}}$ & 	E$_\mathrm{kin}$ & $\mathrm{\dot{E}_{kin}}$	 & $\mathrm{\dot{P}_{out}}$  & $\mathrm{\dot{P}_{out}/\dot{P}_{rad}}$  & Assumptions \\
 	&	(km s$^{-1}$) & (M$_\odot$) & (M$_\odot$ yr$^{-1}$) &  (erg) & (erg s$^{-1}$) & (cm g s$^{-2}$) & 		& 	\\
 	&	 &  $\times$10$^6$ &  &  $\times$10$^{54}$& $\times$10$^{42}$ &$\times$10$^{34}$  & 		& 	\\
 (1) & (2) &  (3)& (4)  & (5) & (6) & (7) & (8) & (9)\\
\hline\noalign{\smallskip} 
$^\star$MEGARA  &    1914.0$\pm$289.3   & 2.7$\pm$0.3  & 4.1$\pm$1.6 & 11.5$\pm$4.8 & 6.4$\pm$2.9 & 4.9$\pm$2.1   & 3.0$\pm$1.2 & RV13; n$_\mathrm{e}$=500\\
$^\star$MEGARA  &    1914.0$\pm$289.3   & 0.5$\pm$0.3  & 0.8$\pm$0.6 & 2.3$\pm$1.6 &  1.3$\pm$0.9 & 1.0$\pm$0.7   & 0.6$\pm$0.4 & RV13; n$_\mathrm{e}$=2500\\
\hline\hline\noalign{\smallskip}
\end{tabular}
}
\begin{minipage}{18.5cm}
\small
{{\bf Notes}: Same notes as in Table~\ref{out_props_Ha} but for the unresolved tertiary component of the [OIII] line. The estimated ionized outflow size is an upper limit given by PSF size, $\mathrm{R_{out}([OIII])}<1.29$ kpc. These values have been derived using the integrated spectrum within the PSF region.}
\end{minipage}
\end{table*}

\begin{table*}[h]
\centering
\begin{small}
\caption{Outflow properties for secondary (resolved) component of the [OIII] line derived from the LR-V setup.}
\label{out_props_OIII_LRV_2c}
\resizebox{\textwidth}{!}{%
\begin{tabular}{ccccccccc} 
\hline\hline\noalign{\smallskip}  
Instrument	& v$_\mathrm{out}$ & M$_\mathrm{out}$	 & $\mathrm{\dot{M}_{out}}$ & 	E$_\mathrm{kin}$ & $\mathrm{\dot{E}_{kin}}$	 & $\mathrm{\dot{P}_{out}}$  & $\mathrm{\dot{P}_{out}/\dot{P}_{rad}}$  & Assumptions \\
 	&	(km s$^{-1}$) & (M$_\odot$) & (M$_\odot$ yr$^{-1}$) &  (erg) & (erg s$^{-1}$) & (cm g s$^{-2}$) & 		& 	\\
 	&	 &  $\times$10$^6$ &  &  $\times$10$^{54}$& $\times$10$^{42}$ &$\times$10$^{34}$  & 		& 	\\
 (1) & (2) &  (3)& (4)  & (5) & (6) & (7) & (8) & (9)\\
\hline\noalign{\smallskip} 
MEGARA 	      &    451.6$\pm$63.9    & 14.9$\pm$1.9   & 9.8$\pm$3.8  & 5.0$\pm$2.1  &  0.9$\pm$0.4   & 2.8$\pm$1.1         & 1.7$\pm$0.7  & RV05; n$_\mathrm{e}$=500\\
{\bf MEGARA}       &    451.6$\pm$63.9 &  3.0$\pm$1.7   & 2.0$\pm$1.4    & 1.0$\pm$0.7    &  0.2$\pm$0.1   & 0.6$\pm$0.4  & 0.3$\pm$0.2   & RV05; n$_\mathrm{e}$=2500\\
\hline\noalign{\smallskip}
$^\star$MEGARA &    602.4$\pm$87.3   & 14.9$\pm$1.9  & 13.1$\pm$5.0      & 5.0$\pm$2.1   &  1.9$\pm$0.9  & 5.0$\pm$2.0    & 3.0$\pm$1.2  & RV13; n$_\mathrm{e}$=500\\
$^\star$MEGARA &    602.4$\pm$87.3   & 3.0$\pm$1.7   & 2.6$\pm$1.8    & 1.0$\pm$0.7   &  0.4$\pm$0.3  & 1.0$\pm$0.7   & 0.6$\pm$0.4  & RV13; n$_\mathrm{e}$=2500\\
\hline\hline\noalign{\smallskip}
\end{tabular}
}
\end{small}
\begin{minipage}{18.5cm}
\small
{{\bf Notes}: Same notes as in Table~\ref{out_props_Ha} but for the [OIII] line. The ionized outflow size derived using this setup is $\mathrm{R_{out}([OIII])=0.70\pm0.26}$ kpc.}
\end{minipage}
\end{table*}
~
\begin{table*}[h]
\centering
\begin{small}
\caption{Outflow properties for tertiary (resolved) component of the [OIII] line derived from the LR-V setup.}
\label{out_props_OIII_LRV_3c}
\resizebox{\textwidth}{!}{%
\begin{tabular}{ccccccccc} 
\hline\hline\noalign{\smallskip}  
Instrument	& v$_\mathrm{out}$ & M$_\mathrm{out}$	 & $\mathrm{\dot{M}_{out}}$ & 	E$_\mathrm{kin}$ & $\mathrm{\dot{E}_{kin}}$	 & $\mathrm{\dot{P}_{out}}$  & $\mathrm{\dot{P}_{out}/\dot{P}_{rad}}$  & Assumption  \\
 	&	(km s$^{-1}$) & (M$_\odot$) & (M$_\odot$ yr$^{-1}$) &  (erg) & (erg s$^{-1}$) & (cm g s$^{-2}$) & 		& 	\\
 	&	 &  $\times$10$^6$ &  &  $\times$10$^{54}$& $\times$10$^{42}$ &$\times$10$^{34}$  & 		& 	\\
 (1) & (2) &  (3)& (4)  & (5) & (6) & (7) & (8) & (9)\\
\hline\noalign{\smallskip}
MEGARA  &    1238.1$\pm$175.1   & 10.3$\pm$0.1  & 27.6$\pm$10.6   & 29.4$\pm$12.4 &  20.8$\pm$9.4  	& 21.5$\pm$8.8    & 12.9$\pm$5.2     & RV05; n$_\mathrm{e}$=500\\
{\bf MEGARA}  &    1238.1$\pm$175.1   &  2.1$\pm$1.2      & 5.5$\pm$3.8   & 5.9$\pm$4.2  &  4.2$\pm$3.0  	& 4.3$\pm$3.0    & 2.6$\pm$1.9 & RV05; n$_\mathrm{e}$=2500\\
\hline\noalign{\smallskip}
$^\star$MEGARA &    1677.8$\pm$246.3   & 10.3$\pm$1.3  & 37.4$\pm$14.5 & 29.4$\pm$12.4  &  43.4$\pm$19.8 & 39.5$\pm$16.4    & 23.7$\pm$9.6       & RV13; n$_\mathrm{e}$=500\\
$^\star$MEGARA &    1677.8$\pm$246.3   & 2.1$\pm$1.2    & 7.5$\pm$5.2   & 5.9$\pm$4.2    &  8.7$\pm$6.3   & 7.9$\pm$5.6     & 4.7$\pm$3.3 & RV13; n$_\mathrm{e}$=2500\\
\hline\hline\noalign{\smallskip}
\end{tabular}
}
\end{small}
\begin{minipage}{18.5cm}
\small
{{\bf Notes}: Same notes as in Table~\ref{out_props_Ha} but for the [OIII] line. The ionized outflow size derived using this setup is $\mathrm{R_{out}([OIII])=0.47\pm0.25}$ kpc.}
\end{minipage}
\end{table*}

\section{Discussion}
\label{sect_discussion}

\subsection{Tracing stratification within the ionized gas phase}

Our analysis of the MEGARA data show that the ionized outflow in IRAS17 is consistent with a multiphase, shock-driven wind, as predicted by theoretical models of AGN `energy-conserving' outflows. A compact, high-velocity phase is detected in both H$\alpha$ (R$_\mathrm{out}$$\sim$1.0 kpc and v$_\mathrm{out}$$\sim$1460 km s$^{-1}$) and [OIII] emission (i.e., tertiary component in the LR-V setup with R$_\mathrm{out}$$\sim$0.5 kpc,  v$_\mathrm{out}$$\sim$1240 km s$^{-1}$), tracing ionized gas compressed and accelerated by the nuclear wind. A secondary [OIII] component with more modest velocities (v$_\mathrm{out}$$\sim$450 km s$^{-1}$ in the LR-V setup) at R$_\mathrm{out}$$\sim$0.7~kpc traces a diffuse, radiatively cooled envelope that has become momentum-dominated. This configuration naturally reproduces a stratified structure, where the inner regions efficiently transfer kinetic energy into the dense ambient medium, while the outer layers represent a slower, cooled phase.

The H$\alpha$ and [OIII] lines trace gas under different physical conditions within the outflow. H$\alpha$ emission arises from denser ionized material at T$\sim$10$^{4}$ K, where recombination is efficient, typically associated with photoionized clumps or shock-compressed regions. In contrast, [OIII] originates from more highly ionized and generally lower-density gas exposed to the AGN radiation field or fast shocks. Owing to its relatively high critical density (n$_\mathrm{e}$$\sim$6.8$\times10^{5}$ cm$^{-3}$), [OIII] is efficiently emitted in diffuse ionized regions, whereas in denser environments the line becomes suppressed by collisional de-excitation.
These different excitation and density requirements naturally lead to distinct radial extents for the two tracers. The more compact [OIII] emission (R$_\mathrm{out}$$\sim$0.5-0.7 kpc) originates where the gas remains sufficiently diffuse and strongly illuminated by the AGN. In contrast, H$\alpha$ emission (R$_\mathrm{out}$$\sim$1 kpc) is detected farther out because denser ionized clumps can survive and cool efficiently at larger distances, including regions where shocks compress the gas within the outflow.

In IRAS17, the detection of higher velocities in H$\alpha$ suggests that this line is tracing denser or partially obscured clumps of ionized gas. These clumps remain visible in H$\alpha$, while the [OIII] emission is partially suppressed due to collisional de-excitation in dense regions and attenuation by dust. The velocities derived from both ionized lines are consistent with those measured for the molecular outflow with NOEMA ($\sim$1000-1800  km  s$^{-1}$ from \citealt{Longinotti23}), supporting a coherent, multiphase outflow powered by the active nucleus (see next section).

As found by \cite{Fiore17} and other works (e.g, \citealt{Harrison18}), the coupling efficiency $\mathrm{\dot{E}_{out}}$/$\mathrm{L_{bol}}$ can cover a large range of values, and depends on the gas phase considered. 
In particular, theoretical AGN feedback models predict a typical coupling efficiency of $\sim$5\% between AGN power and the surrounding dense gas, corresponding to the molecular gas phase (\citealt{DiMatteo05, Hopkins10, Zubovas12, Costa14}; see also Fig. 2 in \citealt{Harrison18}, which compares theoretical predictions to observations). This coupling efficiency is required to reproduce the observed 
M$_{BH}$-$\sigma$ relation in local galaxies. In contrast, a much lower coupling efficiency ($<$1\%, i.e., 0.01-0.1\%) is typically observed in ionized outflows (e.g., \citealt{Carniani15, Bischetti17, Musiimenta23}). Nevertheless, as shown in the simulations of \cite{Hopkins10} and summarized in \cite{Harrison18}, even such low coupling efficiencies in ionized winds can still have a significant impact on the host galaxy.

In IRAS17, the comparison between the kinetic powers of the two [OIII] outflow (i.e., slow and fast) components indicates that both can be powered by the AGN. The fast component dominates the energetics, with $\mathrm{\dot{E}_{fast}}$=0.8\% $\times \mathrm{L_{bol}}$, while the slower component requires only a small fraction of this energy ($\mathrm{\dot{E}_{slow}}$=0.04\% $\times \mathrm{L_{bol}}$, corresponding to $\sim$5\%$\times \mathrm{\dot{E}_{fast}}$). The resulting energy ratio indicates that the AGN has sufficient power to drive both phases simultaneously according to simulations. 
Within this framework, the slow [OIII] outflow may also naturally arise from the interaction of the fast AGN-driven outflow with the surrounding ISM, through shocks and partial energy transfer. 

The inferred kinetic coupling efficiencies for the slow and fast [OIII] outflows (i.e., $\mathrm{\dot{E}_{out}}$/$\mathrm{L_{bol}}$=0.04\%-0.8\%) together with the moderate momentum boosts ($\mathrm{\dot{P}_{out}}$/$\mathrm{\dot{P}_{rad}}$$\sim$0.3-3, respectively) indicate that the ionized outflows operate in a `momentum-driven' or weakly `energy-driven' regime, respectively. In this scenario, only a small fraction of the AGN radiative or mechanical power is converted into bulk kinetic energy of the ionized gas, while the momentum flux of the fast component slightly exceeds the single-scattering limit ($\mathrm{\dot{P}_{out}}$ $\sim$ $\mathrm{\dot{P}_{rad}}$), likely due to shocks and multiple interactions within a clumpy ISM. 
The slow outflow shows a globally blueshifted kinematic pattern and might trace gas at lower velocity within the same ionized outflow due to projection effects, while the fast outflow traces outflowing gas directed more towards the line of sight, therefore showing more extreme blueshifted velocities. Both the slow and fast [OIII] components can be considered as part of a single physical outflow, with kinematic differences arising from projection effects, orientation, and interactions with the surrounding ISM.
The lower-velocity component, with a smaller momentum boost, can naturally arise from the interaction of the fast AGN-driven outflow with the surrounding ISM. Such efficiencies and momentum fluxes are commonly observed in ionized AGN-driven outflows and are consistent with theoretical expectations for inefficient coupling between the AGN and the ambient gas (e.g., \citealt{Harrison18}).

\subsection{Connecting the different outflow phases}

In the model proposed by \cite{Faucher12}, AGN-driven outflows generally operate in the `energy-driven' regime, in which the thermal energy of the shocked wind bubble is efficiently transferred to the swept-up ISM, powering the large-scale expansion (see also \citealt{Zubovas12, King15}). In this scenario, the shocked wind itself remains largely adiabatic, with a cooling time much longer than the dynamical time (t$_\mathrm{cool}$$\gg$t$_\mathrm{dyn}$), so it retains its thermal energy and drives the outflow expansion.
The multiphase outflow forms primarily in the region affected by the forward shock, where the expanding hot bubble encounters and sweeps up the ambient ISM (i.e., \citealt{Zubovas14, Richings18b}). Here, the shocked ISM can cool efficiently if its radiative cooling time is shorter than the dynamical time (t$_\mathrm{cool}$$\lesssim$t$_\mathrm{dyn}$), losing thermal energy and condensing into cooler, denser phases. This process produces a multiphase structure consisting of hot X-ray emitting plasma (from the shocked wind), warm ionized gas (traced by optical lines such as [OIII] and H$\alpha$), and cold molecular material (CO, OH, etc.) originating from the radiatively cooled shocked ISM.
Hydrodynamic instabilities (Rayleigh-Taylor and Kelvin-Helmholtz) at the contact discontinuity between the shocked wind and shocked ISM (see Figure 1 in \citealt{Zubovas12}) promote turbulent mixing, enhancing the coexistence of different temperature and density phases. In contrast, the reverse shock and the inner shocked wind region remain dominated by extremely hot (T$\gtrsim$10$^9$ K) gas, which is largely adiabatic and not directly involved in forming the multiphase material.

Fast molecular outflows may result from the entrainment of pre-existing molecular clouds by a hot, high-velocity AGN wind, although simulations (e.g., \citealt{Scannapieco15, Richings18a}) show that this process is often inefficient due to cloud destruction by hydrodynamical instabilities and thermal evaporation. The destruction of these clouds would also naturally contribute to the observed deficit of molecular gas in the innermost regions of active nuclei (e.g., \citealt{AAH21, Esposito24}), while mass-loading the hot wind and promoting cooling instabilities.

\begin{figure}[!h]
\centering
\includegraphics[width=0.265\textwidth]{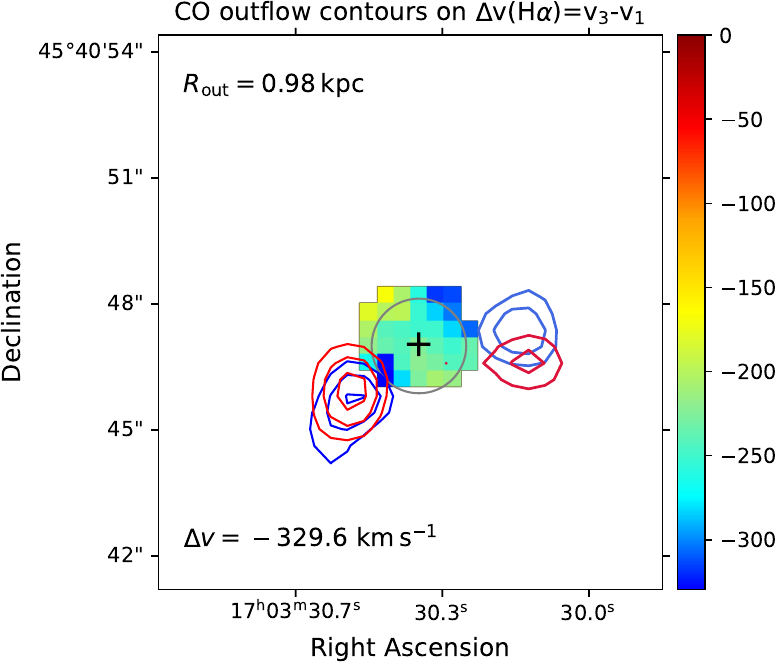}
\includegraphics[width=0.215\textwidth]{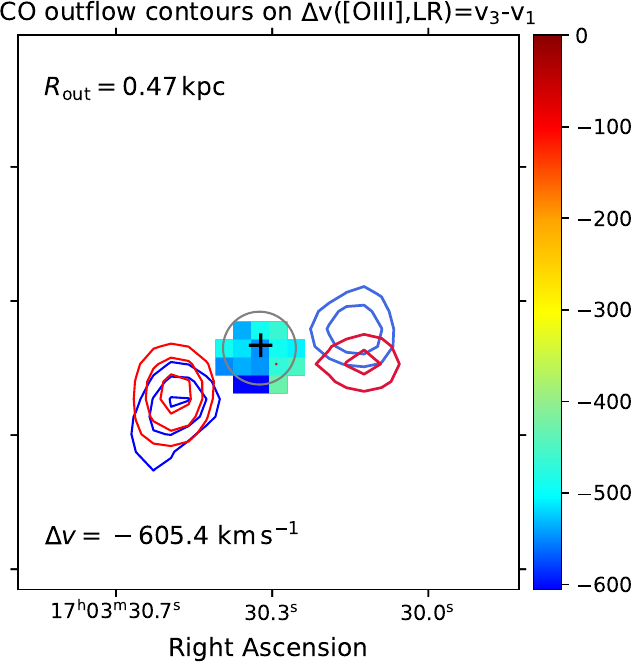}
\caption{Velocity field shift map, $\Delta$v (in km s$^{-1}$), between the outflow and primary components. These maps are derived from the H$\alpha$ tertiary component ({\it left panel}) and from the [OIII] line for the tertiary (resolved) component ({\it right panel}) using LR-V setup. A flux-emission threshold of 10\% of the peak value has been applied to the outflow kinematic maps. The grey circle marks the PSF size at 10\% of the flux peak (PSF$_{10\%}$; see text for details). The blue and red contours trace the approaching and receding molecular outflowing gas, respectively, as presented by \citet{Longinotti23} (their Fig.~4). The ionized-gas emission nicely matches the region of the molecular outflow and shows blueshifted velocities ($\Delta$v$<$0~km~s$^{-1}$; the minimum value is reported in each panel). The intrinsic outflow size for each component is also indicated in the corresponding panel.}
\label{outflow_ion_mol_bis}
\end{figure}

Molecular gas can also form {\it in situ} through the cooling of previously shocked ionized or atomic material, as proposed by \cite{Richings18a, Richings18b}. In these models, the swept-up gas behind the forward shock cools rapidly and may condense into a cold phase during the expansion of the AGN-driven wind. This picture is compatible with `energy-conserving' wind frameworks (e.g., \citealt{King15}, \citealt{Zubovas22, Zubovas24, Zubovas25}), in which the fast nuclear wind creates the conditions required for the gas to cool and form molecules within the outflow itself. Specifically: (1) the outflow begins as a hot, ionized, high-velocity wind launched close to the AGN; (2) shocks arise when this wind interacts with the surrounding ISM and the gas starts to cool; (3) in regions of higher density or enhanced thermal instabilities, the cooling becomes rapid enough to produce a cold phase; (4) thus, while pre-existing molecular clouds may be destroyed early in the interaction, the molecular component observed at larger radii is likely formed directly within the outflow through rapid cooling and condensation. This {\it in situ} formation explains the presence of molecular gas observed at large distances from the nucleus and under conditions where molecules would otherwise be expected to be destroyed, requiring reformation processes. 
This is likely to happen in IRAS17: this scenario is favored by the spatial distribution of the ionized and molecular outflows (see Fig.~\ref{outflow_ion_mol_bis}), where the ionized outflow (R$_\mathrm{ion}$$\sim$1 kpc and $\sim$0.5~kpc for the H$\alpha$ and [OIII] lines) is found within the region of the molecular outflow (R$_{CO}$=2.8$\pm$0.3~kpc; \citealt{Longinotti23}).

\cite{Faucher12} show that the shocked ambient medium cools rapidly behind the forward shock, forming a thin, dense shell prone to instabilities and fragmentation. This shell is the site where cold gas can form within the outflow. This is also supported by the work of \cite{Nims15} who claimed that most of the emission is produced by the forward shock driven into the ambient ISM.

As mentioned in the previous section, the ionized outflow velocities are consistent with the molecular outflow velocities measured with NOEMA (average combined value 1280$\pm$480 km s$^{-1}$ from \citealt{Longinotti23}), indicating that both molecular and ionized gas belong to the same multiphase AGN-driven outflow. These cold, massive clouds can survive and propagate farther from the nucleus because they are less exposed to the AGN radiation field and possess greater inertia compared to the ionized gas. Thus, the observed radial stratification reflects the survival and entrainment conditions of each gas phase rather than a temporal evolutionary pathway (\citealt{Veilleux05, Veilleux20}).
This structure is consistent with the multiphase nature of AGN-driven outflows: the ultra-fast nuclear wind (T$\sim$10$^{7-8}$ K) shocks the surrounding ISM, generating hot gas that then cools and produces ionized phases (T$\sim$10$^{4}$ K), traced by [OIII] and H$\alpha$. Cold molecular gas (i.e., H$_2$, as traced by CO, T$\sim$10-100 K) can arise either through radiative cooling of the shocked gas or via the entrainment of pre-existing molecular material.

In an `energy-conserving' outflow scenario, theoretical models predict that, in general, the velocity of different gas phases should scale inversely with their density, with the fast nuclear wind transferring energy to progressively cooler and denser material as it shocks and expands into the ISM. Accordingly, molecular gas is often expected to move more slowly than the ionized phase, although faster velocities are possible if it is efficiently entrained by the outflow.
However, in IRAS17, we found that the (average) velocity derived from the molecular CO emission is 1280$\pm$480~km~s$^{-1}$ (\citealt{Longinotti23}), which is comparable to the velocities measured in the ionized gas traced by [OIII] of the fast outflow (i.e., 3c component) or H$\alpha$. 
This result suggests that the fast AGN wind may efficiently couple to the cold molecular gas, entrain it, or that projection effects are masking intrinsic differences between the phases. In an {\it in-situ} formation scenario, the molecular gas would inherit the velocities of the shocked ionized gas from which it condensed. These observations support a scenario in which different gas phases are dynamically coupled, leading to similar outflow velocities despite their distinct physical and thermodynamic conditions.

Nowadays, most studies focus on the X-ray and molecular gas phases, while the ionized gas counterpart is often unavailable. This component is crucial for understanding the physical mechanisms driving the outflows. So far, several sources, including extreme quasars, appear to follow a `momentum-conserving' regime (\citealt{Bonanomi23, Baldini24, Zanchettin21, Zanchettin23}, see also \citealt{Marasco20} and references therein), indicating that even powerful quasars can host `momentum-driven' outflows, in contrast to predictions from theoretical models (e.g., \citealt{King15, Zubovas22}).
In particular, \cite{Marasco20} and \cite{Tozzi21} investigated the connection between nuclear X-ray UFOs and the ionized phase (H$\alpha$ and [OIII]) of large-scale outflows in bright nearby and high-$z$ quasars, respectively. They found that the ionized outflows of such quasars follow a `momentum-conserving' regime.
\cite{Travascio24} analyzed MUSE/VLT data of the radio-quiet quasar PDS 456 ($z$$\sim$0.2), a local analogue of powerful Cosmic Noon quasars (see also \citealt{Bischetti19}), which hosts an X-ray UFO and a clumpy molecular outflow extending to $\sim$5~kpc. The extended [OIII] ionized outflow shows a momentum rate consistent with an `energy-conserving' expansion. However, when the multiphase outflow (ionized + molecular) is considered, the total momentum boost indicates that the outflow is not fully in the `energy-conserving' regime. 

On the other hand, \cite{Esposito24}, studied the nearby galaxy, NGC~5506, classified as NLSy1 and hosting an X-ray UFO, as in this work. They found that the molecular and ionized outflows (traced with ALMA and MEGARA/GTC data, respectively) follow a `momentum-conserving' regime. However, the ionized outflow mass was derived only using [OIII], rather than H$\alpha$, which would have allowed a direct comparison with the [OIII] tracer.

Outflows found in nearby galaxies hosting an UFO in the `energy-conserving' regime are rare objects (i.e., IRAS11119+3257 in \citealt{Tombesi15}, Mrk~231 in \citealt{Feruglio15}, IRAS05189+2524 in \citealt{Lutz20}, and the object studied in this work), and their nature warrants further investigation. Understanding their occurrence and physical conditions is essential to constrain feedback models and their impact on galaxy evolution.

\subsection{Identifying the ionization mechanisms in the outflow: `positive feedback' in action?}

In this subsection we present the application of different diagnostic diagrams, such as WHAN (\citealt{CidFernandes11}), WHaD (\citealt{Sanchez24}), and BPT (\citealt{Baldwin81}), to identify the dominant ionization mechanisms in different regions of the galaxy, such as the disk and the outflow.

\subsubsection{WHAN diagram}
\label{positive_section}

To better understand the origin of the ionization source in the inner region of our target, we constructed the WHAN diagram \citep{CidFernandes11}, which compares the equivalent width of the H$\alpha$ line, EW(H$\alpha$), with the [NII]/H$\alpha$ flux ratio. This diagnostic helps disentangle the contributions from star formation and AGN activity, differentiating Seyfert (strong AGN) from LINER (weak AGN) regions.
According to this diagram, galaxies can be classified into different categories\footnote{\cite{CidFernandes11} define the division between weak AGNs and retired/passive galaxies at EW(H$\alpha$) = 3 \AA. Retired or passive galaxies are typically characterized by old stellar populations and exhibit very weak or undetectable emission lines.} based on their log([NII]/H$\alpha$) and EW(H$\alpha$) values: star-forming galaxies, weak and strong AGNs, and retired/passive galaxies. 
Using our spatially resolved data, we derived WHAN diagram for each individual kinematic components (i.e., primary, secondary and outflow; see \citealt{Peralta23}), as shown in Figs.~\ref{WHAN_WHAD_maps} (top panel) and \ref{WHAN_diags}.
We clearly identified that the primary component is dominated by ionization from a strong AGN. The secondary component exhibits a significant contribution from star formation in the inner regions, while AGN activity becomes more prominent in the outer parts. This star-forming component (light green area in Fig.~\ref{WHAN_WHAD_maps}), overlaps with the `butterfly' region of enhanced velocity dispersion observed in the secondary H$\alpha$ component (see Fig.~\ref{fig_LR_R_setup}). The tertiary component, which traces the outflow, is predominantly dominated by star formation throughout most of its spatial extent, with AGN-dominated ionization confined to a small region towards the south.

Deriving the continuum contribution associated with the different components to constrain the EW(H$\alpha$) parameter is challenging in IRAS17, as both star formation and AGN activity contribute to its evolution. To account for this, we corrected the total continuum emission to isolate only the star formation contribution.
In the work by \cite{Salome23} using the CIGALE code, they found that the AGN contributes approximately 43\% to the total continuum emission, leaving 57\% attributable to star formation. This fraction was used to compute the EW(H$\alpha$) values shown in Fig.~\ref{WHAN_diags}.
If the AGN contribution had not been corrected for, the resulting EW(H$\alpha$) would represent a slightly lower value and thus a lower limit; however, this correction would not alter our conclusions.
All these results suggest that the ionized outflow may be enhancing star formation by compressing the gas, thereby triggering {\it positive feedback} in IRAS17. However, based on our data, we cannot distinguish whether the observed star formation occurs {\it within} the outflow itself or is instead due to radiation from stars located in the disk plane (e.g., \citealt{Maiolino17, Gallagher19}).

Furthermore, we also note that the WHAN diagram cannot reliably distinguish between AGN excitation and shock-ionized gas, as both mechanisms can produce similarly high [NII]/H$\alpha$ ratios. Additional diagnostics, such as Baldwin, Phillips \& Terlevich (BPT) line ratios (i.e., \citealt{Baldwin81}), could help in disentangling these contributions.

\subsubsection{WHaD Diagram}

\begin{figure*}[!h]
\centering
\hskip10mm\includegraphics[width=0.91\textwidth]{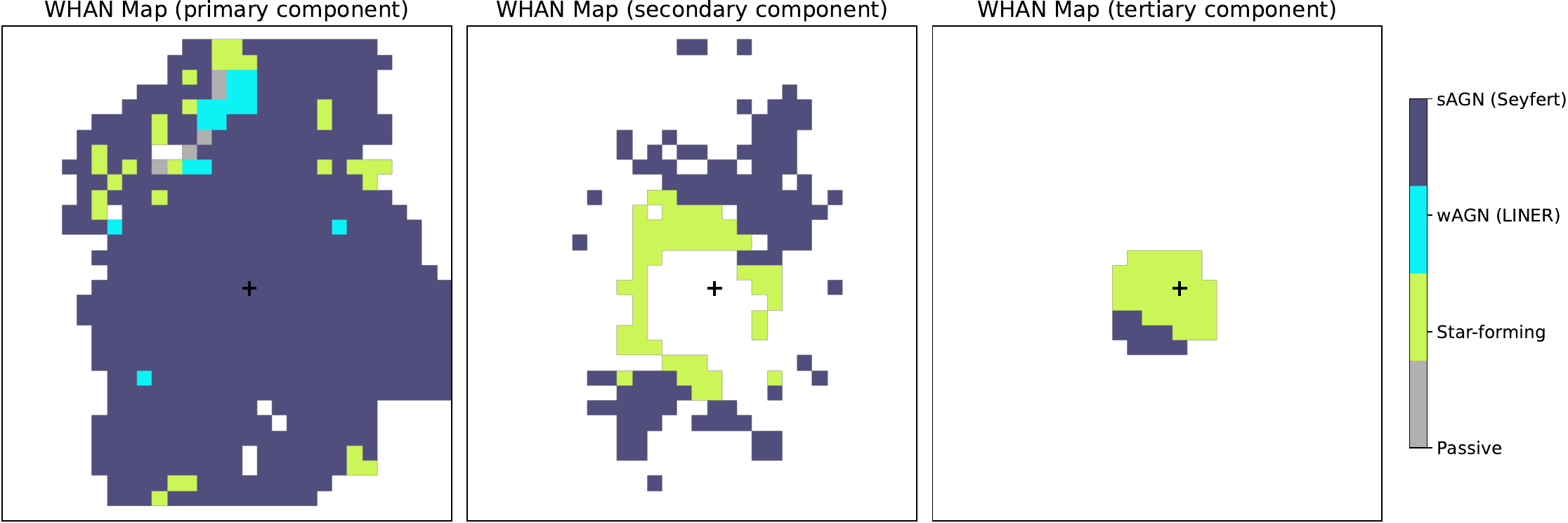}
\vskip3mm
\hskip8mm\includegraphics[width=0.9\textwidth]{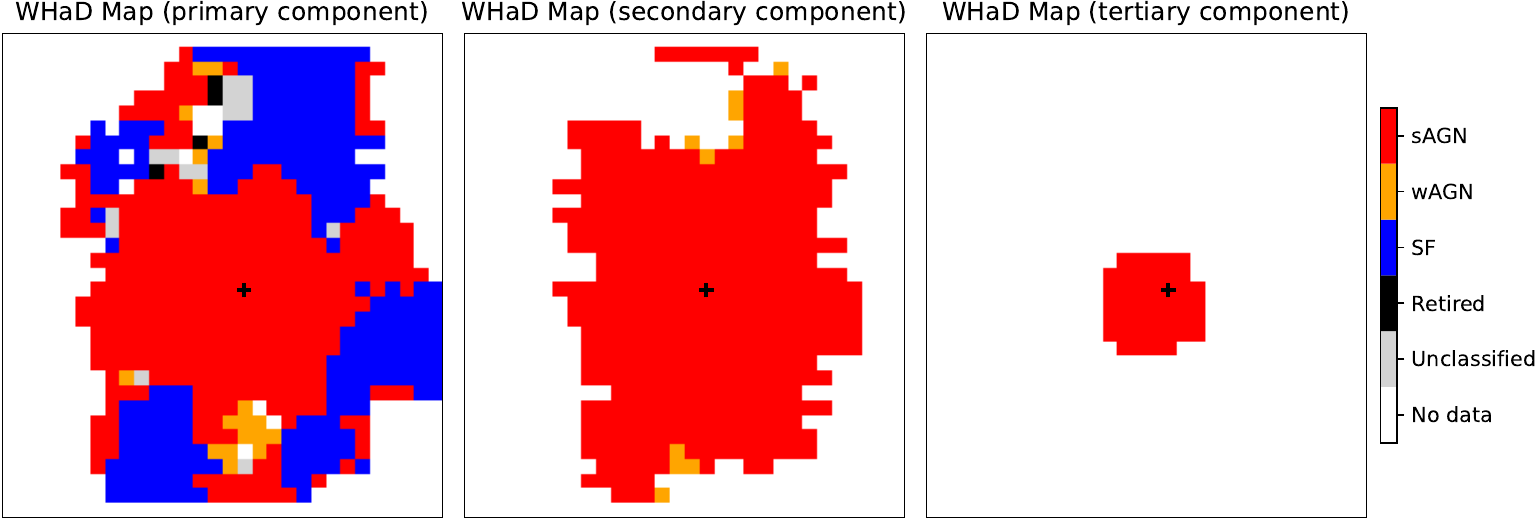}
\caption{{\it Top:} 2D maps corresponding to the WHAN regions according to \cite{CidFernandes11} classification: star-forming galaxies (SF, in light green), weak AGN (wAGN, in cyan), strong AGN (sAGN, in dark blue), and passive and retired galaxies (in grey). {\it Bottom:} 2D maps corresponding to the WHaD regions according to \cite{Sanchez24} classification: star-forming galaxies (SF, in blue), weak AGN (wAGN, in orange), strong AGN (sAGN, in red), and retired and unclassified galaxies (in black and grey).
}
\label{WHAN_WHAD_maps}
\end{figure*}

\begin{figure}[!h]
\centering
\includegraphics[width=0.47\textwidth]{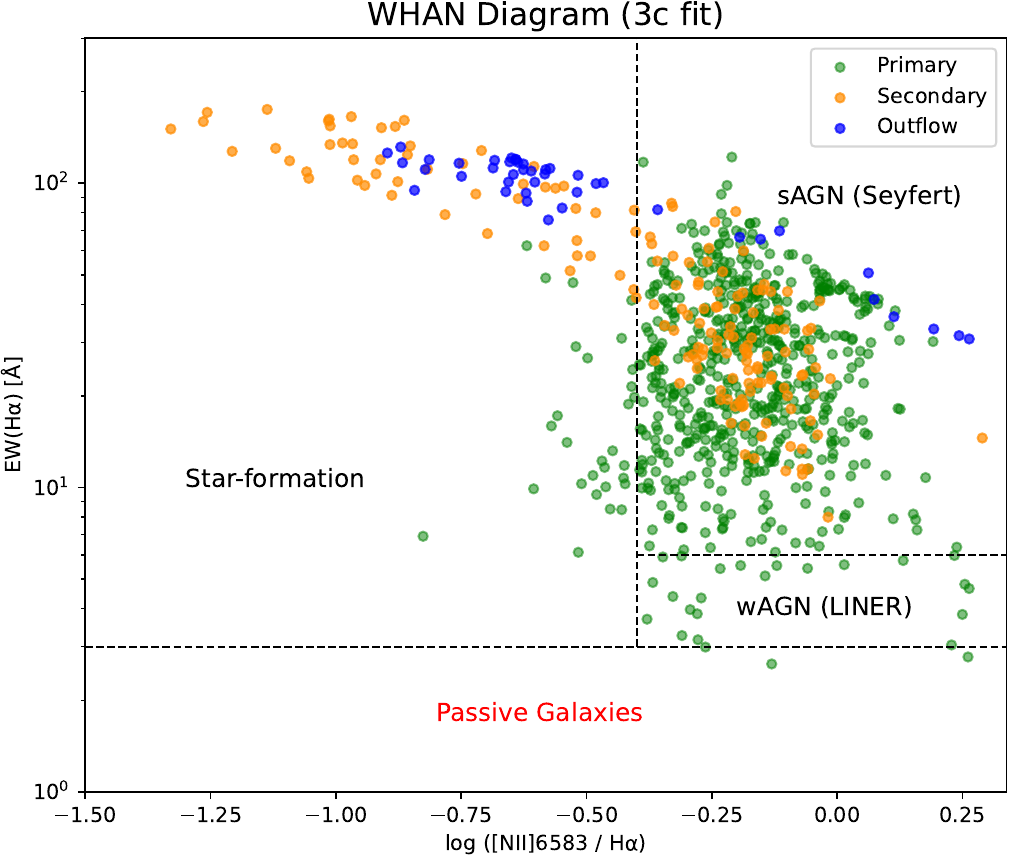}
\caption{WHAN diagram for the primary, secondary and outflow components. The continuum emission has been corrected for the AGN fraction (i.e., f$_{SF}$ = 0.57) as described in \cite{Salome23} to take into account only the contribution from SF. The main results are not affected if this correction is not considered.
}
\label{WHAN_diags}
\end{figure}

\begin{figure}[!h]
\centering
\includegraphics[width=0.5\textwidth, height=0.395\textwidth]{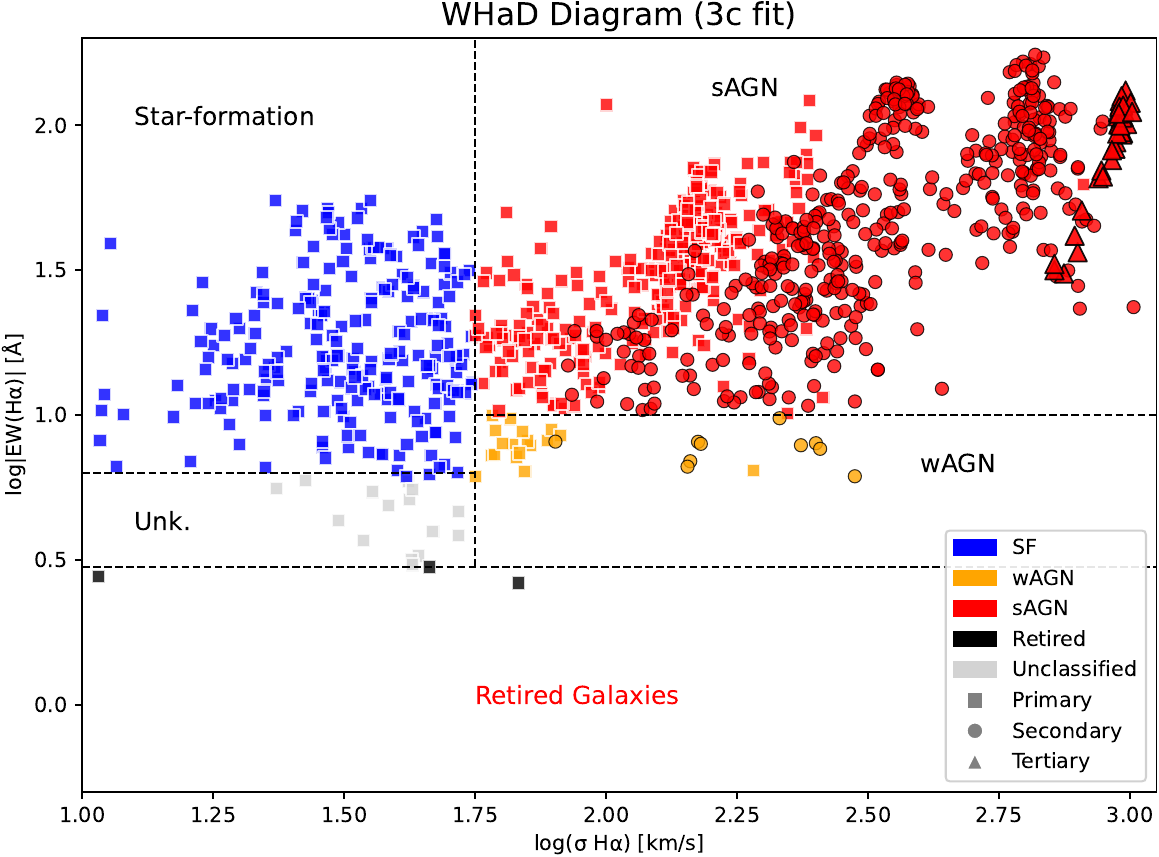}
\caption{WHaD diagrams for the primary, secondary, and outflow components. The continuum emission has been corrected for the f$_{SF}$ = 0.57 from \cite{Salome23} to take into account only the contribution from SF. The main results are not affected if this correction is not considered.
}
\label{WHaD_diags}
\end{figure}

We also explored the WHaD diagram presented by \citet{Sanchez24}, which is used to distinguish between different ionization and kinematic regimes based on the equivalent width and velocity dispersion of the H$\alpha$ line (i..e, EW(H$\alpha$)-$\sigma$(H$\alpha$)). Regions with high EW(H$\alpha$) and low $\sigma$(H$\alpha$) are typically associated with star formation due to young-massive OB-stars, tracing recent star forming activity, where the ionized gas is primarily influenced by thermal motions and turbulence driven by stellar processes. In contrast, areas with low EW(H$\alpha$) and high $\sigma$(H$\alpha$) often indicate AGN-driven outflows or (high-velocity) shock-ionized gas, where the kinematics are dominated by non-thermal processes and the contribution from young stars is minimal. Intermediate values may correspond to composite regions or partially resolved zones where multiple ionization sources coexist (i.e., `Retired Galaxies' (RG), whose ionization is due to hot old low-mass evolved stars, such as post-AGBs).
Our results, shown in Figs.~\ref{WHAN_WHAD_maps} (bottom panel) and \ref{WHaD_diags}, suggest that the H$\alpha$ kinematics in IRAS17 are too extreme to be explained by star formation alone. Consequently, most of the FoV covered by the individual components falls into the ‘AGN-dominated’ region of the diagram. This outcome is not in conflict with our previous findings, as both diagnostics are tracing different physical processes. On the one hand, the ionization indicators (e.g., the [NII]/H$\alpha$ ratio) confirm the presence of star formation in the outflow, consistent with the enhanced star formation rate derived from molecular gas \citep{Salome23}. On the other hand, the H$\alpha$ kinematics reveal that pure star formation cannot account for the observed velocity dispersions, pointing to the additional influence of AGN activity or shocks. As stated in \cite{Sanchez24}, high-velocity shocks can also reproduce the typical values found in sources hosting an AGN.
Overall, these findings indicate a complex interplay between AGN-driven, shocks and star formation processes within the same regions, highlighting the role of feedback in shaping the evolution of the host galaxy.

\subsubsection{BPT diagram(s) for the outflow}

\begin{figure}[!h]
\hskip4mm\includegraphics[width=0.41\textwidth]{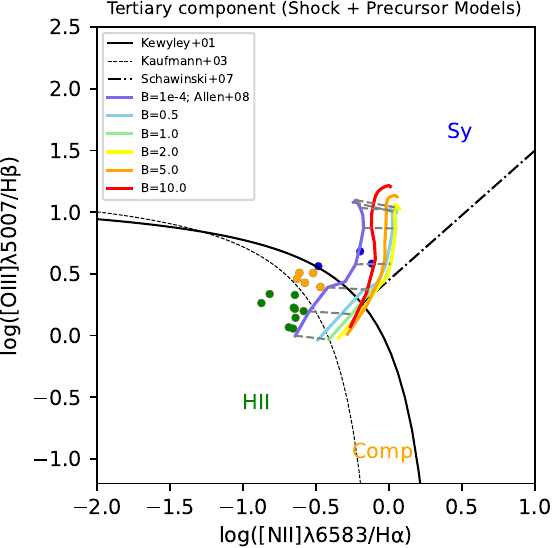}
\vskip4mm
\hskip1mm
\includegraphics[width=0.42\textwidth]{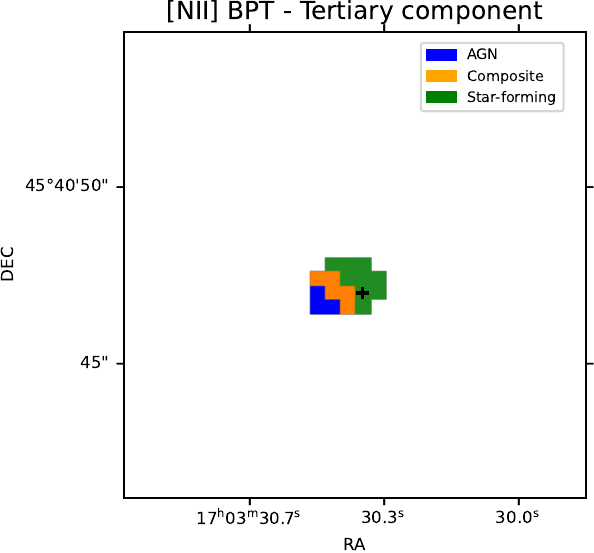}
\caption{{\it Top:} [NII]-BPT diagrams for the outflow component (i.e., tertiary component for all the lines involved). Shock+precursor grid models from \cite{Allen08} are shown assuming Z = Z$_\odot$ and different values of the shock velocities (v = 200-1000 km s$^{-1}$) and magnetic field (B = 10$^{-4}$-10 $\mu$G). The solid curve defines the theoretical upper bound for pure star formation from \cite{Kewley01}, while the dashed one in the [NII]-diagram is the \cite{Kauffmann03} empirical classification. The dot-dashed line represents the demarcation line between Seyfert galaxies and shocks or LINERs from \cite{Schawinski07}.
{\it Bottom:} 2D map derived applying the [NII]-BPT criteria for the outflowing component. The black cross identified the AGN position.
}
\label{NII_BPT_diag}
\end{figure}

Using the [NII]-BPT diagrams (e.g., \citealt{Baldwin81, Kewley01, Kauffmann03, Schawinski07}) to investigate the excitation mechanism in the outflow component (i.e., the tertiary component for all the lines involved; Fig.~\ref{NII_BPT_diag}), we find that the outflowing gas spans all three regions defined by the [NII]/H$\alpha$ and [OIII]/H$\beta$ ratios, corresponding to H II-like, composite, and Seyfert (Sy) emission. This indicates that the ionization close to the AGN is dominated by star formation, transitions through composite regions, and becomes AGN-dominated toward the south-east. When including fast radiative shocks modeled with the MAPPINGS III code (\citealt{Allen08}), we observe significant overlap with all three regions, particularly in the Seyfert-dominated regime (Fig.~\ref{NII_BPT_diag}, top). These results are consistent with those derived from the WHAN diagram and help better distinguish the ionization mechanisms near the AGN.

The observed ionization structure, with H II-like emission in the nucleus and a transition to AGN-dominated emission toward the south-east, is consistent with the extinction map. Low A$_\mathrm{V}$ in the center indicates a relatively dust-free {\it line of sight}, while higher A$_\mathrm{V}$ in the surrounding ring traces dense, dusty gas. These clouds can locally shield the compact AGN radiation, allowing stellar photoionization, which is more spatially distributed, to dominate in some regions, including the nucleus. However, some spaxels with elevated A$_\mathrm{V}$ still show AGN-dominated emission, reflecting anisotropic AGN illumination and the fact that A$_\mathrm{V}$ measures dust along our line of sight rather than the shielding between the AGN and the gas. 

The presence of broad outflowing components and the overlap with fast shock models in the BPT diagrams indicate that the outflow likely drives shocks, which contribute alongside AGN and stellar photoionization to the observed line ratios and ionization structure. Overall, the ionization structure arises from the interplay of gas density, dust distribution, local shielding, shocks, and AGN geometry.

BPT diagrams based on [OI] and [SII], often considered more reliable than [NII]-based BPTs (see discussion in \citealt{Gallagher19}), could only be fitted with a 1c model due to lower S/N in the emission. Nevertheless, they confirm the presence of SF-dominated ionization in the inner regions, supporting our interpretation.

Together, these findings indicate that the outflow simultaneously hosts star-forming regions and AGN-ionized gas, and may also drive shocks. This highlights the combined action of positive feedback (stimulating star formation) and energetic feedback (including gas ionization, shocks, and turbulence) in shaping the ionization and kinematics of the host galaxy.
~

\subsection{Energy budget and feedback mechanisms of the ionized outflow}

The kinetic coupling efficiency, $\varepsilon$, of the ionized outflow is a useful parameter to assess whether the AGN can transfer a significant fraction of its energy to the gas.  
Similarly, the momentum rate of the outflow, $\mathrm{\dot{P}_{out}}$, provides insight into the driving regime of the outflow and allows us to evaluate whether the different outflow phases (i.e., molecular and ionized) are dynamically linked.

\subsubsection{Kinetic coupling efficiency and energy transfer}

As presented in \cite{Fiore17}, an important parameter essential for understanding the strength and impact of AGN feedback is the kinetic coupling efficiency defined as the ratio between the kinetic power of the outflow and the bolometric luminosity of the source: $\varepsilon = \mathrm{\dot{E}_{out}} / \mathrm{L_{bol}}$. This parameter tells us how efficiently the energy output from the central source (AGN or starburst) is transferred into the kinetic energy of the outflowing gas. It quantifies the fraction of AGN luminosity that goes into driving the outflow. 
A high efficiency ($\varepsilon$$\geq$1\%) means the AGN is very effective at transferring its energy into feedback on the host galaxy, identifying a strong coupling. The outflow can have the potential to regulate star formation and impact galaxy evolution: this means that a significant fraction of the AGN's energy is transferred to the gas, heating it and/or expelling it. This process can disperse the cold molecular gas, heat it, or prevent its collapse, thereby suppressing star formation (quenching).
On the other hand, a low $\varepsilon$ ($\lesssim$1\%) identifies a weak coupling, where the outflow doesn't strongly impact host ISM.

In our case, we found that $\varepsilon$ ranges between $\sim$3\%-15\% for the H$\alpha$ line (using the formula by \citealt{Rupke05} for the v$_\mathrm{out}$ and considering the two electron densities), while showing a lower value for the tertiary component of the [OIII] line (i.e., from the LR setup, $\varepsilon\sim$1\%-4\%)\footnote{For the secondary component (i.e., slow outflow) observed in the LR-V setup we found a coupling efficiency $\varepsilon<0.1\%$.}, a value high enough to enable efficient energy transfer from the outflow to the ISM, potentially impacting the host galaxy (i.e., quenching star formation for $\varepsilon$$>$1\%, see \citealt{Harrison18}). 
Moreover, strong turbulence, as seen in the H$\alpha$ maps, with $\sigma$$>$120~km s$^{-1}$, can inject additional energy into the gas, preventing it from cooling and collapsing under gravity. 

These findings suggest a scenario in which AGN activity can produce both negative (suppressing star formation) and positive feedback (potentially triggering star formation in some regions; see Sect.~\ref{positive_section}), acting on different regions and timescales across the host galaxy.


\subsubsection{Momentum transfer and outflow driving mechanisms}

Another important parameter is the ratio between the momentum rate of the outflow and that supplied by the AGN (i.e., $\mathrm{\dot{P}_{out}/\dot{P}_{rad}}$), which is a powerful diagnostic of the driving mechanism and efficiency of AGN feedback. The AGN radiative momentum rate is defined as $\mathrm{\dot{P}_{rad}=L_{bol}/c=1.7\times10^{34}}$ g cm s$^{-2}$, being L$_\mathrm{bol}$=5.2$\times$10$^{44}$ erg s$^{-1}$ (\citealt{Giroletti17, Longinotti18}). 
Values close to unity ($\lesssim$1) suggest `momentum-conserving' outflows, likely driven by direct radiation pressure, while significantly higher values (a factor $\gtrsim$10) indicate `energy-conserving' outflows, in which an AGN-driven wind efficiently transfers energy to the ISM via shocks.
Ratios well below unity may imply inefficient coupling or that other processes, such as stellar winds or supernovae, are contributing to the observed outflow. Thus, this diagnostic helps distinguish between different feedback regimes and assess the AGN’s impact on galaxy evolution.
In our case, we found $\mathrm{\dot{P}_{ion}/\dot{P}_{rad}}$$\sim$13 (see Table~\ref{out_props_Ha} and Fig.~\ref{fig_energy_all}), which suggests an `energy-conserving' regime, where the AGN-driven wind shocks the ISM, forming a hot, adiabatic expanding bubble that accelerates the surrounding gas. Such momentum boosts are consistent with theoretical models of energy-driven feedback and imply that the AGN is efficiently transferring energy to the ISM, potentially regulating star formation through either negative or positive feedback (\citealt{King15} and reference therein). In `energy-conserving' outflows, the kinetic energy from the AGN is efficiently transferred through a hot wind that drives the cooler gas phases. This results in a coordinated evolution of all phases, forming a dynamically coupled, multiphase outflow. In contrast, momentum-conserving outflows may exhibit weaker coupling and more independent evolution of each phase due to rapid energy dissipation.
In our source, the comparison of the momentum rates among phases reveals that the ionized and molecular components exceed the momentum rate of the UFO by factors of $\sim$5-27 and $\sim$20-30, respectively (see Table~\ref{out_props_Ha}). These large momentum boosts reinforce the picture of an `energy-conserving' outflow where the AGN's energy is efficiently deposited into the larger-scale ISM. This implies that the different phases of the outflow are dynamically linked, supporting the idea of a sustained and powerful AGN feedback capable of shaping the host galaxy’s evolution.

\subsection{Gas depletion and Feedback timescales in the molecular and ionized outflows: witnessing `negative' feedback?}

We compared the depletion times and the mass loading factor of the ionized and molecular gas phases in IRAS17. The former parameter is defined as the ratio between the outflow mass and the mass outflow rate, $\mathrm{\tau_{depl} = \frac{M_{out}}{\dot{M}_{out}}}$, and represents the time it would take for the outflow alone to remove all the available gas (if no new gas is accreted or formed). A short depletion time implies that the outflow can rapidly deplete the galaxy's gas reservoir. 
The latter parameter is the ratio between the mass outflow rate and the SFR, $\mathrm{\eta = \dot{M}_{out}/SFR}$, and measures the strength or efficiency of the outflow relative to the galaxy's star formation activity. A value of $\eta>1$ indicates that the galaxy is ejecting gas at a rate higher than it is forming stars, suggesting strong feedback capable of regulating or even quenching star formation.
Our results may be affected by several factors, such as the assumed wind geometry and the luminosity-to-mass conversion factors, which could introduce uncertainties of up to an order of magnitude (see also \citealt{Herrera20}).

As presented in \cite{Salome23}, NLSy1 galaxies, including IRAS17, clearly show very short depletion times, $\mathrm{t_{dep, SF} = M({H_2})/SFR<0.1}$ Gyr, compared to typical star-forming disks (i.e., $\sim$2 Gyr; see \citealt{Leroy13}).
On one hand, they proposed a nuanced view of AGN feedback: AGNs can initially promote star formation, particularly through mechanical compression from jets or outflows. However, this effect may be temporary, as AGN activity eventually depletes the cold gas reservoir, leading to long-term quenching. These findings fit within a `two-phase feedback' framework, where positive and negative AGN feedback coexist but act on different timescales. On the other hand, AGN feedback may be weak, ineffective, or simply not yet acting. This challenges the traditional view that AGNs always strongly suppress star formation in their host galaxies.
Finally, as argued by \cite{Cresci18} and others, AGN feedback may not act instantaneously: instead, there could be a delay between AGN ignition and the onset of feedback effects, particularly those on larger scales, such as gas removal or heating.

For what concerns the depletion time of the molecular outflow, we derived a depletion time, $\mathrm{\tau_{depl}(H_2) = \frac{M_{out}(H_2)}{\dot{M}_{out}(H_2)}}$, in IRAS17 of $\sim$0.7 Myr from \cite{Longinotti23}, based on a molecular gas mass of $\sim$10$^8$ M$_\odot$ and a mass outflow rate of 139 M$_\odot$ yr$^{-1}$. This value is shorter than the dynamical timescales of the outflow, calculated as t$_\mathrm{dyn}$=R$_\mathrm{out}/$v$_\mathrm{out}$=2.8 ~kpc/1280~km~s$^{-1}$$\sim$~2.1~Myr, suggesting that the molecular gas can be largely expelled before the outflow propagates through the entire observed region. For the ionized gas, we computed a depletion time of $\sim$0.61~Myr, using the H$\alpha$ ionized gas mass of $0.4\times10^7$​~M$_\odot$ and an outflow mass rate of 6.4 M$_\odot$ yr$^{-1}$. 
Its dynamical timescales is comparable, $\mathrm{t_{dyn}\sim0.66~Myr}$ (i.e., from the ratio $\mathrm{0.98~kpc /1461~km ~s^{-1}}$). 
The comparable values of $\tau_\mathrm{depl}$ and $t_\mathrm{dyn}$ indicate that the outflow can efficiently remove the ionized gas from the host region within its dynamical timescale.

The fact that depletion times are comparable to or shorter than the dynamical times indicates high efficiency in gas removal from the galaxy's central regions. In the molecular phase, depletion occurs on timescales shorter than the gas travel time, indicating that dense, star-forming gas is preferentially expelled, favoring negative feedback. In the ionized phase, the close match between depletion and dynamical times reinforces that the outflow is effective across gas phases. Combined with ionization diagnostics showing coexisting star formation, these results support a dual-feedback scenario: the outflow both removes a substantial gas reservoir (negative feedback) and compresses pockets of gas to trigger star formation (positive feedback).

Furthermore, the presence of line diagnostics indicating star formation in parts of the ionized gas suggests that the molecular and ionized phases are not independent. They are likely co-spatial or dynamically coupled: ionized winds may entrain molecular clouds or compress them, launching molecular gas outward or triggering localized star formation. In this scenario, while the molecular phase is more vulnerable to depletion, the ionized phase acts as the primary conduit for AGN energy, coupling it into the multiphase outflow and regulating the balance between suppression and triggering of star formation.

The mass loading factor, $\mathrm{\eta}$, derived using a SFR=26~M$_\odot$ yr$^{-1}$ (see \citealt{Salome21}), highlights the dominant role of the cold gas. We derived $\mathrm{\eta_{mol} \sim 5.3}$ for the molecular gas, indicating that the outflow is removing gas at more than five times the SFR, while the ionized gas shows a lower $\mathrm{\eta_{ion} \sim 0.3}$, insufficient by itself to regulate star formation. These results suggest a scenario in which AGN feedback initially affects the ionized and/or atomic gas, which can subsequently cool and form molecular clouds. The later expulsion of this molecular gas by the outflow represents the primary negative feedback mechanism, effectively suppressing star formation in the host galaxy.

In agreement with our results, \cite{Herrera20} derived the mass loading factors for different gas phases in the nearby ULIRG, IRAS F08572+3915. They found that the molecular phase dominates the mass loading compared to the atomic and ionized phases, and is the only phase where the gas ejection rate exceeds the rate of molecular gas consumption, yielding $\mathrm{\eta_{mol} > 1}$. Overall, in their source, the molecular gas is the dominant component driving the outflow and regulating star formation, whereas the ionized phase contributes less to both the mass budget and feedback efficiency, as similarly observed for IRAS17 in this study.

\subsection{Can the outflow(s) escape the galaxy?}

To assess whether the ionized outflow can effectively remove gas from the galaxy, we modeled the kinematics of the primary (systemic) H$\alpha$ emission using {\it kinemetry} (e.g., \citealt{Krajnovic06}; see also \citealt{Bellocchi12}). This approach allowed us to determine the disk inclination and derive the intrinsic circular velocity, $v_\mathrm{{circ}}$, which in turn was used to estimate the escape velocity, $v_\mathrm{{esc}}$, as a function of radius. By comparing the measured outflow velocities from H$\alpha$ with $v_\mathrm{{esc}}$, we evaluated whether a significant fraction of the ionized gas can reach or exceed the escape velocity, thus revealing the capability of the AGN-driven wind to expel gas from the galaxy.

{\it Kinemetry} models the observed line-of-sight velocity field by expanding it along concentric elliptical annuli using a Fourier series. The zeroth-order term, $A_0$, represents the systemic velocity, while the first-order terms, $A_1$ and $B_1$, describe the bulk rotational motion. The amplitude of the first harmonic, $k_1 = \sqrt{A_1^2 + B_1^2}$, provides a direct measure of the observed rotational velocity, $v_\mathrm{{obs}}$, projected along the line of sight. Following \cite{Bellocchi12}, we employed a ten-term harmonic expansion in our analysis and adopted a COVER parameter of 0.75\footnote{The COVER parameter of 0.75 indicates that the program stops if fewer than 75\% of the points along an ellipse are covered by data.}.
The geometry of each annulus is defined by its position angle, $\Gamma$, and flattening, $q = b/a$, where $a$ and $b$ are the semi-major and semi-minor axes, respectively. Under the assumption of a thin disk, the flattening is related to the disk inclination as $\cos i$$\sim$q.
The intrinsic circular velocity is then obtained by correcting for inclination as $v_\mathrm{circ}(R) = k_1(R)/\sin i = k_1(R)/\sqrt{1 - q^2}$. The resulting inclination-corrected rotation curve is used to estimate the escape velocity of the galaxy, $v_\mathrm{{esc}}$, assuming a spherically symmetric potential modeled as a singular isothermal truncated sphere (e.g., \citealt{Bischetti19, Veilleux20, Marconcini25}): $v_\mathrm{esc}$(R)$\approx$$\sqrt{2}\, v_\mathrm{circ}$(R). This provides a first-order assessment of whether the ionized gas motions remain gravitationally bound.

\begin{figure}[!h]
\includegraphics[width=0.48\textwidth]{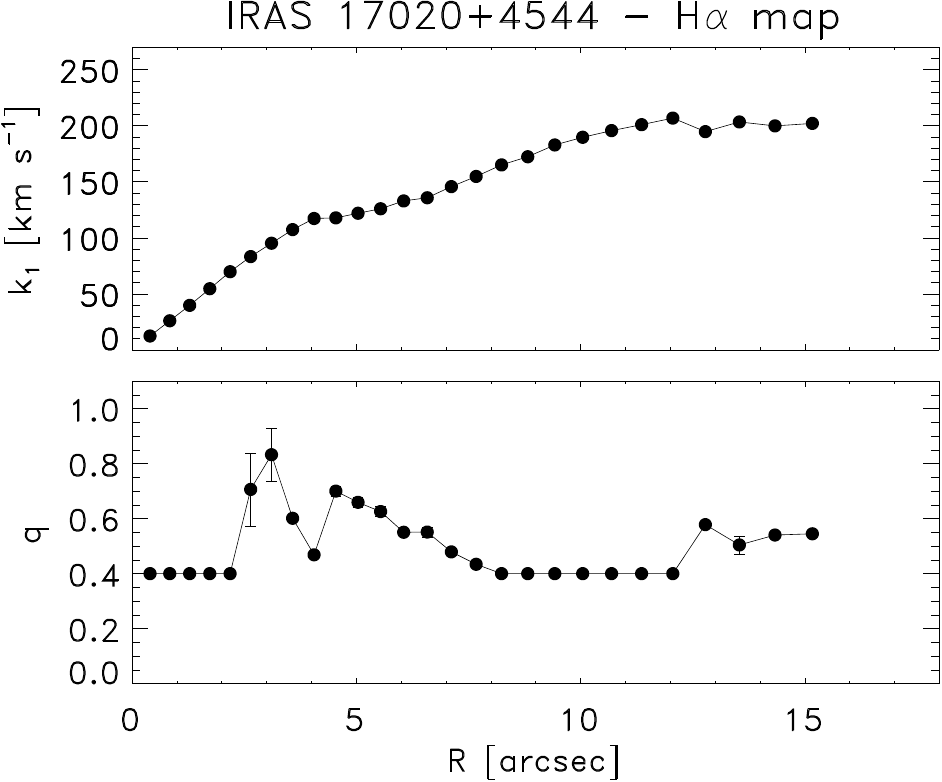}
\caption{Observed rotational curve parameter, $k_1$ (top), and the flattening, $q$, of each ellipse (bottom), extracted at different radii using \textit{kinemetry}. Error bars are smaller than the symbol size.}
\label{kinemetry_fig}
\end{figure}

In our case, we considered the last five data points of the first harmonic amplitude, $k_1$, which traces the observed rotational velocity and gives an observed circular velocity of 201.4 km s$^{-1}$, and we then corrected it for the mean inclination derived using such values, from which we derived a mean flattening of $q$$\sim$0.5; see Fig.~\ref{kinemetry_fig}). This yields an intrinsic circular velocity of $v_\mathrm{circ}$$\sim$268.5 km s$^{-1}$. Assuming a simple spherical potential (\citealt{Veilleux20}), this corresponds to an escape velocity of $v_\mathrm{esc}$$\sim$380~km s$^{-1}$. 
The outflow velocity traced by H$\alpha$ reaches $\sim$1450 km s$^{-1}$, with a comparable value derived from [OIII]. These velocities are well above $v_\mathrm{esc}$ (nearly four times higher), suggesting that a substantial fraction of the ionized gas is likely unbound ($v_\mathrm{out}$$\gg$$v_\mathrm{esc}$). These results demonstrate that the outflow is sufficiently powerful to remove gas from the galaxy's potential.

\section{Summary and main conclusions}
\label{sect_summary}

We presented new GTC/MEGARA optical IFU observations of the nearby NLSy1 galaxy IRAS~17020+4544, one of the few known sources exhibiting both an X-ray ultra-fast outflow (UFO; \citealt{Longinotti15}) and a molecular outflow (\citealt{Longinotti23}) consistent with an `energy-conserving' regime. This galaxy also shows the presence of an elongated VLBI radio jet (\citealt{Giroletti17}) that may be inducing shocks in the ISM.

Using MEGARA/GTC data in both low-resolution (LR; R$\sim$6000) and medium-resolution (MR; R$\sim$12000) modes, we characterized the ionized outflow in the galaxy's central region by analyzing the H$\alpha$ and [OIII] emission lines. The ionized outflow is confined within the more extended molecular outflow. By deriving its kinematic and energetic properties (velocity, mass, kinetic power, and momentum), we assessed the energy budget of the outflow and compared it with that of the molecular and X-ray phases. 

The main results of this work are summarized below:

\begin{itemize}

\item The H$\alpha$ and [OIII] emission line profiles are well fitted with up to three Gaussian components, primarily detected in the central regions where the ionized outflow is located. We identify three components:

(1)~a primary (or narrow) component, present across the entire line emitting region, identified with the systemic component, characterized by relatively low velocity dispersions ($\sigma_{\rm mean}$$\sim90$-100 km s$^{-1}$) and a clear rotational pattern. Peculiar features in the H$\alpha$ velocity dispersion map, such as an `X-shaped' enhancement, may indicate turbulence driven by the outflow or by the jet-ISM interactions. (2)~a secondary `broad' component of the H$\alpha$ line covers roughly the same area as the systemic component and exhibits complex kinematic structures, with high velocity dispersions ($\sigma_{\rm mean}$$\sim$380 km s$^{-1}$). In [OIII], this component traces an ionized outflow with modest velocities within an intrinsic radius of $\mathrm{R_{out}}\sim0.7$~kpc.
(3) a tertiary `very broad' component identifies the most extreme ionized outflow, present in both H$\alpha$ and [OIII]. In H$\alpha$, its extension is $\mathrm{R_{out}}$=0.94$\pm$0.32 kpc, while in the (LR-V) [OIII] emission it is more compact ($\mathrm{R_{out}}\sim0.5$~kpc).

\item The fast ionized outflow traced by the H$\alpha$ and [OIII] lines (tertiary component) and the slower ionized outflow detected only in [OIII] (secondary component) are confined within the extent of the molecular outflow observed in CO ($\mathrm{R_{CO}}$=2.8$\pm$0.3 kpc). The [OIII] outflow associated with the secondary component is resolved in both MR-G and LR-V setups, while the faster outflow is only resolved in the LR-V setup. The presence of multiple [OIII] velocity components indicates a stratified ionized gas structure: fast emission originates from dense, shock-compressed gas near the wind-ISM interface, whereas slower emission arises from more diffuse, partially cooled gas at larger radii. This stratified structure was previously observed in IRAS17 in the X-ray band, confirming a multi-phase, multi-component wind.

\item We confirm that the ionized gas phase follows an `energy-conserving' regime, as previously found for the molecular phase (\citealt{Longinotti23}). Both the H$\alpha$ and fast [OIII] outflows exhibit large momentum boosts, consistent with efficient energy deposition from the AGN into the large-scale ISM. These results suggest that different gas phases are dynamically linked, indicating powerful AGN feedback capable of significantly influencing the host galaxy's evolution.

\item Diagnostic diagrams, depletion timescales, and mass-loading factors consistently point to a dual-feedback scenario. On the one hand, ionizing radiation detected in the outflow and in the `butterfly' region, together with the enhanced SFR inferred from the molecular gas, indicates localized, outflow-triggered star formation (positive feedback). This suggests that the outflow compresses the surrounding gas, triggering small-scale episodes of star formation in IRAS\,17020+4544. On the other hand, the extreme kinematics of the ionized gas, which cannot be explained by stellar processes alone, together with the short depletion times and high molecular mass-loading factor ($\eta>$1), reveal efficient removal or heating of the cold gas reservoir (negative feedback). The high velocities and energetics of both the ionized and molecular outflows indicate that the outflow is primarily AGN-driven, with only a minor contribution from star formation.

While positive feedback may occur locally, our results show that negative feedback dominates globally and on longer timescales, effectively suppressing star formation across the host galaxy. This dual behaviour underscores the complex and multifaceted role of AGN-driven winds in regulating galaxy evolution.

\item Comparing the H$\alpha$ outflow velocity with the escape velocity derived from its primary component, we found that the ionized outflow is capable of leaving the galaxy, since $v_\mathrm{out} \gg v_\mathrm{esc}$. This result supports a scenario in which AGN-driven winds can deplete the host galaxy's ISM and regulate star formation.

\end{itemize}

Our results shed light on the complex feedback processes occurring in this extreme and peculiar source. To better understand the interplay between different gas phases and spatial scales (i.e., UFOs, ionized outflows, and molecular outflows), studies of a larger, more complete sample of NLSy1 galaxies hosting X-ray UFOs are needed. Such a sample will allow testing whether these systems are consistent with theoretical models of AGN feedback (e.g., \citealt{Zubovas12, King15, Zubovas22}).

\begin{acknowledgements}

We thank the referee for a constructive report that helped improve the paper. EB, AGP, CCT, ACM acknowledge support from the Spanish grants PID2022-138621NB-I00 and PID2021-123417OB-I00, funded by MCIN/AEI/10.13039/501100011033/FEDER, EU. ALL acknowledges support from DGAPA-PAPIIT IA103625. YK acknowledges support form UNAM-PAPIIT grant 102023. 
This research has made use of the NASA/IPAC Extragalactic Database (NED), which is funded by the National Aeronautics and Space Administration and operated by the Jet Propulsion Laboratory, California Institute of Technology, under contract with the National Aeronautics and Space Administration. 
This paper is based on observations made with the Gran Telescopio Canarias (GTC), installed in the Spanish Observatorio del Roque de los Muchachos of the Instituto de Astrofísica de Canarias, in the island of La Palma. 
This work is based on data obtained with MEGARA instrument, funded by European Regional Development Funds (ERDF), trough Programa Operativo Canarias FEDER 2014-2020.

\end{acknowledgements}

\bibliographystyle{aa} 
\bibliography{biblio}

\clearpage

\appendix

\section{Deriving the extinction map, A$_\mathrm{V}$}
\label{more_info_AV}   

We calculated the visual extinction, A$_\mathrm{V}$, from the Balmer decrement H$\alpha$/H$\beta$, assuming the \citet{Calzetti00} attenuation law with R$_\mathrm{V}$ = 4.05, typical of starburst galaxies. An intrinsic ratio of H$\alpha$/H$\beta$ = 3.1 and a fixed electron temperature of T$_\mathrm{e} = 10^4$~K were adopted. Based on the interpretation of the kinematic components presented in Tables~\ref{tab_Ha}, \ref{tab_MR}, and \ref{tab_LR}, the extinction map was derived using the H$\beta$ flux from the MR-G data. This configuration allows us to resolve multiple kinematic components, and the H$\beta$ line is well covered within the observed wavelength range. In contrast, in the LR-V setup only a single component could be fitted, as the line lies near the blue edge of the spectrum, making its kinematics less reliable.

\begin{figure*}[!h]
\centering
\includegraphics[width=0.85\textwidth]{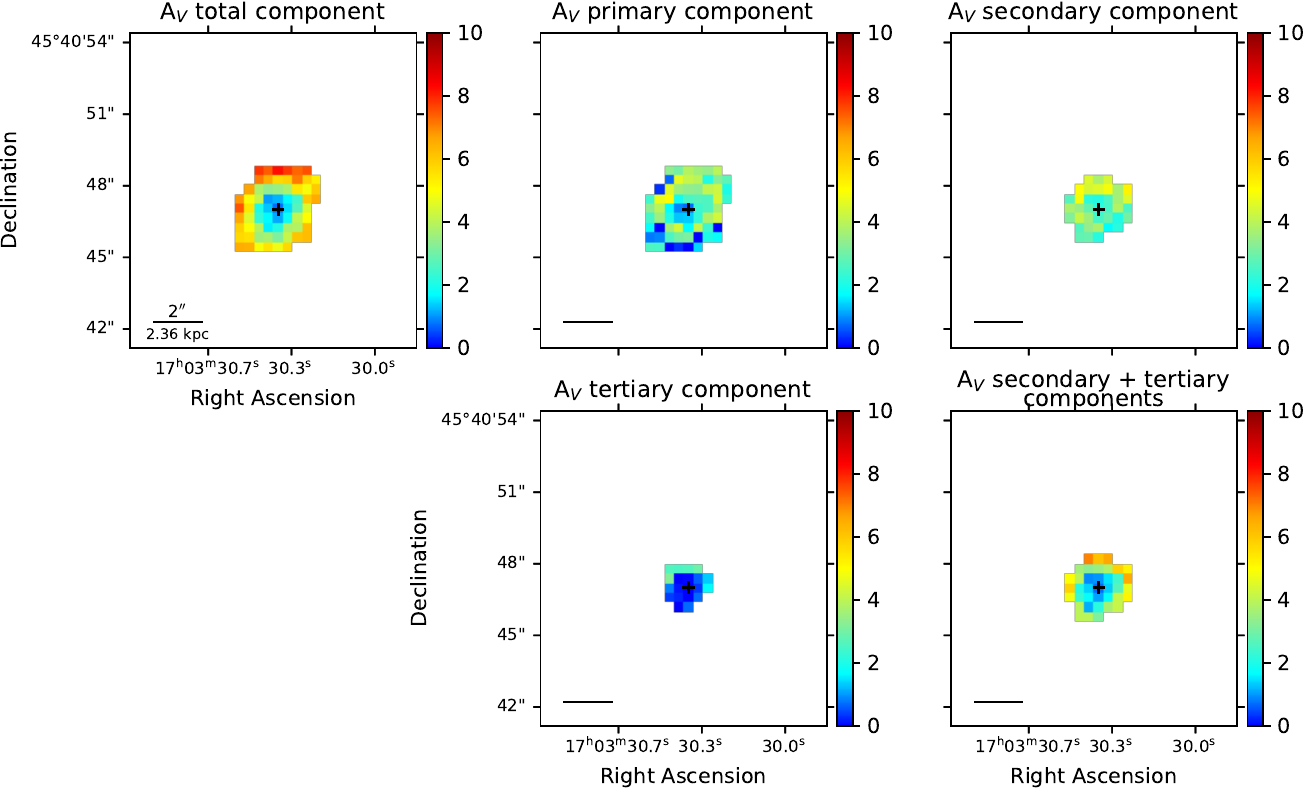}
\caption{Extinction maps, A$_\mathrm{V}$, derived from the H$\alpha$ and H$\beta$ emission-line maps. {\it From left to right, top to bottom:} A$_\mathrm{V}$ obtained from the total, primary, secondary, tertiary, and combined (secondary and tertiary) components. The black cross marks the continuum peak emission. The combined map (i.e., $\mathrm{A_V}$ secondary+tertiary components) was finally adopted for estimating the outflow visual extinction, A$_\mathrm{V}$.}
\label{AV_map}
\end{figure*}

We derived the extinction of each component using the integrated flux maps of H$\alpha$ and H$\beta$ for the primary (systemic), secondary and tertiary (outflow) components. We also computed A$_\mathrm{V}$ using the total flux of the three components and by combining the H$\alpha$ and H$\beta$ fluxes of the secondary and tertiary components only (i.e., $\mathrm{A_V}$(secondary+tertiary); see bottom right panel in Fig.~\ref{AV_map}).
The results show moderately different extinction values across components: the primary and secondary components display relatively high extinctions (A$_\mathrm{V}$ $\sim$2.8 and 3.5 mag, respectively), while the tertiary (outflowing) component exhibits a significantly lower value (A$_\mathrm{V}$=1.24 mag). When combining the secondary and tertiary components or using the total flux, we obtained values comparable to that of the primary component (A$_\mathrm{V}$=2.73 mag).

The primary and secondary components present similar, relatively high extinctions, suggesting that both are located behind comparable dust columns within the host galaxy or circumnuclear region. In contrast, the tertiary (outflowing) component shows much lower extinction, possibly indicating that the outflow lies on the near side of the dust distribution or has partially cleared its line of sight. However, the lower A$_\mathrm{V}$ of the outflow may also reflect the reduced SNR of its H$\beta$ flux and uncertainties in line decomposition, continuum placement, or aperture coverage. Consequently, this value should be regarded as an indicative lower limit rather than a precise measurement.

Studies on dust extinction in outflows {\it versus} galactic disks yield mixed results. Some works (e.g., \citealt{Rose18, Mingozzi19, Fluetsch21}) found that outflows are less affected by extinction (as in our case), while others (e.g., \citealt{Holt11, VM14, Rodriguez19}) reported higher extinction in the outflows.
Indeed, in some systems, outflows may be intrinsically dustier than the disk. \cite{Fluetsch21} found a correlation between the difference in extinction between the outflow and the disk and the ionized outflow gas mass: the most massive ionized outflows tend to have comparatively more dust than the surrounding disk.
In such cases, the preferential expulsion of dusty gas may enhance the extinction in the outflow while simultaneously lowering it in the disk. This provides further evidence that dust plays a key role in galactic outflows, particularly in the most massive ones. In particular, this finding supports models in which galactic winds are driven by radiation pressure on dusty clouds (e.g., \citealt{Fabian12, Ishibashi18}).
  
All A$_\mathrm{V}$ maps reveal a clear ring-like structure with high extinction (A$_\mathrm{V}$ $\gtrsim$~4 mag), likely tracing regions of enhanced dust content. The inner region of the A$_\mathrm{V}$ total map, however, shows extinction values similar to those of the outflowing (tertiary) component.
We finally adopt A$_\mathrm{V}$ = 2.73 mag as a representative extinction value for the ionized outflow.

Although the stellar continuum was not subtracted from our spectra, we acknowledge that this could affect the derivation of the extinction A$_V$, particularly the flux of the H$\beta$ line, which is more strongly affected by underlying stellar absorption than H$\alpha$. This leads to an overestimated H$\alpha$/H$\beta$ ratio and, consequently, to an artificially high extinction correction (A$_\mathrm{V}$). 
In our case, we verified that using the extinction derived from the tertiary (outflowing) component (A$_\mathrm{V}$=1.24 mag), instead of the higher value obtained by combining the secondary and tertiary flux emissions (A$_\mathrm{V}$ = 2.73 mag), would reduce the estimated outflow momentum rate ratio (i.e., $\mathrm{\dot{P}_{\rm out}/\dot{P}_{\rm rad}}$) from $\sim$13-61 to $\sim$4-20, depending on the adopted electron density (n$_\mathrm{e}$ = 2500 and 500 cm$^{-3}$, respectively).

However, given the uncertainties on our measured quantities, the resulting values remain consistent within the errors and do not affect the main physical interpretation or conclusions of our analysis.

\section{Additional kinematic maps}
\label{more_maps}

\begin{figure*}[!h]
\centering
\includegraphics[width=0.85\textwidth]{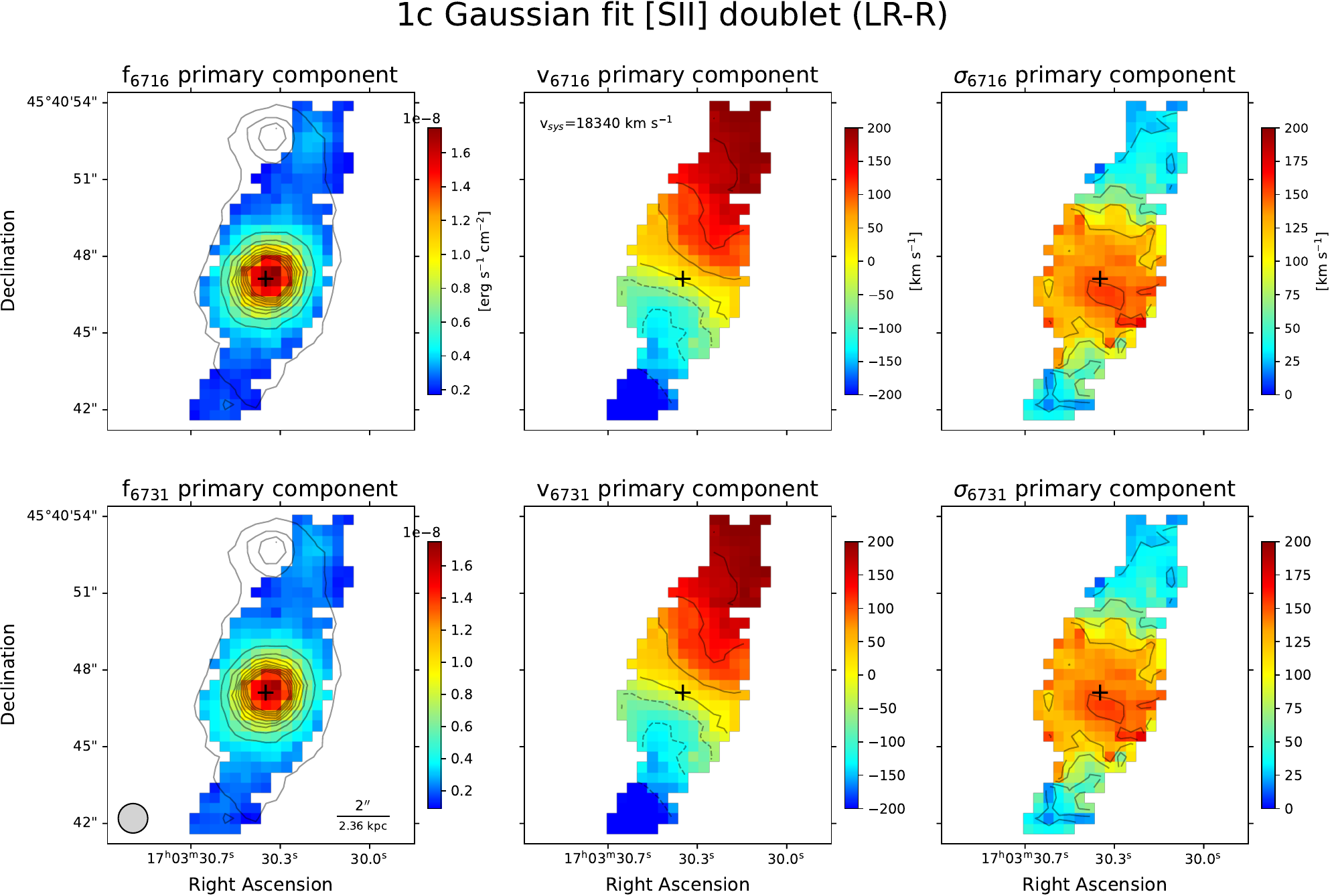}
\vskip1.5cm
\includegraphics[width=0.85\textwidth]{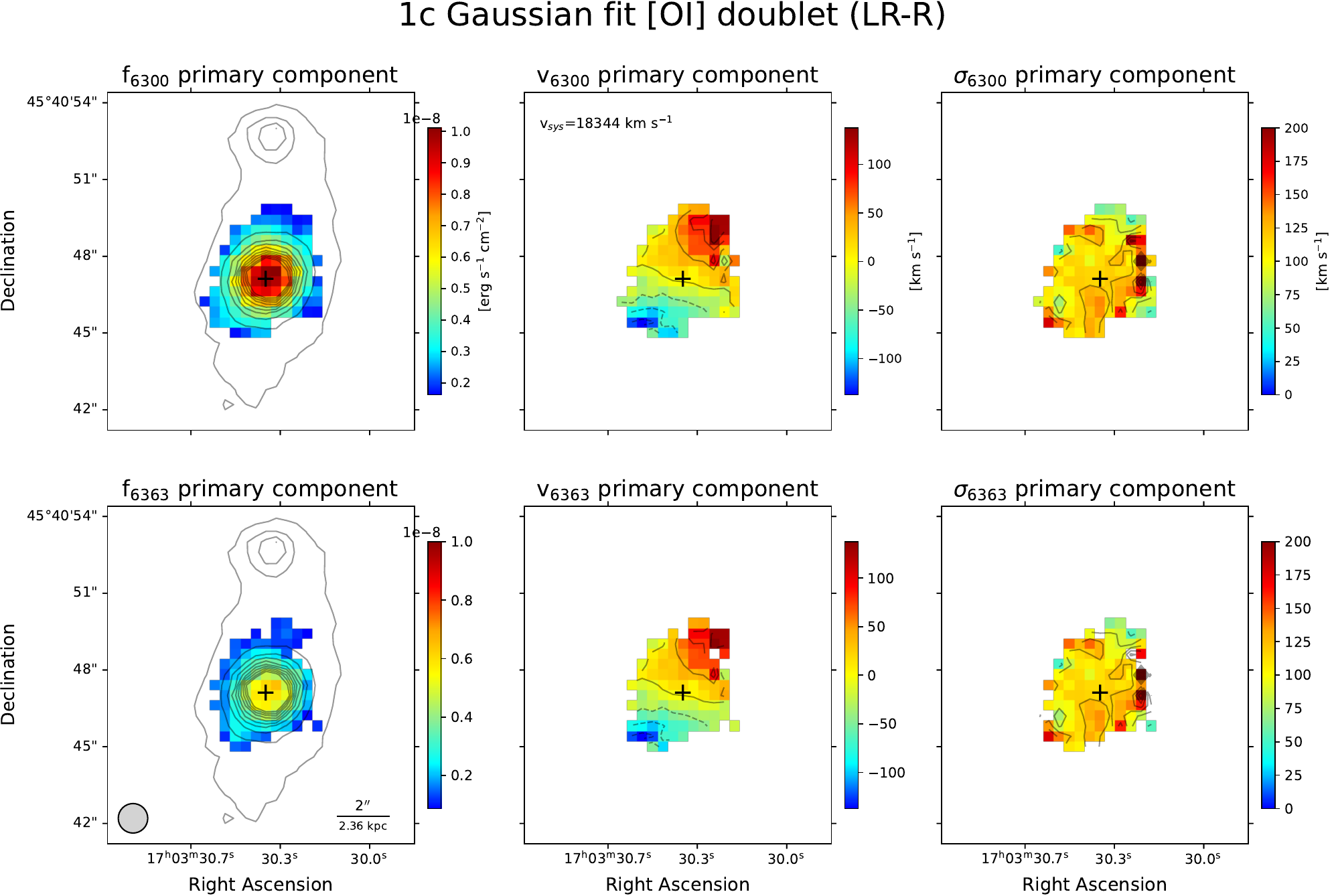}
\caption{Kinematic maps of the [SII]6716,6731 (top) and [OI]6300,6363 (bottom) doublets. Same figure caption as in Fig.~\ref{fig_LR_R_setup}.}
\label{LR_R_S2_O1}
\end{figure*}

In this appendix, we present the kinematic maps derived from the one component (1c) fits to the [SII] and [OI] doublets observed in the LR-R setup.
The results are shown in Fig.~\ref{LR_R_S2_O1}.

For the [SII] doublet, both the velocity field and the velocity dispersion maps (Fig.~\ref{LR_R_S2_O1}, top) follow the same kinematic pattern observed for the molecular gas traced by CO(1-0) (\citealt{Salome21}). An enhanced velocity dispersion is found in the inner region, particularly along the rotational axis, and a clear rotational pattern is also visible in the velocity field.
Some spaxels in the companion galaxy also show [SII] emission, suggesting the presence of high-ionization regions, possibly associated with star-forming activity or shocks. AGN activity appears unlikely, given the absence of [OIII] emission in that region.

For the [OI] doublet (Fig.~\ref{LR_R_S2_O1}, bottom), its emission is detected only in the nuclear region, where it follows the same rotational pattern seen in the [SII] maps, with a velocity dispersion distribution similar to that of the [SII] doublet.
~

\section{Revealing ionized gas emission from the companion galaxy}
\label{app_companion}

\cite{Salome21}, using NOEMA data, found that IRAS17 is interacting with a small companion galaxy located at a similar redshift. This interaction had not been previously identified, as IRAS17 appears to be an undisturbed barred spiral galaxy in optical images (\citealt{Ohta07}). It remains unclear whether the companion’s flux emission corresponds to the ‘comma-like’ (tidal tail) feature visible in the NOT optical image, located just near the foreground star, as shown in Fig.~\ref{cont_maps}.

With our optical data, it is difficult to disentangle the contribution of the companion from that of IRAS17 using the H$\alpha$ line alone. However, guided by the CO(1-0) systemic flux emission reported by \cite{Salome21}, we identified the region likely associated with the companion and performed a two-component fit to the H$\alpha$ spectra. The companion shows a slightly higher redshift than IRAS17, with $z \sim 0.0618$ as derived by our MEGARA/GTC data in the H$\alpha$ emission. A secondary H$\alpha$ component is detected toward the southwest, redshifted relative to the systemic component and exhibiting a low velocity dispersion ($\sigma \lesssim 100$~km~s$^{-1}$), possibly indicating a connection with IRAS17 and a low level of turbulence (see discussion in Sect.~\ref{description_Ha}).

The companion also exhibits emission in other lines, such as [OIII], H$\beta$ (detected only in the LR-V setup), and [SII]. A tentative BPT classification based on the [SII]/H$\alpha$ ratio suggests that the companion is dominated by star formation. Based on our kinematic results, we propose that the companion may have interacted with IRAS17 in the past, and that we are now witnessing the residual effects of this merger in the ionized gas phase, which does not appear to be strongly perturbed. A more detailed investigation of this source is beyond the scope of this work.

\end{document}